\documentclass[iop]{emulateapj}
\usepackage{enumerate}
\usepackage{bm}
\usepackage{amsmath}
\usepackage{color,bm}
\usepackage{CJK}
\usepackage{calligra}

\newcommand{\tr}[1]{\textrm{#1}}

\def\bfnabla{{\mbox{\boldmath $\nabla$}}}

\renewcommand\bv{{\mbox{\boldmath $v$}}}
\newcommand\bb{{\mbox{\boldmath $B$}}}

\newcommand\bn{{\mbox{\boldmath $n$}}}

\newcommand\bF{{\mbox{\boldmath $F$}}}
\newcommand\bfr{{\sf\boldmath f}}
\newcommand\bI{{{\sf\boldmath I}}}

\newcommand\bq{{\mbox{\boldmath $q$}}}
\renewcommand\bf{{\mbox{\boldmath $f$}}}
\def\<{\,\langle\langle}
\def\>{\,\rangle\rangle}

\DeclareMathOperator{\atantwo}{atan2}
\DeclareMathOperator{\erfc}{Erfc}

\begin{document}
\begin{CJK*}{UTF8}{gbsn}

\shortauthors{Jiang \& Oh}
\author{Yan-Fei Jiang(姜燕飞)\altaffilmark{1} \& S. Peng Oh\altaffilmark{2}}
\affil{$^1$kavli institute for theoretical physics, Kohn Hall, University of California, Santa Barbara 93106, USA} 
\affil{$^2$Department of Physics, University of California, Santa Barbara, CA 93106, USA}

\title{A New Numerical Scheme for Cosmic Ray Transport}

\begin{abstract}

Numerical solutions of the cosmic-ray (CR) magneto-hydrodynamic equations are dogged by a powerful numerical instability, which arises from the constraint that CRs can only stream down their gradient. The standard cure is to regularize by adding artificial diffusion. Besides introducing ad-hoc smoothing, this has a significant negative impact on either computational cost or complexity and parallel scalings. We describe a new numerical algorithm for CR transport, with close parallels to two moment methods for radiative transfer under the reduced speed of light approximation. It stably and robustly handles CR streaming without any artificial diffusion. It allows for both isotropic and field-aligned CR streaming and diffusion, with arbitrary streaming and diffusion coefficients. CR transport is handled explicitly, while source terms are handled implicitly. The overall time-step scales linearly with resolution (even when computing CR diffusion), and has a perfect parallel scaling. It is given by the standard Courant condition with respect to a constant maximum velocity over the entire simulation domain. The computational cost is comparable to that of solving the ideal MHD equation. We demonstrate the accuracy and stability of this new scheme with a wide variety of tests, including anisotropic streaming and diffusion tests, CR modified shocks, CR driven blast waves, and CR transport in multi-phase media. The new algorithm opens doors to much more ambitious and hitherto intractable calculations of CR physics in galaxies and galaxy clusters. It can also be applied to other physical processes with similar mathematical structure, such as saturated, anisotropic heat conduction. 
	

\end{abstract}


\maketitle

\section{Introduction}

In recent years, there has been a strong resurgence of interest in the impact of cosmic-rays (CRs) on galactic scales. CRs have long been known to be important in ISM and proto-planetary disks, by ionizing gas in otherwise neutral regions shielded from photoionizing radiation, thus affecting both chemistry and coupling to magnetic fields. However, the fact that their local energy density is comparable to the thermal energy density also implies that they could be an important source of heat and momentum driving. For instance, in the intracluster medium, CRs have been invoked as a means of stemming a cooling flow catastrophe \citep{loewenstein91,GuoOh2008,fujita11, pfrommer13, jacob17a, ruszkowski17}. In the ISM, CR wave heating potentially provides the required supplemental heating to explain observed line ratios \citep{reynolds99} in the warm ionized medium of the Galaxy \citep{wiener13-ISM}. Most recent activity, however, has focused on the exciting possibility that CRs could drive the galactic winds we observe, both directly by exerting a force, and indirectly by heating the gas and affecting thermal pressure gradients. Star formation feedback is a key outstanding problem in galaxy formation, with no consensus solution. Cosmic-ray mediated feedback is a very attractive possibility because unlike thermal driving, there is no danger of rapid dissipation via radiative cooling. It also allows for winds which are significantly cooler than thermal winds, in better agreement with observations, which see a significant amount of warm ionized, $\sim$10$^{4}$K gas \citep{chen10,rubin10,tumlinson11}. Furthermore, in agreement with observations (cf \citealt{steidel10}), and in contrast to models of thermally driven winds, CR winds have velocities that rise with distance. Thus, although the possibility of CR driven winds was noted early on \citep{ipavich75,breitschwerdt91}, these arguments have motivated a great deal of recent work on CR feedback, both analytically \citep{socrates08,samui10,recchia16}, and numerically with isotropic \citep{uhlig12,booth13,salem14,simpson16,Wieneretal2017} and field-aligned \citep{hanasz13,girichidis16,pakmor16,ruszkowski17} CR transport (for an excellent recent review, see \citealt{zweibel17}). However, all of these works are subject to important caveats about assumptions in CR transport.

Despite their small collisional cross-section, cosmic rays do not stream out of our Galaxy at the speed of light. Instead they are remarkably isotropic (to several parts in $\sim 10^{4}$), despite the discreteness and transience of supernova sources. Analysis of CR spallation products and radio nuclide abundances indicate residence lifetimes of $\sim 10^{7}$ yr (at 1 GeV), orders of magnitude larger than the light travel time across the Galaxy. These facts led to the `self-confinement' theory for CRs, whereby they scatter collisionlessly off magnetic irregularities in the plasma generated by a powerful resonant streaming instability \citep{kulsrud69,wentzel74}, which kicks in whenever CRs stream faster than the Alfv\'en speed.  They can also potentially be scattered by MHD turbulence \citep{yan02}. The growth rate of the streaming instability is proportional to the CR abundance. For the scenarios we are interested in, where the CRs are abundant enough to be dynamically important, self-generated turbulence should dominate by far. The CR generated waves damp in the plasma, resulting in momentum and energy transfer from the CRs to the plasma. 

The mean free path of CRs, $r_{\rm L}/(\delta B/B)^{2}$ (where $r_{\rm L}$ is the Larmor radius and $\delta B/B$ is the rms fractional B-field perturbation) is usually much smaller than any other lengthscale of interest. Thus, in most situations it is appropriate to treat the CRs as a fluid. High wave-particle scattering rates render CRs almost isotropic in the frame of the Alfven waves (which has velocity $\bv + \bv_A$, the sum of the local gas and Alfven velocities). It therefore is most expedient to evaluate the Vlasov equation in the wave frame, and expand the distribution function in inverse-powers of the CR scattering rate. Expanding to second order and averaging over pitch angle yields the advection diffusion equation \citep{Skilling1971}: 
\begin{equation}
\begin{split}
\frac{\partial  f_\tr{p}}{\partial t}+(\bv+\bv_s)\cdot\nabla  f_\tr{p}&=\nabla\cdot(\kappa_\tr{p} \bn\bn\cdot\nabla  f_\tr{p})\\ &+\frac{1}{3}p\frac{\partial  f_\tr{p}}{\partial p} \nabla\cdot (\bv+\bv_s)+ Q.
\end{split}
\label{eqn:crevol}
\end{equation}
Here, $ f_\tr{p}(\mathbf{x},p,t)$ is the cosmic ray distribution function (isotropic in momentum space), $\bv$ is the gas velocity, $\bv_s= - {\rm sgn}(\bb\cdot \nabla P_{\rm c}) \bv_A$ is the streaming velocity (equal in magnitude to the Alfven velocity, but pointing down the CR gradient along the B-field), $\bn$ is a unit vector pointing along the magnetic field, and $ Q$ is the cosmic ray source function. Equation \ref{eqn:crevol} is straightforward to interpret. The left-hand side is the total time derivative, including advection with the Alfven waves. Due to the finite wave-particle scattering rate, the CR distribution function has finite anisotropy, which to second order results in diffusion relative to the wave frame (first term on RHS). Since there are no electric fields in the wave frame, CRs evolve adiabatically\footnote{This is of course not true in the lab frame, where CRs irreversibly transfer energy to the gas.}, as expressed in the second term on the RHS. Taking moments of equation \ref{eqn:crevol} then yields the CR hydrodynamic equation (equation \ref{CR_oldeq}), which requires an expression for the first moment, the flux $\bF_c$. This is customarily expressed as the sum of streaming and diffusion terms (equation \ref{CR_oldflux}), though we shall soon see that this is flawed.  

In practice, solving equation \ref{CR_oldeq} suffers from two difficulties. The first is physical, and stems from our limited knowledge of diffusion coefficients. The standard practice in galaxy formation simulations is to adopt a constant $\kappa$ independent of plasma properties, scaled to empirical Milky Way values. This is unsatisfactory for obvious reasons. The next level of sophistication is to solve for $\kappa$ analytically in quasi-linear theory, by balancing wave growth and damping rates, and finding the equilibrium wave amplitude \cite{kulsrud69}; see \cite{wiener13} for some recent examples). We discuss this further in the Appendix. While the resultant wave amplitudes are usually small ($\delta B/B \sim 10^{-3}$ for ISM-like conditions), suggesting that quasi-linear theory should be valid, this calculation has yet to be performed numerically. The quasi-linear estimates appear inconsistent with the low observed CR anisotropy at $E > 100$ GeV, unless some other source of scattering is present \citep{farmer04}. At high energies, extrinsic turbulence likely dominates the scattering rate, and the situation is much more uncertain\footnote{At the $\sim$GeV energies which dominate the CR energy budget, the self-confinement scenario is on fairly secure ground, though there could be surprises.}. The cross-field diffusion rate is also highly uncertain. Unresolved magnetic structure can also contribute to effectively diffusive behavior, through field-line wandering. Further theoretical progress -- likely fueled by particle-in-cell simulations -- is needed here. 

The second stumbling block, however, is purely numerical. It arises from the unusual advection velocity $\bv_s= - {\rm sgn}(\bb\cdot \nabla P_{\rm c}) \bv_A$, which depends on the direction of the CR gradient. This reflects the physical constraint that net streaming of CRs only occurs down their density gradient.\footnote{It is possible for CRs to propagate up their density gradient if they are scattered by extrinsic turbulence. In this case, energy is transferred from the waves to the CRs rather than vice-versa, as in second-order Fermi acceleration.} Unstable resonant growth of the Alfv\'en waves which scatter and advect CRs only occurs when they co-propagate with CRs; otherwise they are damped. Numerically, the sharp step-function-like discontinuity in the streaming velocity near extrema, where it flips sign, leads to grid-scale oscillations due to overshoot as CRs stream away from maxima or toward minima. These rapidly grow and swamp the true solution. This numerical instability can be regularized by adding a numerical diffusion term, with an ad-hoc smoothing parameter \citep{Sharmaetal2009}, similar to adding explicit viscosity to enable shock capturing in hydrodynamics. An explicit solver of course requires $\Delta t \propto (\Delta  x)^{2}$ for diffusive behavior, but an implicit solver in principle only requires a timestep scaling $\Delta t \propto \Delta x$. Thus far, this is the only method known to tame this powerful numeric instability, and it has been used in all calculations which attempt to model CR streaming. 

In practice, this has several dissatisfactory features. It adds an uncontrolled amount of numerical diffusion to our solutions, rather than implementing the physical diffusion we wish to model. A linear timestep scaling is not realized in practice; due to the larger diffusion rates at extrema in higher resolution calculations, significantly shorter timesteps are required. Indeed, the amount of smoothing required to obtain converged results is only obtained by trial and error, requiring convergence tests with respect to both resolution and smoothing parameter. The maximum allowable smoothing is a function of the size of the simulation domain, making the results box-size dependent \citep{Sharmaetal2009}. Furthermore, the large matrix inversions required for implicit methods introduce significant additional complexity. In short, regularization is expensive and of uncertain reliability, particularly in stiff situations where CR scale heights have a wide dynamic range. Thus, for instance, there are no published time-dependent simulations of streaming in multi-phase gas with resolved boundary layers. In practice, researchers has hitherto solved CR transport only in the limit of either pure streaming or pure diffusion. As of this writing, there are no galaxy scale calculations of CR transport which solve the advection diffusion equation in full generality, introducing significant caveats into any conclusions drawn.  

In this paper, we exploit parallels between CR transport and radiative transfer, using established, well-tested methods for solving the radiative transfer equation in full generality \citep{Jiangetal2012} to do likewise for the CR transport. The key new feature is that instead of treating CR energy density as the only independent variable, we now solve time-dependent equations for both CR energy density and flux. This is similar to the step forward from the traditional diffusion approximation for radiative transfer to the two moment approach as in \cite{Jiangetal2012}.
 In a sense, this is hardly surprisingly, since both are relativistic fluids, although there are some subtleties in the CR case which differ from the radiative transfer case. This enables fast, robust calculations with a standard Courant time-step with no ad-hoc artificial diffusion and simultaneous calculation of both streaming and physical diffusion. This enables hitherto intractable calculations. 

The outline of this paper is as follows. In \S\ref{sec:equation}, we review the original equations used for CR transport, and then introduce the new, revised set of equations we solve. In \S\ref{sec:implementation}, we describe how we implement our algorithm in the public available radiation MHD code {\emph Athena++}. In \S\ref{sec:test}, we show results from an extensive series of tests of the new method. Finally, we discuss implications in 
\S\ref{sec:discussion} and conclude in \S\ref{sec:conclusion}. In the Appendix, we discuss physical values for diffusion coefficients derived from quasi-linear theory.

\section{Governing Equations}
\label{sec:equation}
\subsection{The Original CR Transport Equation}

In the traditional approach, only the CR energy density is evolved by assuming that CR flux is always a function of CR energy density itself. The standard equations are \citep{mckenzie82,breitschwerdt91}: 
\begin{eqnarray}
\frac{\partial E_c}{\partial t}=\left(\bv+\bv_s\right)\cdot\nabla P_c-\bfnabla\cdot\bF_c+Q.
\label{CR_oldeq}
\end{eqnarray}
Here $E_c$ and $P_c$ are the CR energy density and pressure respectively. The streaming velocity $\bv_s$ is usually calculated as 
\begin{eqnarray}
\bv_s=-\bv_A\frac{\bb\cdot\bfnabla P_c}{|\bb\cdot\bfnabla P_c|}.
\label{eq:vs}
\end{eqnarray}
Here $\bv_A=\bb/\sqrt{4\pi\rho}$ is the Alfv\'en velocity. The streaming velocity $\bv_s$ is along the magnetic field lines and takes the direction down the CR pressure gradient. It has the magnitude of the local Alfv\'en speed. 
The CR pressure $P_c$ is usually assumed to be isotropic 
and taken to be $P_c=E_c/3$ as CRs are ultra relativistic particles. The term $Q$ is to account for additional 
energy source of CRs such as shock acceleration in addition to the compression or expansion that is already 
included by the term $\left(\bv+\bv_s\right)\cdot\nabla P_c$. 

The CR energy flux $\bF_c$ is usually written as
\begin{eqnarray}
\bF_c=\left(E_c+P_c\right)\left(\bv+\bv_s\right)-\bn\left(\bn\cdot\bfnabla P_c\right)/\sigma_c^{\prime}.
\label{CR_oldflux}
\end{eqnarray}
The first part is the advective flux due to flow velocity $\bv$ and streaming velocity $\bv_s$ while the second term is  
the diffusive flux along the direction of local Alfv\'en velocity $\bn\equiv \bv_A/|\bv_A|$ with a diffusion coefficient 
$\sigma_c^{\prime}$. For a length scale $l$, $1/(l\sigma_c^{\prime})$ is the CR diffusion speed, which is equivalent to the photon diffusion speed $c/\tau$ with an optical depth $\tau$ for radiative transfer, where $c$ is the speed of light. 

The CR transport equation \ref{CR_oldeq} is very similar to the radiative transfer equation under diffusion approximation \citep{TurnerStone2001}. The additional terms that are unique to CRs are the terms related to $v_s$. The CR energy flux $\bF_c$ can also be 
compared with the lab frame radiation flux, which is the sum of the diffusive flux in the co-moving frame of the fluid and the advective flux. 
CRs do work on the fluid, which causes energy exchange between CRs and gas, when they move down the gradient of $P_c$.  

\subsection{New Equations for CR Transport}
As explained in the introduction, the above equations are obtained by taking moments of the CR distribution function $f_\tr{p}(\mathbf{x},p,t)$ over the momentum space. It is an approximate approach because the time dependent term of the first moment $\bF_c$ is completely neglected, which is usually a good assumption when there are strong interactions between CRs and magnetic fields. This is very similar to the diffusion approximation for radiative transfer, which only works in the optically thick regime for photons. It is usually assumed that this will simplify the numerical simulations as we only need to solve one equation for $E_c$. However, it turns out the advection term associated with the streaming velocity $\bv_s$ causes numerical problems, as explained in the introduction. \cite{Sharmaetal2009} proposes adding a regularization term, which introduces artificial diffusion near the region where $\bfnabla P_c$ is close to zero. 
The amount of diffusion in this scheme is a free parameter chosen to suppress the  unphysical oscillations. 

Given the above numerical issue with the traditional equation (\ref{CR_oldeq}), we take one step back to 
the full equations describing the  CR energy density and flux, without 
assuming that CR flux takes its steady state value. Inspired by the two moment equations of radiative transfer \citep{Stoneetal1992,SekoraStone2010,Jiangetal2012}, we solve the time dependent equation 
for both $E_c$ and $\bF_c$ as follows: \footnote{This set of equations can also be derived following \cite{Skilling1971} by taking moments of the Boltzmann equation and explicitly evaluating the interaction coefficient in quasi-linear theory (E. Ostriker, private communication).}
\begin{eqnarray}
\begin{split}
\frac{\partial E_c}{\partial t}+\bfnabla\cdot\bF_c&=\left(\bv+\bv_s\right)\cdot\left(\bfnabla\cdot {\sf P_c}\right)+Q,\\
\frac{1}{V_m^2}\frac{\partial \bF_c}{\partial t}+\bfnabla\cdot{\sf P_c}&=
-{\sf \sigma_c}\cdot\left[\bF_c-\bv\cdot\left(E_c{\sf I}+{\sf P_c}\right)\right].
\end{split}
\label{neweq}
\end{eqnarray}
where we shall define the interaction coefficient $\sigma_{\rm c}$ in the ensuing discussion. The left hand side of equation \ref{neweq} describes the transport of CRs in the lab frame, which comes from the direct quadrature of $f_\tr{p}$ in the momentum space. In the absence of source terms on the right hand hand, they represent conservation of CR energy and momentum respectively. The maximum velocity CRs can propagate is $V_m$, which is a constant through the entire simulation domain. 
In principle, this should be the speed of light. However, we shall demonstrate that as long as $V_{\rm m}$ is much larger than any other signal speed in the simulation, our results are not sensitive to the exact value of $V_{\rm m}$. We thus instead choose $V_m$ to be a value much smaller than the speed of light but still much larger than the maximum Alfv\'en and flow velocities in the simulations. This is important because the time step of our numerical scheme is determined by the CFL condition for $V_m$. This is also similar to the reduced speed of light approximation used in radiative transfer equations \citep{GnedinAbel2001,SkinnerOstriker2013}. 
More discussion of the effects of $V_m$ is in \S\ref{sec:discussion}. 

The right hand side of equation \ref{neweq} represents source and sink terms for the CR energy density $E_{\rm c}$ and flux $F_{\rm c}$ respectively. 
The term for the energy density is obvious: besides explicit energy sources $Q$, there is also an overall sink term since CR forces do work on the gas at a rate $\bv \cdot \nabla P_c$ and heat the gas at a rate $\bv_s \cdot \nabla P_{\rm c}$. However, the appropriate source term $S_{\rm F}$ for the flux is less obvious. Physically, it must reduce to two limits: in steady state, when $\partial \bF_c/\partial t =0$, it must give the canonical expression for $\bF_c$ given in equation \ref{CR_oldflux}. Also, when there is no CR scattering, either because there is no wave growth (e.g., when $\nabla P_{\rm c}=0$), or wave damping is extremely strong, it must asymptote to zero -- without scattering, the flux equation is conservative. 

An obvious ansatz which satisfies the first requirement is: 
\begin{equation}
S_{\rm F} = - \sigma_c^{\prime} \left[\bF_c-\left(\bv+\bv_{\rm s} \right) \cdot\left(E_c{\sf I}+{\sf P_c}\right) \right].
\label{eqn:wrong_source_term}
\end{equation}
While our motivation for writing this is simply to reproduce equation \ref{CR_oldflux} when $\partial \bF_c/\partial t =0$, this source term is actually physically well-motivated, and indeed takes the same form as the equivalent source term for momentum flux density in the two moment radiative transfer equations. It consists of an interaction coefficient $\sigma_{\rm c}^{\prime}$ multiplied by the flux in the frame of the waves, $\left[\bF_c-\left(\bv+\bv_{\rm s} \right) \cdot\left(E_c{\sf I}+{\sf P_c}\right) \right]$. The interaction coefficient $\sigma_{\rm c}^{\prime}$ is computed in the same wave frame (note that in equation \ref{CR_oldflux}, $\sigma_{\rm c}^{\prime}$ parametrizes diffusion relative to the wave frame). Since scattering is nearly isotropic in the wave frame, it makes sense to evaluate the scattering rate (and hence diffusion coefficient) there. Note that the LHS of the flux equation is evaluated in the lab frame, but if we use equation \ref{eqn:wrong_source_term} as the source term, then the RHS is evaluated in the wave frame. This can be justified rigorously, and such a mixed frame approach is common in radiation hydrodynamics \citep{SekoraStone2010}, where the equations are correct to $O(v/c)$. 

However, this form of the source term does not obviously satisfy the second requirement that $S_{\rm F} \rightarrow 0$ as $\nabla P_{\rm c}\rightarrow0$. In fact, if one calculates the diffusion coefficient in quasi-linear theory for self-excited waves,  then $\sigma_c^{\prime} \propto \nabla P_{\rm c}/P_{\rm c} \Gamma_{\rm damp} \rightarrow 0$ (see equation \ref{eq:kappa}), which also satisfies the second requirement. By contrast, calculations which naively assume a constant $\kappa$ (e.g., based on empirical diffusion coefficients measured in the Milky Way), clearly do not. 
Using quasi-linear diffusion coefficients is the most straightforward way to cast the two moment equations. In this paper, we have chosen to write the flux equation in a slightly more general form which does not explicitly invoke quasi-linear theory, but still obeys the two limits mentioned above. 

In particular, note that evaluating $S_{\rm F}$ in the wave frame does not make sense at extrema: the wave frame does not exist, since there are no waves! (or at least, an equal number of forward and backward propagating waves). Since we can pick any frame to evaluate $S_{\rm F} = \sigma_{\rm c} \cdot {\mathbf F}_{\rm c}$, it makes most sense to do so in the fluid frame, which is always defined. The scattering rate (and hence diffusion term) in the fluid frame is more complicated, and is in general a tensor. However, one can find the diagonal terms by rearranging terms from equation \ref{CR_oldflux}:
\begin{equation}
\bF_c - \bv\left(E_c+P_c\right)=- \left[\frac{1}{\sigma_c^{\prime}} - \frac{(E_{\rm c}+P_{\rm c})\bv_s}{\bfnabla P_c}\right]\bfnabla P_c,
\label{CR_newflux}
\end{equation}
where we can identify the term on the LHS as the flux in the fluid frame, and the term in the brackets on the RHS as the effective interaction coefficient in this frame. 
This form of the equations satisfies both asymptotic limits 
mentioned, and explicitly enforces these limits even when diffusion coefficients which do not depend on $\bfnabla P_{\rm c}$ (e.g., $\sigma_{\rm c}^{\prime}=$const) are used. This has the useful property that our code remains stable in such limits, which is necessary when comparing against certain previous calculations. 

However, note that this is not a strict hyperbolic conservation law for the flux, with particles propagating at speed $V_{\rm m}$. As the CR gradient flattens, the source term $ S_{\rm F} \propto \bfnabla P_{\rm c} (\bF_c - \bv\left(E_c+P_c\right))$ goes to zero at the same rate as the divergence term on the LHS, and thus must be retained. The main new important feature in the two-moment equations is that for: 
\begin{equation}
\frac{c \Delta t}{L_{\rm z}}  < \frac{v_{\rm A}}{c} 
\label{eq:breakdown}
\end{equation}  
where $L_{\rm z} \equiv P_{\rm c}/|\nabla P_{\rm c}|$ and $\Delta t$ is the time step, the time-dependent term in equation \ref{neweq} cannot be ignored. This will always happen in the neighborhood of extrema, where $L_{\rm z} \rightarrow \infty$. A reduced speed of light $V_{\rm m}$ simply increases the size of the region where this happens, easing timestep and resolution requirements. 

Thus, there are two ways of viewing the failure of the canonical set of equations near extrema. The first is to note that the wave frame is undefined, since no waves are excited when $\nabla P_{\rm c} = 0$ and CRs do not stream in a preferred direction; thus, the standard approach of calculations with respect to the wave frame fails. Alternatively, one can note that the standard equation for the flux, equation \ref{CR_oldflux}, implies CR propagation at a velocity $\sim O(v_{\rm A})$ even as $\nabla P_{\rm c} \rightarrow 0$.\footnote{This is obvious for $\sigma_{\rm c}^{\prime}$=const, and even in quasi-linear theory, the drift relative to the Alfven frame is independent of $\nabla P_{\rm c}$ (equation \ref{eq:drift_velocity}) for linear damping mechanisms.} In steady state, a current with drift speed $v_{\rm A}$ implies a spatial anisotropy of $O(v_{\rm A}/c)$, which is inconsistent with spatial isotropy as $\nabla P_{\rm c} \rightarrow 0$. However, once the inequality in equation \ref{eq:breakdown} is satisfied, equation \ref{CR_oldflux} is no longer valid. 


In principle we need a closure relation to determine the CR pressure tensor ${\sf P_c}$ to close this set of equations. For the regimes we are most interested in, we assume CRs are close to be isotropic so that ${\sf P_c}\equiv P_c{\sf I}=E_c{\sf I}/3$, where 
${\sf I}$ is the unit tensor. Relaxing this assumption in different regimes will be considered in future work but can be easily included in our framework.

In summary, we solve the following set of ideal MHD equations \citep{Stoneetal2008} with CR transport as
\begin{eqnarray}
\begin{split}
\frac{\partial\rho}{\partial t}+\bfnabla\cdot(\rho \bv)&=0,\\
\frac{\partial( \rho\bv)}{\partial t}+\bfnabla\cdot({\rho \bv\bv-\bb\bb+{{\sf P}^{\ast}}}) &=\\
{\sf \sigma_c}\cdot\left[\bF_c-\bv\cdot\left(E_c{\sf I}+{\sf P_c}\right)\right]&, \\
\frac{\partial{E}}{\partial t}+\bfnabla\cdot\left[(E+P^{\ast})\bv-\bb(\bb\cdot\bv)\right]&=\\
-\left(\bv+\bv_s\right)\cdot\left(\bfnabla\cdot{\sf P_c}\right)&,  \\
\frac{\partial\bb}{\partial t}-\bfnabla\times(\bv\times\bb)&=0, \\
\frac{\partial E_c}{\partial t}+\bfnabla\cdot\bF_c&= \\
\left(\bv+\bv_s\right)\cdot\left(\bfnabla\cdot {\sf P_c}\right)+Q&,\\
\frac{1}{V_m^2}\frac{\partial \bF_c}{\partial t}+\bfnabla\cdot{\sf P_c}&= \\
-{\sf \sigma_c}\cdot\left[\bF_c-\bv\cdot\left(E_c{\sf I}+{\sf P_c}\right)\right]&. 
\end{split}
\label{eq:crmhd}
\end{eqnarray}
Here ${\sf P}^{\ast}$ is the sum of gas ($P_g$) and magnetic pressure with magnetic permeability $\mu=1$. The gas total energy $E$ is the sum of gas kinetic energy and internal energy $E_g$, which is related to the gas pressure with the adiabatic index $\gamma$ as $E_g=P_g/(\gamma-1)$.
Motivated by our previous discussions, we take the interaction coefficient ${\sf \sigma_c}$ to be
\begin{eqnarray}
{\sf \sigma_c}^{-1}={\sf \sigma_c^{\prime}}^{-1}+\frac{\bb}{|\bb\cdot(\bfnabla\cdot{\sf P_c})|}\bv_A\cdot\left(E_c{\sf I}+{\sf P_c}\right),
\label{eq:source}
\end{eqnarray}
where ${\sf \sigma_c^{\prime}}^{-1}$ is the classical diffusion coefficient to account for the CR diffusion, which is equivalent to $\kappa_p$ in equation \ref{eqn:crevol}. Notice that the second term is $-\bv_s\cdot\left(E_c{\sf I}+{\sf P_c}\right)$ following the definition of $\bv_s$ in equation \ref{eq:vs}. Therefore, the equation for CR energy density $E_c$ is exactly the same as equation \ref{CR_oldeq} as desired when the term $(1/V_m^2)\partial \bF_c/\partial t$ can be neglected.
In this framework, the total conserved energy and momentum are $E+E_c$ and 
$\rho\bv+\bF_c/V_m^2$ respectively, since the energy and momentum source terms for gas and CRs have exactly the same values but with opposite signs. 

\section{Numerical Implementation}
\label{sec:implementation}
Our new algorithm for CR transport has been implemented in the recently developed, public available radiation MHD code {\sf Athena++} 
(Stone et al. 2017, in preparation) \footnote{http://princetonuniversity.github.io/athena/}. For comparison, we have also implemented an explicit scheme to solve the original CR transport equation \ref{CR_oldeq} with a regularization term, which is described below. 

\subsection{The Classical Method with Regularization}
\label{sec:regularization}
In order to avoid the numerical oscillation caused by the term $\bfnabla\cdot\left[\bv_s\left(E_c+P_c\right)\right]$ in the original CR transport equation \ref{CR_oldeq}, \cite{Sharmaetal2009} proposed modifying this term to avoid a singularity in $\bv_s$ when $\bfnabla P_c$ is close to zero. For instance, in 1D, and for $P_c=E_c/3$, if we ignore diffusion and CR wave heating, equation \ref{CR_oldeq} is ${\partial E_c}/{\partial t} + {\partial(4 E_{\rm c} v_{\rm s})/3}/{\partial x}=0$. \citet{Sharmaetal2009} regularize this by modifying it to: 
\begin{eqnarray}
\frac{\partial E_c}{\partial t}-\frac{\partial }{\partial x}\left[\frac{4|v_A|}{3}E_c\tanh\left(\frac{\partial E_c/\partial x}{\epsilon}\right)\right]=0,
\label{eq:regularize}
\end{eqnarray} 
Here $\epsilon$ is a free parameter, which is usually taken to be a small number and adjusted for each simulation. Effectively, this regularization scheme adds an additional diffusive term to the original equation \ref{CR_oldeq} to damp numerical oscillations as long as the time step is small enough. When $(1/\epsilon)\partial E_c/\partial x$ is small, the flux can be expanded as 
$4|v_A|E_c/(3\epsilon)\partial E_c/\partial x$ and the above equation becomes a diffusion equation with an energy dependent diffusion coefficient $4|v_A|E_c/(3\epsilon)$.
\cite{Sharmaetal2009} shows that for explicit scheme, the necessary time step criterion for numerical stability is $\Delta t \lesssim \Delta x^2\epsilon/2E_c|v_A|$ because of the diffusive nature of the regularization term. This constraint usually results in a very small time step, which also drops much faster with increasing resolution compared with the classical CFL condition. We implement this regularization scheme with explicit upwind method as described by \cite{Sharmaetal2009} in {\sf Athena++}. In principle, an implicit method can also be used with this regularization scheme to allow for a much larger time step. However, because this term involves spatial gradient of $E_c$, an implicit method usually requires matrix inversion over the whole simulation domain, which can be slow and significantly impact the parallel scaling. More discussion on numerical performance is given in \S\ref{sec:discussion}.

\subsection{Implementation of the New Algorithm}
The basic numerical algorithm for the ideal MHD part is the same as the original {\sf Athena} \citep{Stoneetal2008} and 
{\sf Athena++} codes, which will not be repeated here. We will only describe how we solve the CR transport equation \ref{neweq} 
and add the momentum and energy source terms to the gas. The numerical scheme is inspired by the algorithm to solve the radiative transfer equation as in \cite{Jiangetal2014b}. We split the transport and source terms, which are described in the following sections separately. Here, we first summarize the basic procedure of the CR transport algorithm based on the two step van-Leer integrator. 

\begin{enumerate}[I]
	\item Calculate the time step $\Delta t$ based on the cell sizes and the maximum velocity $V_m$. Calculate the effective interaction coefficient 
	$\sigma_c$ based on $\bb^n,\rho^n,E_c^n$ at time step $n$. 
	\item 	\label{step II} Reconstruct the left and right states for $E_c^n$ and $\bF_c^n$ at the cell interfaces for the three coordinate directions.  
	\item \label{step III} Calculate the flux based on the \emph{HLLE} Riemann solver and apply the flux divergence for half time step $\Delta t/2$. Calculate the energy source term and apply the same amount but with opposite signs to the gas and CRs. The MHD part is also updated for half time step simultaneously. 
	\item \label{step IV} Rotate the flow velocity $\bv^n$ and CR flux $\bF_c^n$ at time step $n$ to a coordinate system that $\bb$ is along the $x$ direction at each cell. Add CR source terms for half time step $\Delta t/2$ implicitly. Apply an inverse rotation of the updated CR flux $\bF_c^{n+1/2}$ to the original coordinate direction. Calculate the momentum changes of CRs during this step and apply the exactly the same amount but with opposite signs to the gas momentum at time step $n+1/2$.
	\item Update $\sigma_c$ with $\bb^{n+1/2},\rho^{n+1/2},E_c^{n+1/2}$ at time step $n+1/2$. 
	\item Repeat steps \ref{step II} to \ref{step III} but using $E_c^{n+1/2}$ and $\bF_c^{n+1/2}$ for reconstruction and \emph{HLLE} Riemann solver. Apply the flux divergence to $E_c^n$ and $\bF_c^n$ for a full time step $\Delta t$. The MHD quantities are also updated for a full time step at the same time.
	\item Calculate the CR source terms implicitly for a full time step $\Delta t$ and add the source terms to the gas quantities at time step $n+1$.
	\item Update the time $t_{n+1}=t_n+\Delta t$ and repeat these steps for the next cycle.
	
\end{enumerate}

\subsubsection{The Transport Step}
The transport step solves the left hand side of equation \ref{neweq} as 
a set of standard hyperbolic equations 
\begin{eqnarray}
\frac{\partial \bq}{\partial t}+\bfnabla\cdot\bfr=0,
\label{eq:transport}
\end{eqnarray}
where the conserved quantities and their fluxes are
\begin{eqnarray}
\bq=
\begin{bmatrix}
E_c \\
F_{c,x}/V_m^2 \\
F_{c,y}/V_m^2 \\
F_{c,z}/V_m^2
\end{bmatrix}
,
\bfr=
\begin{bmatrix}
F_{c,x} & F_{c,y} & F_{c,z} \\
P_{c,xx} & P_{c,xy} & P_{c,xz} \\
F_{c,yx} & P_{c,yy} & P_{c,yz} \\
F_{c,zx} & P_{c,zy} & P_{c,zz} 
\end{bmatrix}.
\end{eqnarray}
These are standard hyperbolic equation with a single signal speed, which are exactly the same as the moment equations of radiative transfer \citep{SekoraStone2010,Jiangetal2012}. In general, if the three diagonal components of ${\sf P_c}$ are related to $E_c$ as 
$f_{xx}E_c,f_{yy}E_c,f_{zz}E_c$, the signal speeds along the three directions are $\sqrt{f_{xx}}V_m,\sqrt{f_{yy}}V_m,\sqrt{f_{zz}}V_m$ 
respectively. As we solve the transport term \emph{explicitly}, the time step $d t$ is the minimum of $\Delta x/\sqrt{f_{xx}}V_m$, 
$\Delta y/\sqrt{f_{yy}}V_m$, and $\Delta z/\sqrt{f_{zz}}V_m$ over the whole simulation domain, where $\Delta x$, $\Delta y$ 
and $\Delta z$ are the sizes of grid cells along three directions. Because $V_m$ is usually chosen to be larger than any other signal speed in the system, $d t$ should also be the time step for the MHD part. 

For given cell centered values of $E_c^n(k,j,i)$, $\bF_c^n(k,j,i)$ at time step $n$ and spatial location $(k,j,i)$, we calculate the flux at the cell interface based on the \emph{HLLE} Riemann solver. First, we apply the piecewise linear (second-order) reconstruction \citep{Stoneetal2008} for $E_c,\bF_c,\bv$ to get the left ($\bq_L$) and right ($\bq_R$) states of the conserved quantities $\bq$. Then the \emph{HLLE} fluxes at cell interfaces $(k,j,i-1/2)$ are constructed as
\begin{eqnarray}
\bF^{HLLE}_{i-1/2}=\frac{v^+\bf_{L,i-1/2}-v^-\bf_{R,i-1/2}}{v^+-v^-}\nonumber\\
+\frac{v^+v^-}{v^+-v^-}\left(\bq_i-\bq_{i-1}\right),
\end{eqnarray}
where $\bf_{L,i-1/2}$ and $\bf_{R,i-1/2}$ are $x$ components of the flux tensor $\bfr$ calculated based on the left and right 
states $\bq_L$ and $\bq_R$. Fluxes along other coordinate directions are calculated in the similar way. 

The maximum and minimum wave speeds $v^+$ and $v^-$ need special care. If we completely split the transport and source steps, $v^+$ 
and $v^-$ will be $\sqrt{f_{xx}}V_m$ and $-\sqrt{f_{xx}}V_m$ for the hyperbolic equation \ref{eq:transport}. However, as noted by \cite{Jiangetal2013b} (see the Appendix A) for the radiative transport equation, this can generate too large numerical diffusion in the regime that the source term is expected to balance $\bfnabla\cdot\bfr$. Following \cite{Jiangetal2013b}, for a cell size $\Delta l$, we define an effective optical depth $\tau\equiv \Delta l\sigma_c/V_m$ and reduction factor ${\calligra R}=\sqrt{(1-\exp(-\tau^2))/\tau^2}$. Then the maximum wave speed is chosen to be $\tr{min}(V_m, {\calligra R}V_m/3^{1/2})$ while the minimum wave speed is $V^-=-V^+$. This is necessary to have enough numerical dissipation to stabilize the scheme, but it is still small enough so that physical flux can be captured, particularly in the regime that the diffusion flux is small.

\begin{figure*}[htp]
	\centering
	\includegraphics[width=0.49\hsize]{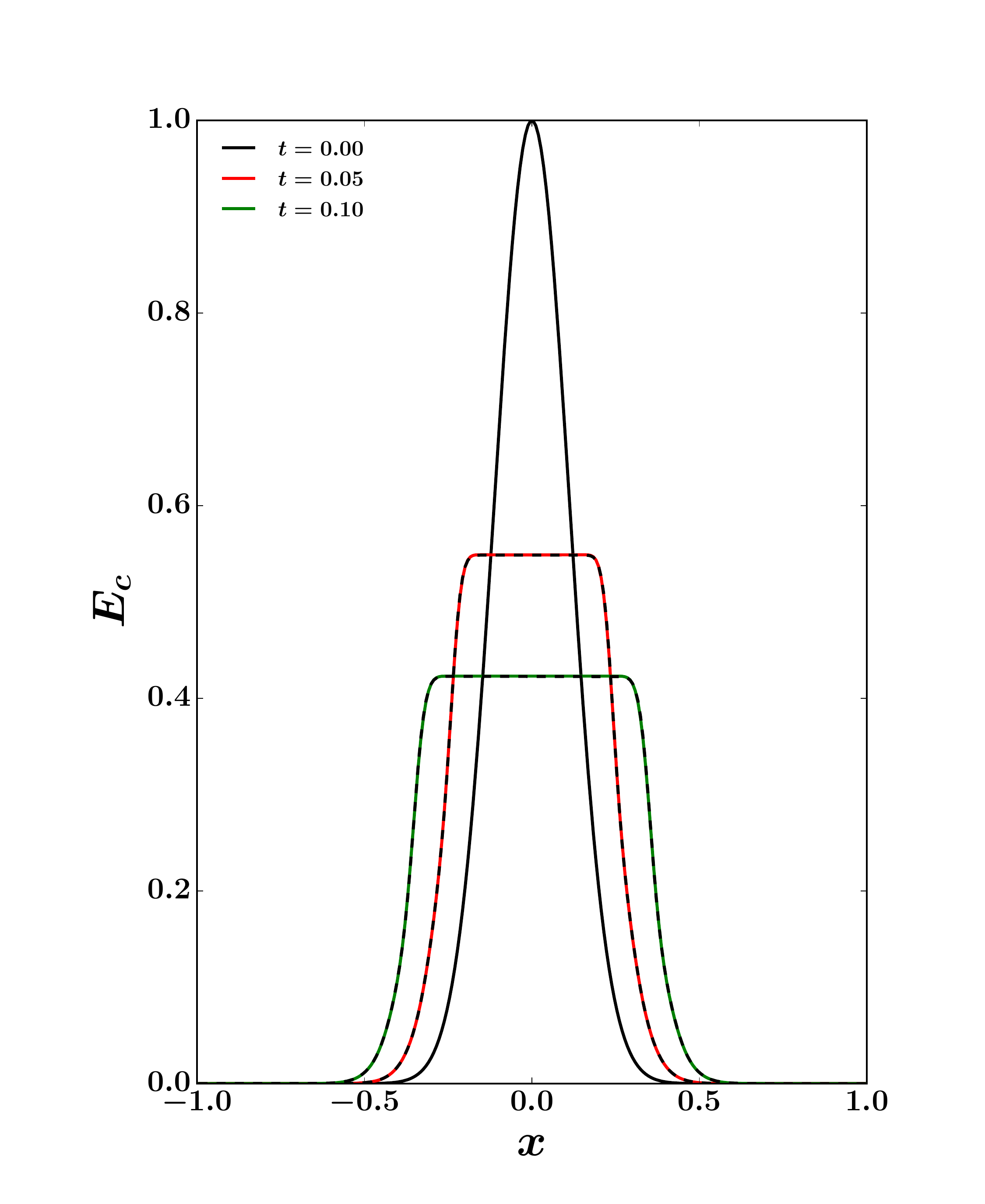}
	\includegraphics[width=0.49\hsize]{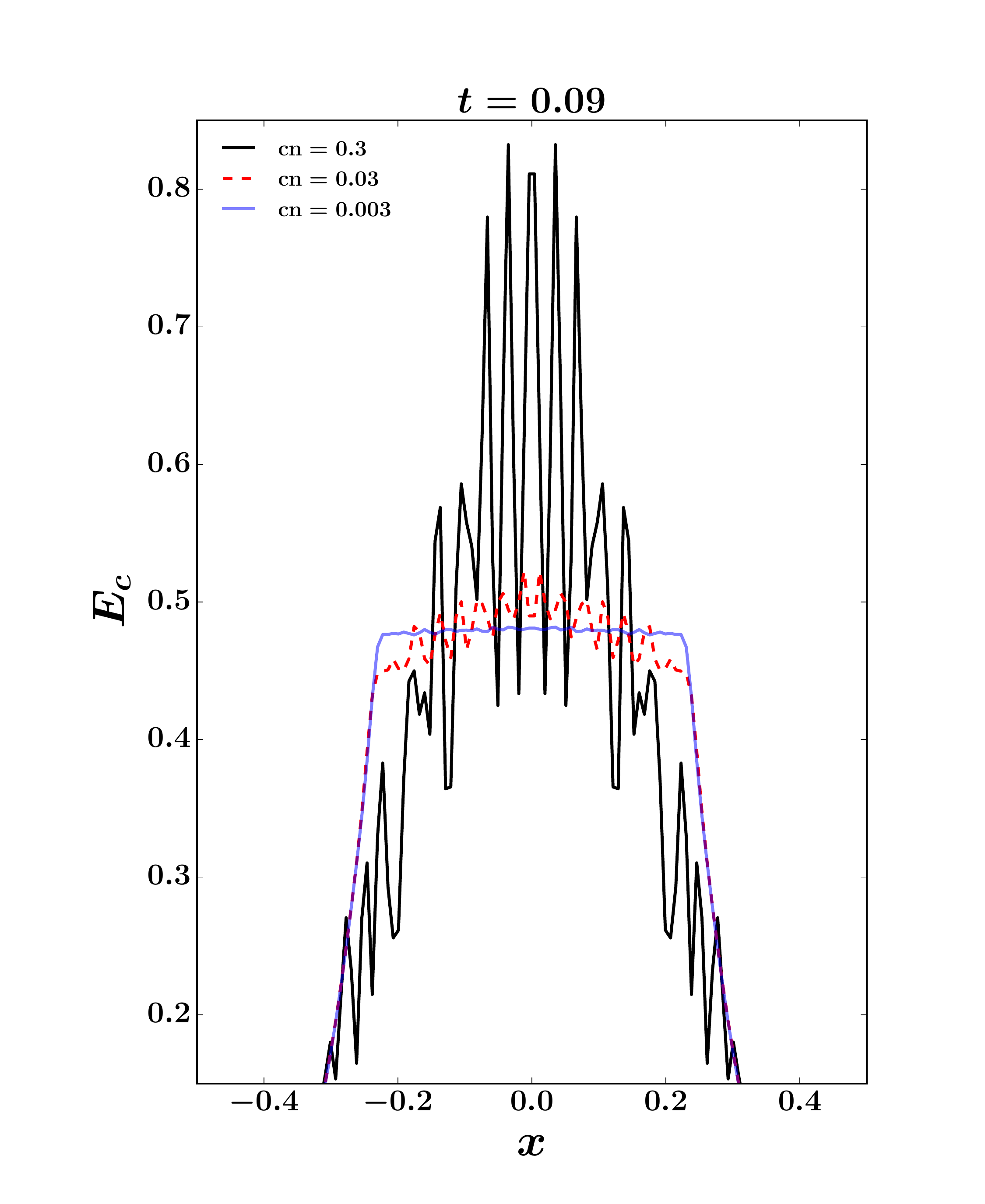}
	\caption{Comparison of CR streaming between our new numerical algorithm (left panel) and the traditional method (right panel) for the test described in Section \ref{sec:streaming}. In the left panel, the solid black line is the initial Gaussian profile while the red and green lines are solutions at time $t=0.05$ and $t=0.1$ respectively. The dashed black lines in the left panel are for the solution with $V_m=200$, which are almost identical compared with the solution with the default value $V_m=100$. The right panel shows the solutions at time $t=0.09$ by solving the traditional equation without any regularization but with different CFL numbers (thus different time steps). The solution with a normal CFL number $0.3$ shows very strong numerical oscillations. Amplitudes of he oscillations get smaller with reduced time step but they do not go away completely even with time step smaller by a factor of 100.   }
	\label{streamingtest}
\end{figure*}


\begin{figure}[htp]
	\centering
	\includegraphics[width=1.0\hsize]{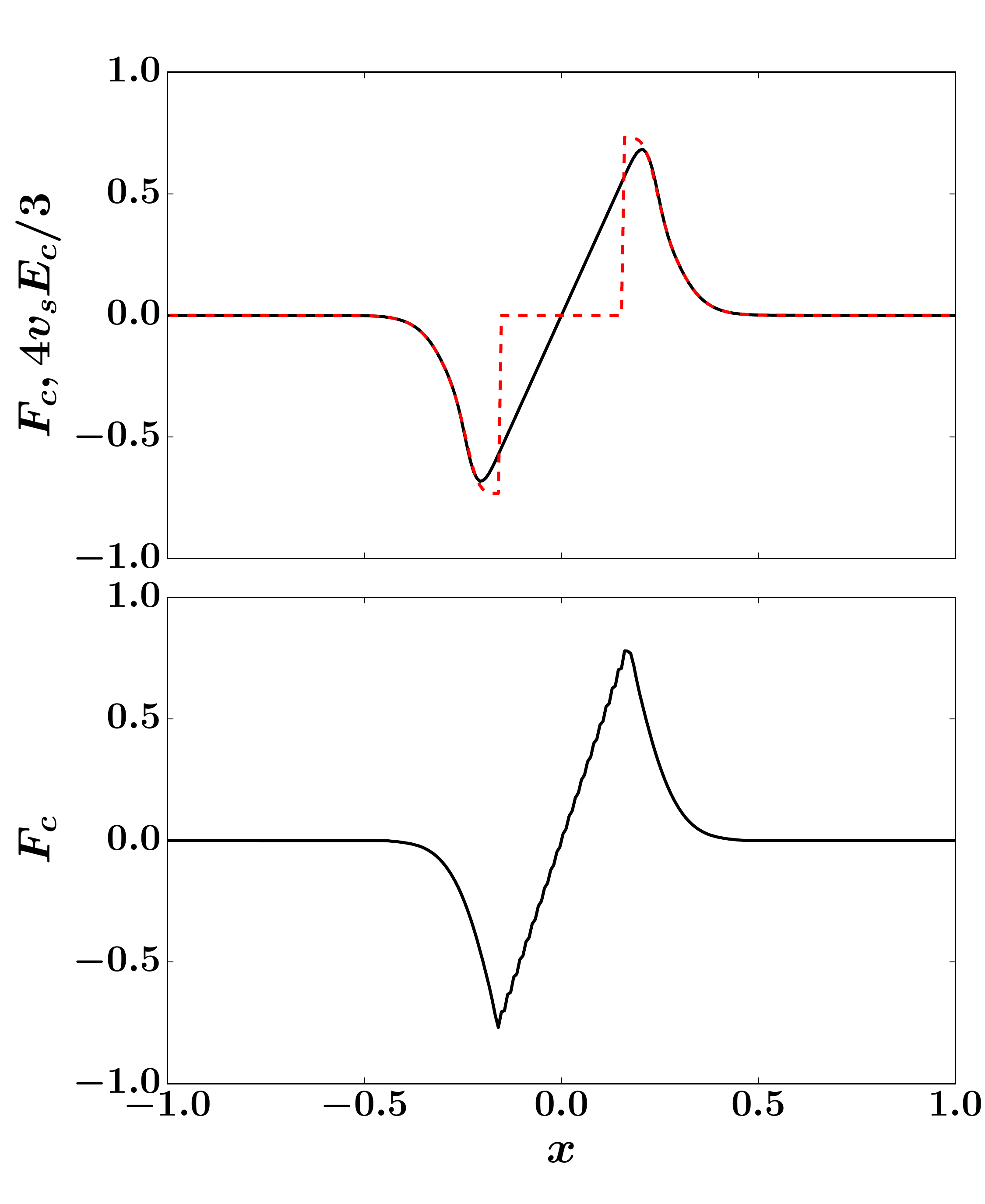}
	\caption{The top panel shows the profile of CR flux $F_c$ (the black line) calculated with our new algorithm at time $t=0.05$ for the CR streaming test described in Section \ref{sec:streaming}. The dashed red line is the steady-state flux $4v_sE_c/3$, which agrees with $F_c$ very well in the region when CR pressure gradient is not zero. The CR flux $F_c$ in the bottom panel is from the solution to the regularized equation \ref{eq:regularize} with $\epsilon=0.1$ at time $t=0.05$.}
	\label{crflux}
\end{figure}

\subsubsection{The Source Terms}
\label{sec:source}
Because the streaming velocity $\bv_s$ is along the magnetic field lines and the diffusion coefficient ${\sf \sigma_c^{\prime}}$ can also have different values for directions along and perpendicular to the magnetic field lines, 
the interaction coefficient ${\sf \sigma_c}$ should be a tensor in general. In order to satisfy equation \ref{eq:source}, we first rotate the coordinate system at each cell such that the magnetic field lines $\bb$ are along one coordinate $x$. Specifically, we apply the following rotation operators ${\sf R_2}{\sf R_1}$ to all the vectors $\bF_c,\bv$ and the tensor ${\sf P_c}$ at each cell
\begin{eqnarray}
{\sf R_1}&=&
\begin{bmatrix}
\cos\phi_B & \sin\phi_B & 0 \\
-\sin\phi_B & \cos\phi_B & 0 \\
0  & 0  & 1
\end{bmatrix},\nonumber\\
{\sf R_2}&=&
\begin{bmatrix}
\sin\theta_B & 0 & \cos\theta_B \\
0 & 1 & 0\\
-\cos\theta_B & 0 & \sin\theta_B 
\end{bmatrix}.
\end{eqnarray} 
Here the rotation angles $\theta_B,\phi_B$ are determined as
\begin{eqnarray}
\cos\phi_B\equiv\frac{B_x}{B_x^2+B_y^2},\ \sin\phi_B\equiv\frac{B_y}{\sqrt{B_x^2+B_y^2}}.
\end{eqnarray}
The scalars such as $E_c,\rho$ and $\bb\cdot\left(\bfnabla\cdot{\sf P_c}\right)$ are unchanged. After this rotation, the magnetic field $\bb$ is along the first coordinate axis, which significantly simplifies the interaction coefficient ${\sf \sigma_c}$ as it becomes a vector. At each time step $n$, we calculate the interaction coefficients based on the fluid and CR variables at the beginning of the time step after the rotation as
\begin{eqnarray}
\frac{1}{\sigma_{c,1}}&=&\frac{1}{\sigma_{c,1}^{\prime}}+\frac{B}{|\bb\cdot\left(\bfnabla\cdot{\sf P_c}\right)|}\bv_A\cdot\left(E_c{\sf I}+{\sf P_c}\right),\nonumber\\
\sigma_{c,2}&=&\sigma_{c,2}^{\prime},\nonumber\\
\sigma_{c,3}&=&\sigma_{c,3}^{\prime},
\end{eqnarray}
where $1,2,3$ represent the three components of the diffusion coefficients along the coordinate axes. 
The source terms are then added implicitly as
\begin{eqnarray}
\begin{split}
\frac{E_c^{n+1}-E_c^n}{\Delta t}&=\sum_{i=1}^3\left(v_i^{n+1/2}+v_{s,i}^{n+1/2}\right)\left[\bfnabla\cdot{\sf P_c}\right]^{n+1/2}_i,\\
\frac{1}{V_m^2}\frac{F_{c,i}^{n+1}-F_{c,i}^n}{\Delta t}&=\\
-\sigma_{c,i}&\left[F_{c,i}^{n+1}-\sum_{k=1}^3v_k^n\left(E_c^{n+1}+P_c^{n+1,k,i}\right)\right].
\end{split}
\end{eqnarray}
Here the pressure gradient is calculated during the transport step since $\bfnabla\cdot{\sf P_c}$ is just the flux for $\bF_c$ returned by the \emph{HLLE} Riemann solver. 
After this step, we apply the inverse rotation ${\sf R_1^{-1}R_2^{-1}}$ to $\bF_c^{n+1}$ get the new CR flux in the original coordinate system. The amount of energy and momentum change during this step are
$E_c^{n+1}-E_c^n$ and $(\bF_c^{n+1}-\bF_c^{n})/V_m^2$, which are added to the gas total energy $E$ and 
momentum $\rho\bv$ with the opposite signs. Because the source term is treated implicitly, we retain a linear timestep scaling with resolution even when computing CR diffusion.

\section{Tests}
\label{sec:test}

We implement the algorithm in the public available radiation MHD code {\sf Athena++}\footnote{http://princetonuniversity.github.io/athena/}, which has 
been carefully tested for the basic MHD part. In this section, we will show 
series of tests to compare the properties of our new algorithm and the 
traditional method for CR transport. 

\subsection{Test of the CR subsystem}
In this section, we will show test problems designed for the CR subsystem only. All the MHD variables are fixed to be constant. The CR energy and momentum source terms  in equation \ref{neweq} are not added back to the MHD part. The CR source term $Q$ is set to be zero. These test problems are generic and are not for any specific physical system. Therefore, we will only quota dimensionless numbers without specifying the units. 

In these tests, we ignore the energy source term, so that we solve the conservative system: 
\begin{equation}
\frac{\partial E_{\rm c}}{\partial t} = - \nabla \cdot \bF_c.
\label{eq:conservative}
\end{equation}
This form of the energy equation was also studied by \citet{Sharmaetal2009}; it isolates the numerical instability associated with streaming down a gradient in its simplest form. 

\subsubsection{CR Streaming in 1D}
\label{sec:streaming}

\textbf{Gaussian profile.} To test how our new algorithm works for the CR streaming term $\bv_s\cdot\left(E_c\sf{I}+\sf{P_c}\right)$, we set up a 1D 
Gaussian profile of CR energy density $E_c=\exp\left(-40x^2\right)$ with simulation domain $x\in\left(-1,1\right)$. The Alfv\'en velocity is fixed to be $1$ so that the streaming velocity $v_s=-{\rm sgn}(\partial E_c/\partial x)$, which changes sign at $x=0$. The flow velocity $v$ is fixed to be $0$. We set the diffusion coefficient $\sigma_c^{\prime}=10^{8}$ so that the diffusive flux is completely negligible in this test problem. The default value for the maximum velocity $V_m$ is chosen to be $100$ and we also vary this value to see how the solutions are affected. Outflow boundary conditions for all variables at both ends of the simulation domain are used, which means $E_c$ and $F_c$ are just copied from the last active zones to the ghost zones. We use $256$ grid zones for the 1D simulation domain.  A similar test is also performed at the Appendix of \cite{Wieneretal2017}.

Profiles of CR energy density $E_c$ at two different time snapshots are shown in the left panel of Figure \ref{streamingtest}. The initial Gaussian profile spreads due to CR streaming along CR pressure gradient. The central region is completely flat as it should be. We have also evolved the solution with $V_m=200$ and $V_m=50$ while keeping all the other parameters fixed. The solution with $V_m=200$ is shown as the dashed black lines in the left panel of Figure \ref{streamingtest}, which is almost identical to the  solution with $V_m=100$. We find the same result with $V_m=50$. 

To compare with the solution to the traditional equation, we have also solved equation \ref{CR_oldeq} for the same setup with and without regularization. The old solutions without any regularization at time $t=0.09$ are shown in the right panel 
of Figure \ref{streamingtest}, which demonstrates that it is numerically unstable and the 
oscillations get smaller with smaller time step \citep{Sharmaetal2009}. 
We have also solved the regularization equation with $\epsilon=0.1,0.05,0.02$ respectively.  
The numerical oscillation is almost gone with the regularization term by design. With smaller values of $\epsilon$, the central part is flatter because the additional diffusive term is smaller. But the time step is also reduced with smaller values of $\epsilon$. Larger value of $\epsilon$ causes more diffusive behavior at the region where $E_c$ should be flat. For this test problem, even with $\epsilon=0.1$, the time step is a factor of 10 smaller than the time step in our new algorithm with $V_m=100$.

\begin{figure*}[htp]
	\centering
	\includegraphics[width=0.49\hsize]{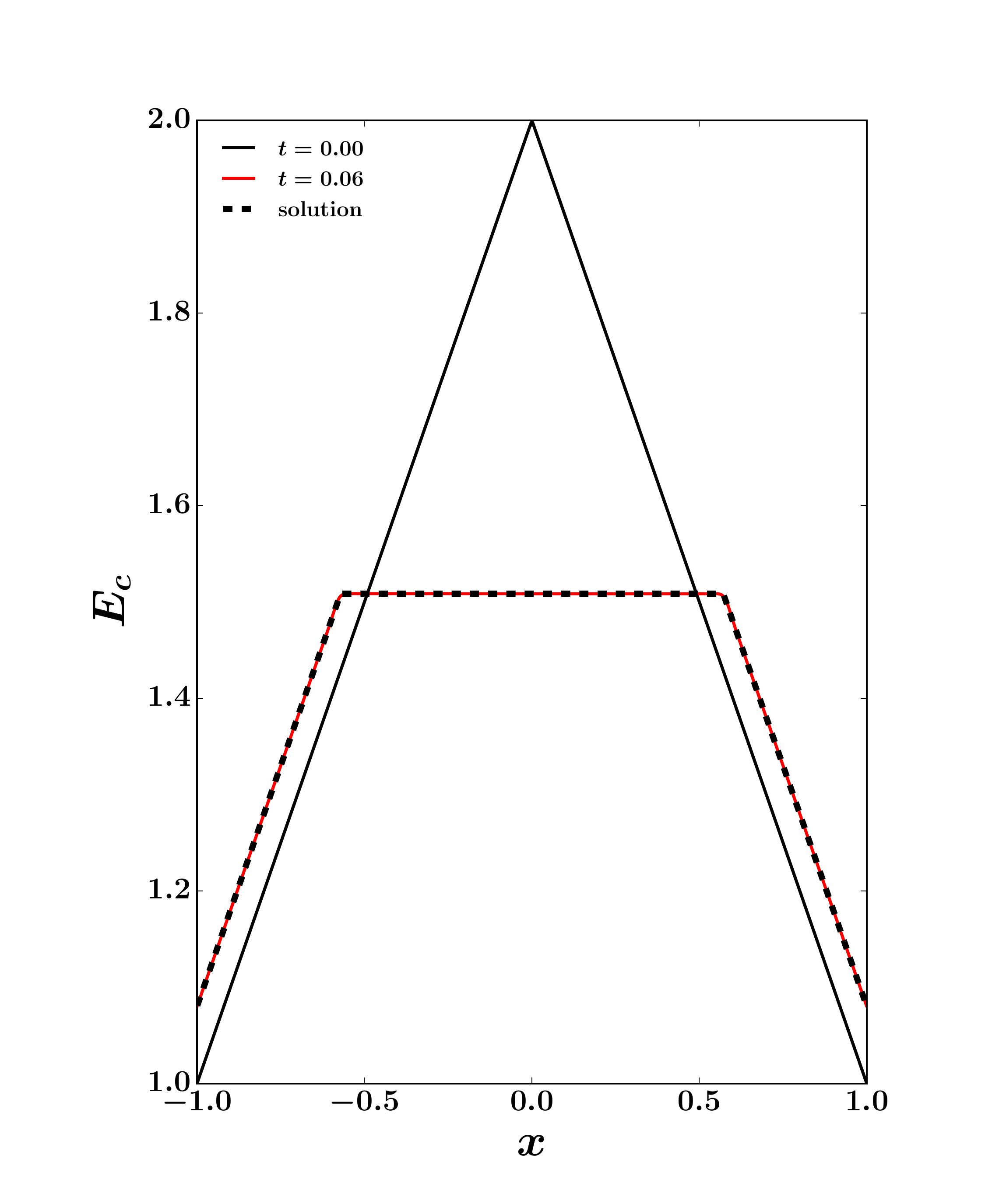}
	\includegraphics[width=0.49\hsize]{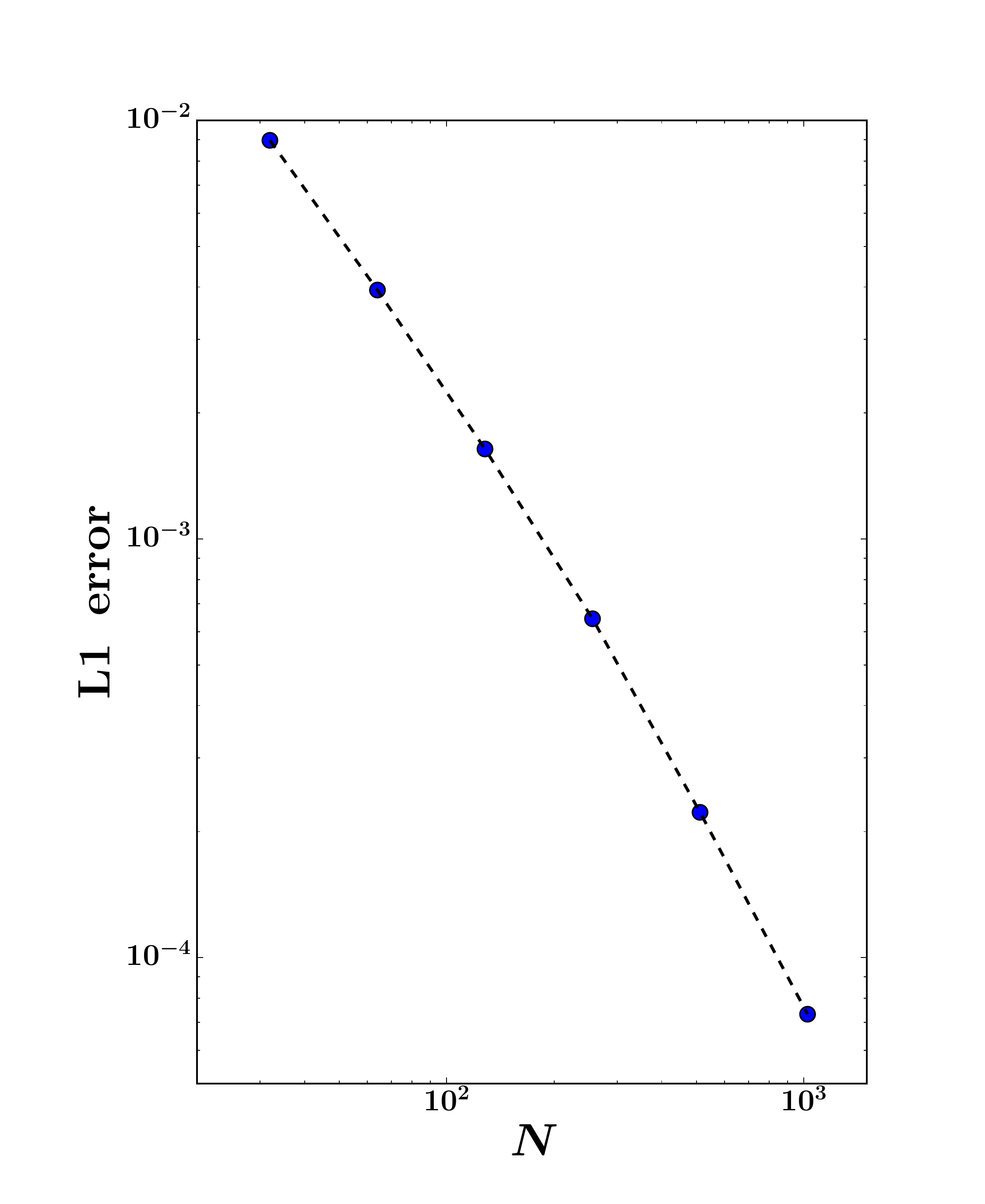}
	\caption{Streaming of CRs with an initial triangle profile. The Alfv\'en velocity is a constant $V_A=1$. \emph{Left:} the solid black line is the initial profile while the red line is the numerical solution at time $t=0.06$. The dashed black line is the analytical solution at the same time. \emph{Right:} the $L1$ error, which is calculated for the solution at time $t=0.06$, decreases as the resolution increases with the rate $L1\propto N^{-1.3}$. }
	\label{conv_triangle}
\end{figure*}

Profile of CR flux $F_c$ at time $t=0.05$ as calculated by our new algorithm is shown in the upper panel of Figure \ref{crflux}. We also over-plot the expected streaming flux given by the steady-state expression (equation \ref{CR_oldflux}) $4v_sE_c/3$ as the dashed red line. We set $v_s$ to be zero when the difference between $E_c$ in the neighboring zones is smaller than $5\times 10^{-5}$ for the expected streaming flux. In the region $|x|\gtrsim 0.2$ where CR pressure gradient is not close to zero,  $F_c$ agrees with the traditional streaming flux nicely. When $\bfnabla P_c$ approaches zero, the interaction coefficient $\sigma_c$ also becomes zero. In this regime, the term $\left(1/V_m^2\right)\partial F_c/\partial t$ becomes important, which is also the main difference between our new algorithm and the traditional approach. The equation becomes hyperbolic as in the optically thin regime of radiative transfer equation. The CR flux $F_c$ does not have the singularity suffered by the steady-state expression given by equations \ref{eq:vs} and \ref{CR_oldflux}. CR flux for the solutions to the regularized equation \ref{eq:regularize} also does not have the singularity as shown in the bottom panel in the left column of Figure \ref{crflux}. Although this is achieved via an artificial diffusion term, the profile of $F_c$ is actually very similar to to the profiles of $F_c$ from our new algorithm. 

The linear profile of $F_c$ can be understood. At extrema, and in the absence of sources, as CRs stream outward, profiles develop a flat top. CRs cannot stream further to produce an inverted profile, as that would require CRs to stream up their gradient. 
In order to maintain the flat top in equation \ref{eq:conservative}, we require $\dot{E}_{c}(x,t) = f(t)$, where $f(t)$ is some function independent of position throughout the flat top. From equation \ref{eq:conservative}, this implies $\bF_c \propto x$, i.e., a linear interpolation between the fluxes at either end of the flat top, with $\bF_c(0)=0$ in the absence of sources. 
In the old method, $F = 4/3 v_{\rm A} E_{\rm c} \nabla E_{\rm c}/\epsilon = C x$ (where C is a dimensional constant) implies $E_{\rm c} \sim {\rm const} + (C \epsilon/v_{\rm A})^{0.5} x$; thus, the departure from flatness scales as $(\epsilon/v_{\rm A})^{0.5}$.

\begin{figure*}[htp]
	\centering
	\includegraphics[width=0.33\hsize]{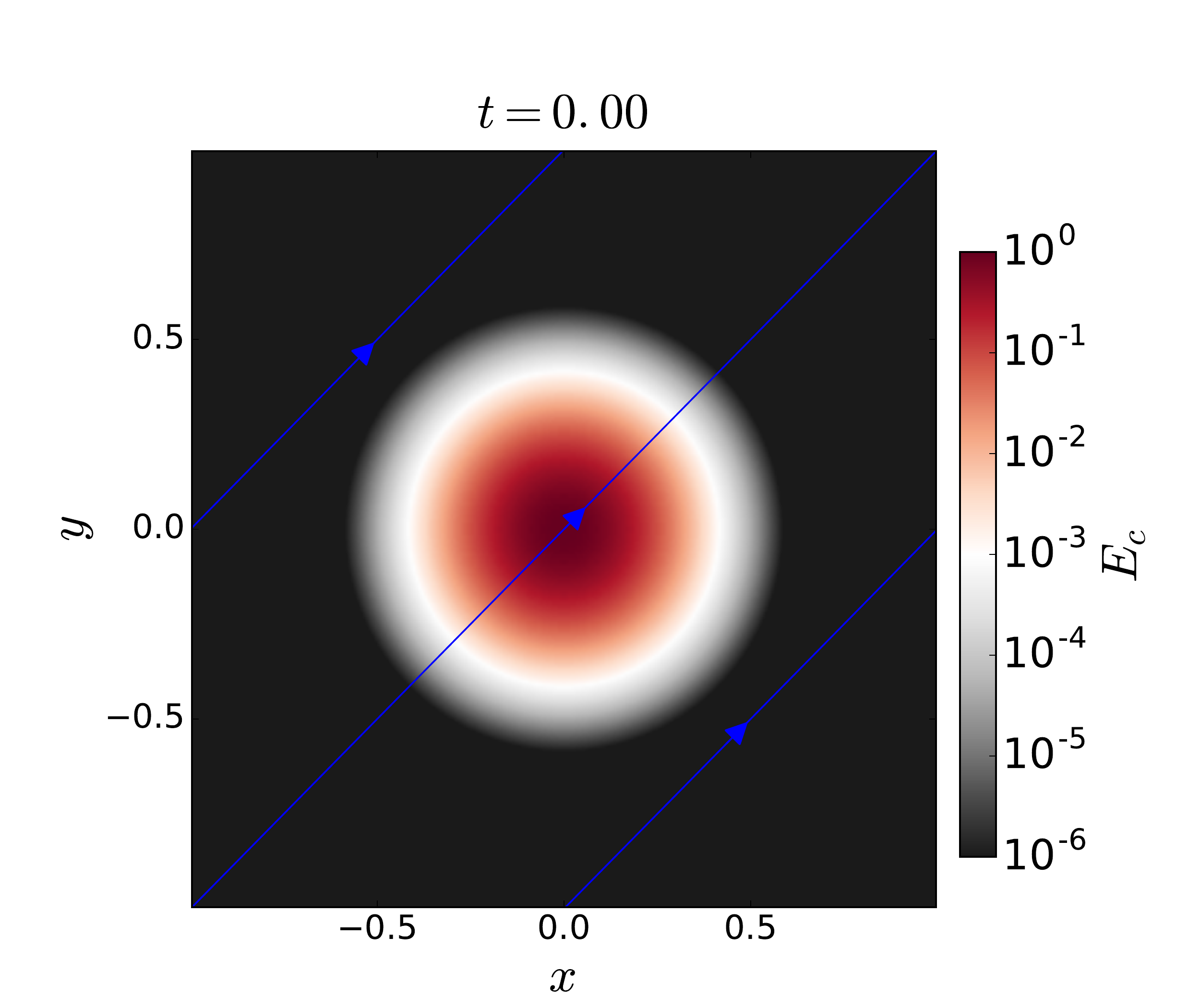}
	\includegraphics[width=0.33\hsize]{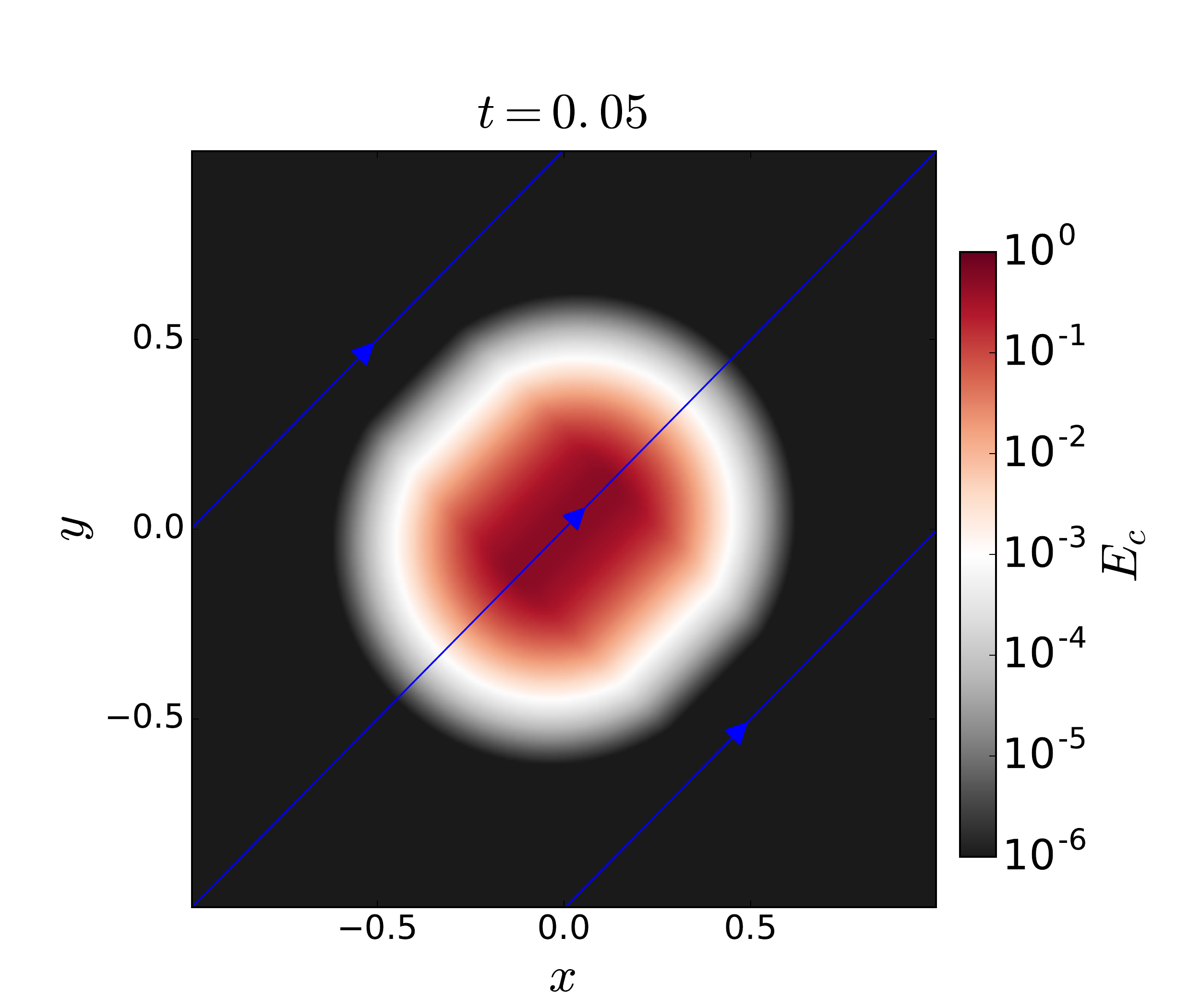}
	\includegraphics[width=0.33\hsize]{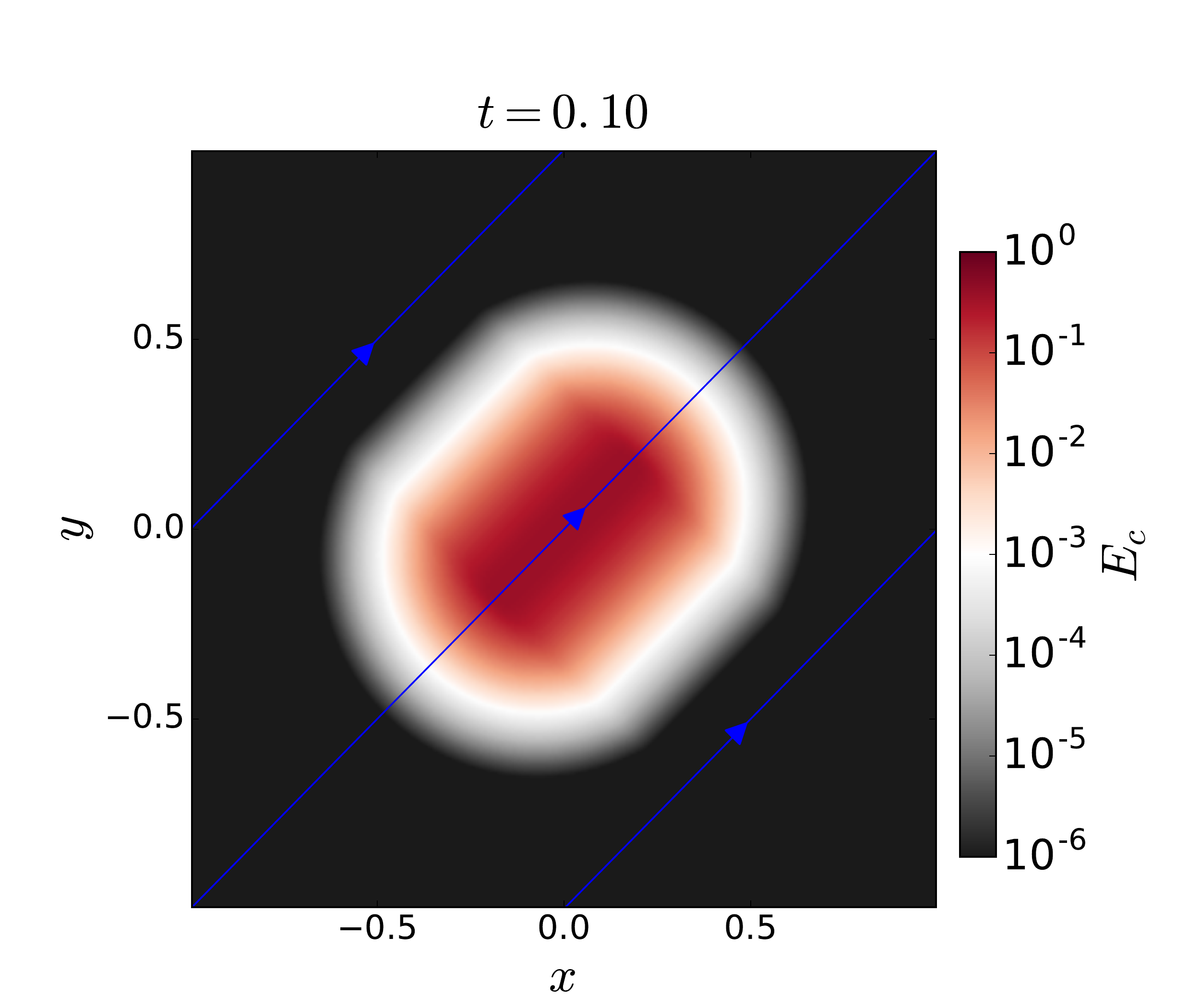}
	\caption{Test of CR streaming along magnetic fields in 2D as described in Section \ref{sec:2Dstreaming}. Uniform magnetic fields are along the direction as indicated by the blue lines. CRs only stream along the magnetic fields as in the 1D test shown in Figure \ref{streamingtest} while CR profiles perpendicular to the magnetic fields do not change. From left to right, the panels display  
		snapshots of $E_c$ at time $t=0$, $0.05$ and $0.1$ respectively.}
	\label{2Dstream}
\end{figure*}

\textbf{Triangular profile}. It is also interesting to consider a triangular initial profile $E_c=2-|x|$, as shown in the left panel of Figure \ref{conv_triangle}, which has a discontinuous first derivative at $x=0$ in the initial conditions. Furthermore, since $\nabla E_{\rm c}=0, $ const in the flat and non-flat parts respectively, it is straightforward to find analytic solutions to test convergence rates.   

We fix all fluid variables to be constant and set the Alfv\'en velocity to $v_A=1$. Since we have ignored the CR energy source term $v_A \cdot \nabla P_{\rm c}$, the total CR energy is only lost through the left and right boundaries. We expect that at time $t$, $E_c$ should be a constant for the central part from $-x_m$ to $x_m$. The value of $E_c$ at the left and right boundaries should be $1+4 v_{\rm A} t/3$ because the non-flat parts just move with the speed $4v_A/3$ but keep the same slope, which means $E_c(x,t)=2+4v_At/3-|x|$ 
for $x\in (-1,-x_m)$ and $x\in(x_m,1)$. The fluxes at the left and right boundaries are then $F_c(\pm1,t)=\pm (4v_A/3)(1+4v_At/3)$. 
The value of $x_m$ is determined by energy conservation $\int_{-1}^{1}E_c(x,t)dx+2\int_0^t(4v_A/3)(1+4v_At/3)dt=3$, 
which is 
\begin{eqnarray}
x_m=\sqrt{\left(1+\frac{4}{3}v_At\right)^2+\frac{8v_At}{3}-1}.
\end{eqnarray}
The analytical solution agrees with the numerical solution very well as shown in the left panel of Figure \ref{conv_triangle}. The convergence rate is intermediate between linear and second order, as in the current numerical scheme, the advection step is second order accurate, but the treatment of the source term (particularly, the momentum source term here) is only first order accurate. 
The $L1$ error between the analytical solution at time $t=0.06$ and the numerical solution with $V_m=1000$ decreases with 
increasing resolution as $N^{-1.3}$, which is demonstrated in the right panel of Figure \ref{conv_triangle}. If we use a smaller 
$V_m=100$, $L1$ error saturates at $10^{-3}$ when resolution $N$ is larger than 256. This is because the analytical solution completely neglects the term $(1/V_m^2)\partial F_c/\partial t$, which becomes more important with smaller $V_m$.  In particular, the transition from the flat top to the constant slope is not abrupt but affected by the time-dependent term (and in particular the choice of $V_{\rm m}$). From equation \ref{eq:breakdown}, this should happen when $\Delta x < v_{\rm A}/V_{\rm m} L_{\rm z} = 0.01$ for $V_{\rm m}=100$, or $N > 200$, as we have found. Unless very high accuracy in transition regions is required, our choice of $V_{\rm m} \gg V_{\rm A}$ does not significantly affect the accuracy of solutions. 

When we include the energy source term $\bv_A\cdot(\bfnabla\cdot{\sf P_c})$, we can get an approximate analytical solution by neglecting the CR pressure gradient at $x_m$, where $E_c$ changes from the flat top to the linear profile. In this case, $E_c$ in the regions from $-1$ to $-x_m$ and $x_m$ to $1$ moves with the speed $v_A=1$. The value of $x_m$ is determined by the following equation 
\begin{eqnarray}
-2x_m\frac{d x_m}{d t}-\frac{2v_A}{3}x_m+\frac{8}{3}v_A(1+v_At)+\frac{8v_A}{3}=0.
\end{eqnarray}
At time $t=0.06$, the solution is $x_m=0.56$. The numerical solution also agrees with the analytical solution very well in this case.

\subsubsection{Streaming along Magnetic Fields in Multi-dimensions}
\label{sec:2Dstreaming}
In 2D and 3D simulations, magnetic field lines are generally not aligned with the coordinate axes while CRs only stream along the  magnetic fields. To test the performance of our scheme with anisotropic source terms, we setup uniform  magnetic fields along the diagonal of the 2D domain $(x,y)\in(-1,1)\times (-1,1)$, as shown in Figure \ref{2Dstream}. The Alfv\'en velocity is $v_{\rm A}=1$. The initial distribution of CR energy density is $E_c=\exp\left[-40(x^2+y^2)\right]$. We use $256^{2}$ grids and outflow boundary conditions. The fluid variables are kept fixed in this test and flow velocity is always 0. We use a large normal diffusion coefficient $\sigma_c^{\prime}=10^{8}$ so that diffusion is negligible. The default value $V_m=100$ is used as in the 1D test done in the last section.

The initial distribution of $E_c$ as well as snapshots at time $t=0.05$ and $0.1$ are shown in Figure \ref{2Dstream}. As expected, CRs only stream along the direction of magnetic fields. When we take the profiles of $E_c$ across the line $x=y$, 
the time evolution is the same as shown in the left panel of Figure \ref{streamingtest}. This is because along that direction, this test problem is basically the same as the 1D case. If we take the profiles of $E_c$ along the direction perpendicular to the magnetic field lines, the shape of $E_c$ does not change except that the amplitude becomes smaller as it should. Thus, our code is fully capable of handling anisotropic streaming. 

\subsubsection{Bottleneck Effect: balance between CR streaming and heating terms}
\label{sec:bottleneck}
When CR diffusion is negligible and CR pressure gradient is not close to $0$, in steady state, CR energy density with a static background flow should satisfy $\bfnabla\cdot\left[\bv_s\cdot\left(E_c{\sf \bI} + {\sf P_c}\right)\right]=\bv_s\cdot\left(\bfnabla\cdot{\sf P_c}\right)$. If we assume ${\sf P_c}=E_c/3{\sf I}$ and if $E_c$ is monotonic in 1D, this reduces to:
\begin{eqnarray}
4\frac{\partial v_A}{\partial x}E_c+3v_A\frac{\partial E_c}{\partial x}=0.
\label{eq:EcvArelation}
\end{eqnarray}
As long as the spatial gradients of $v_A$ and $E_c$ are not zero, this gives the conserved integral \citep{breitschwerdt91}: 
\begin{eqnarray}
E_cv_A^{4/3}={\rm constant}.
\label{eq:CR_integral}
\end{eqnarray}

Warm clouds can also have a more subtle effect \citep{Skilling1971,begelman95,Wieneretal2017}, known as the `bottleneck effect'. It exploits two important features of CR streaming: CRs can only stream down their density gradient, and the streaming instability which couples the CRs to the gas is only triggered when the bulk drift speed $v_{\rm D} > v_A$. Since $v_{\rm D} \sim v_A \propto \rho^{-1/2}$, a cloud of warm ($T\sim 10^{4}$K) ionized gas embedded in hot ($T\sim 10^{6}$K) gas results in a minimum in drift speed. This produces a `bottleneck' for the CRs; CR density is enhanced as CRs are forced to slow down, akin to a traffic jam. Since CRs cannot stream up a gradient, the system readjusts to a state where the CR profile is flat up to the minimum in $v_A$; thereafter the CR pressure falls again as $v_A^{-4/3}$. 

We can test whether our algorithm will recover this solution in steady state. We set up the following 1D density profile that is also used to test the bottleneck effect by \cite{Wieneretal2017b}
\begin{eqnarray}
\rho(x)=\rho_h+\left(\rho_c-\rho_h\right)\left[1+\tanh\left(\frac{x-x_0}{\Delta x}\right)\right]\nonumber\\
\times\left[1+\tanh\left(\frac{x_0-x}{\Delta x}\right)\right].
\label{eq:cloud}
\end{eqnarray}
This density profile is designed to mimic a high density, cold cloud with maximum density $\rho_c=1$ embedded in the low density hot background 
with minimum density $\rho_h=0.1$. The cloud is centered at $x_0=200$ with width $\Delta x=25$. This density profile is slightly different from equation (12) of \cite{Wieneretal2017b} because in our algorithm, we do not need a density gradient after the cold cloud to keep CR flux leaving from the right boundary. 
We use a uniform background magnetic field $B_x=1$ so that $v_A$ has a minimum at the density peak. Note that in this special case of a uniform field, from equation \ref{eq:CR_integral}, $E_{\rm c} \propto \rho^{2/3}$. The simulation domain covers the 1D range $x\in(0,1000)$ with 512 grid points. For the ghost zones at the left boundary, we fix $E_c=3$ and set $F_{c}$ to be the same as the value in the last active zone but with the opposite sign (reflecting boundary). Outflow boundary conditions are used for the right boundary, which means all the quantities are just copied from the last active zone to the ghost zones. The whole simulation domain is initialized with $E_c=10^{-6}$ and $F_{c}=0$. The background flow is static and all the gas quantities are kept fixed in this test. We set the diffusion coefficient $\sigma_c^{\prime}=10^{6}$ so that diffusion is negligible. 
 
 \begin{figure}[htp]
 	\centering
 	\includegraphics[width=1\hsize]{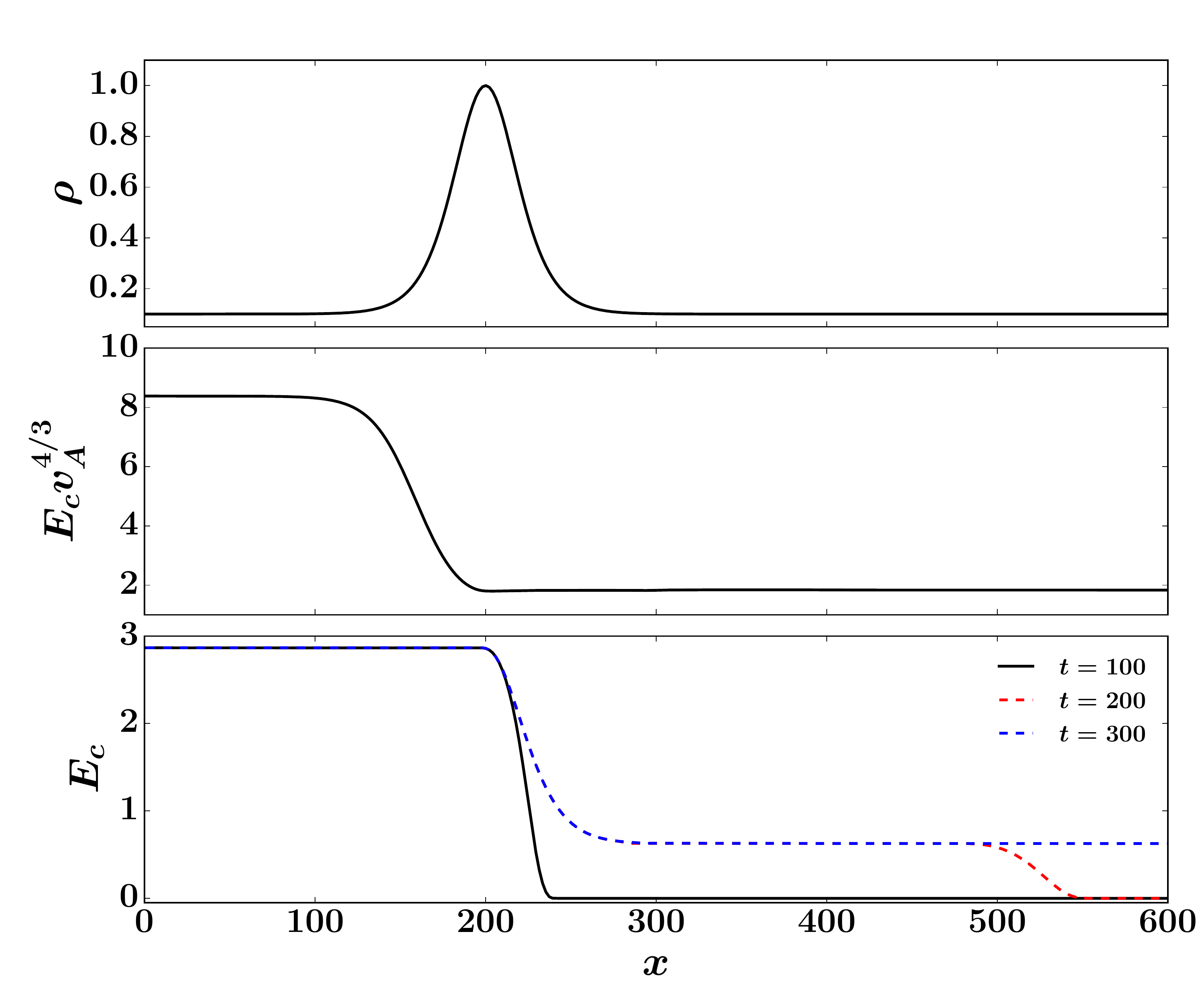}
 	\caption{\emph{Top}: density profile used in Section \ref{sec:bottleneck}. \emph{Middle}: profile of the quantity $E_cv_A^{4/3}$ after 
 	the solution reaches steady state. \emph{Bottom}: profiles of $E_c$ at three different times $t=100,200,1000$ respectively.}
 	\label{CRcloud}
 \end{figure}

Density profile, profiles of $E_c$ at three different time snapshots and the quantity $E_cv_A^{4/3}$ for the steady state solution are shown in Figure \ref{CRcloud}. After the simulation starts, CRs propagate into the simulation domain from the left boundary with positive streaming velocity $\bv_s$ because of the large $E_c$ set at $x=0$. For $x<200$, $v_A$ decreases with increasing $x$ because of the increasing density. If $E_c$ has non-zero spatial gradient during the steady state in this region, equation \ref{eq:EcvArelation} implies that $E_c$ needs to increase with $x$. However, this is inconsistent with the requirement that $\bv_s$ needs to be down the $P_c$ gradient. Therefore, the only allowed steady state solution is a constant $E_c$ at $x<200$. The minimum in $v_{\rm A}$ at $x=200$ serves as a `bottleneck'.In this region, $E_cv_A^{4/3}$ is also not a constant. For the region $x>200$, 
$v_A$ increases with increasing $x$ until it becomes flat at $x\approx 280$. The steady state solution now has a decreasing $E_c$ with $x$. 
The profiles of $E_c$ and $v_A$ also satisfy the relation \ref{eq:EcvArelation}  very well as shown in the middle panel of Figure \ref{CRcloud}. 
Beyond $x\approx 280$, all the quantities become flat again. 
 
\subsubsection{CR Diffusion in 1D}
\label{sec:diffusion}
\begin{figure*}[htp]
	\centering
	\includegraphics[width=0.49\hsize]{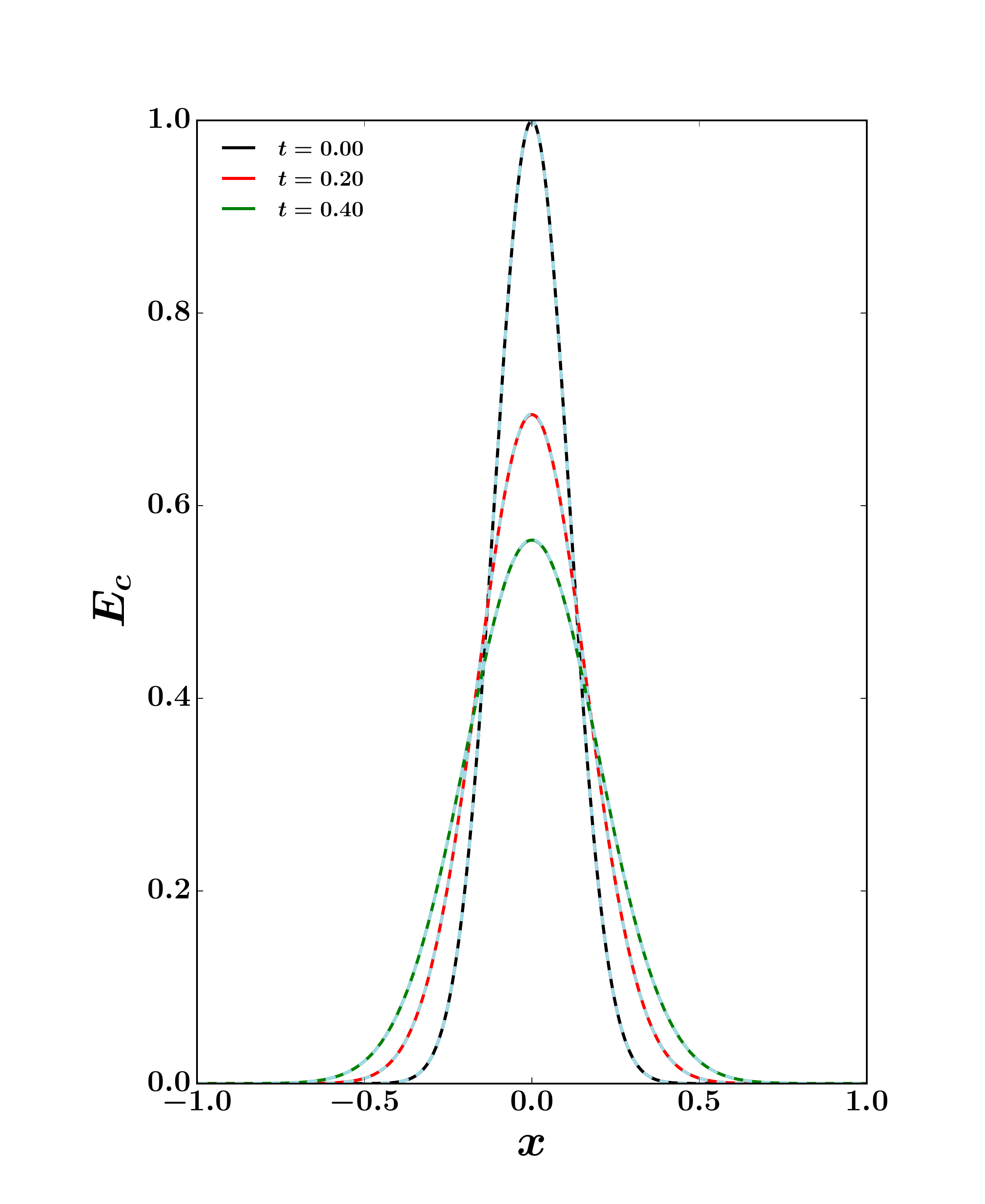}
	\includegraphics[width=0.49\hsize]{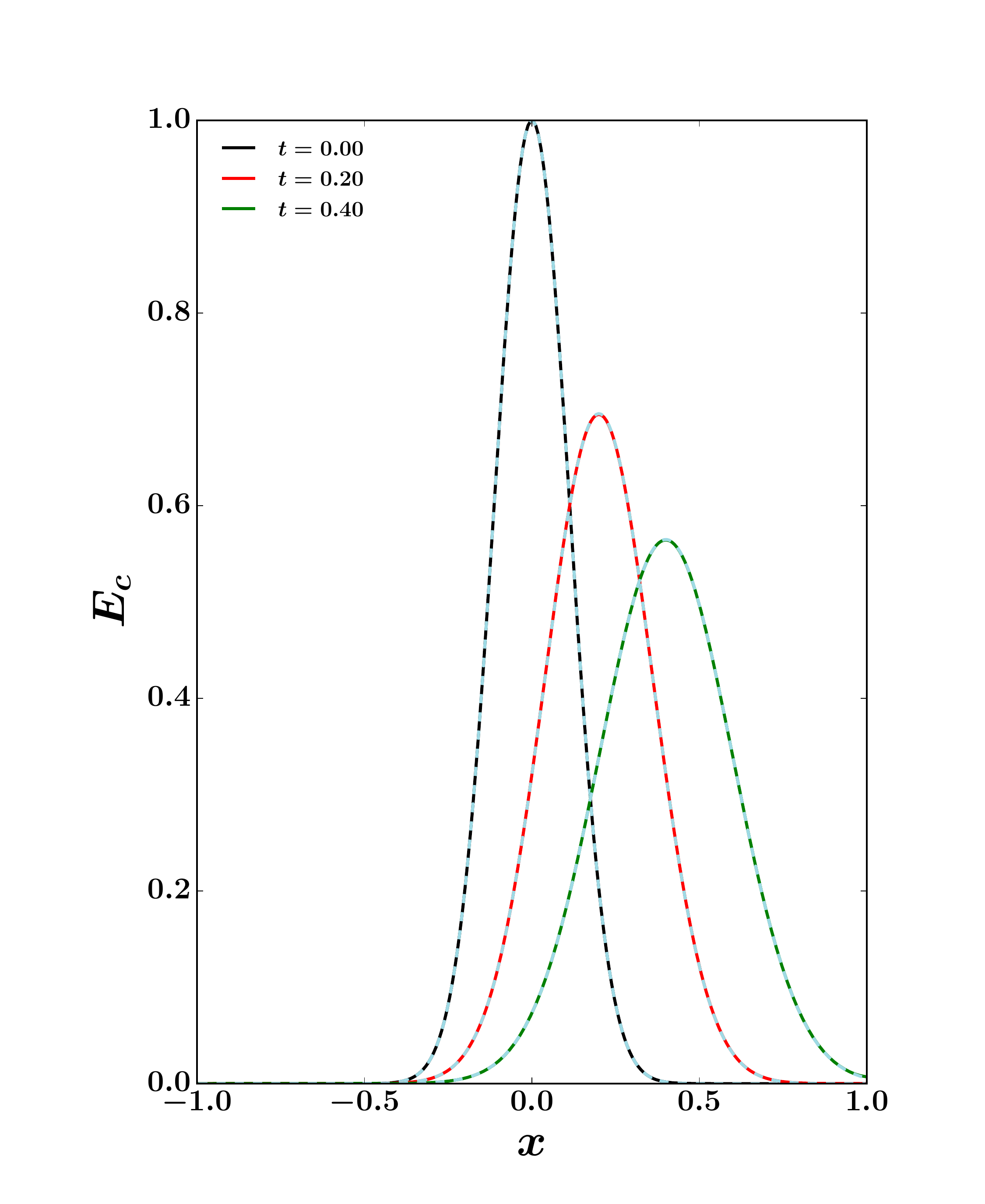}
	\caption{Test diffusion of CR energy density in a static (left panel) and moving (right panel)  fluid as described in Section \ref{sec:diffusion}. The solid lines are solutions from our new algorithm at three different time snapshots as indicated in the figure while the dashed lines are analytical solutions. The moving fluid in the right panel has a constant velocity $v=1$.
	The numerical solutions agree with the analytical solutions very well. }
	\label{diffusiontest}
\end{figure*}

To test the diffusion term in our new algorithm, we set the streaming velocity $v_s=0$ and fix the diffusion coefficient 
$\sigma_c^{\prime}=10$. The 1D simulation domain, boundary condition and resolution are all the same as in the Gaussian test described in the last section. The initial profile of $E_c$ is also taken to be the Gaussian profile $E_c=\exp\left(-40x^2\right)$. For a static fluid, evolution of $E_c$ is described by the analytical solution 
\begin{eqnarray}
E_c(t)=\frac{1}{1+160t/\left(3\sigma^{\prime}\right)}\exp\left[\frac{-40x^2}{1+160/\left(3\sigma^{\prime}\right)}\right].
\label{eq:diffsolution}
\end{eqnarray}
The numerical solutions at time $t=0.2$ and $t=0.4$ are compared with the analytical solutions in the left panel of Figure \ref{diffusiontest}, which agree very well. 

The same test can also be done for a moving fluid with a constant velocity $v=1$, while all the other parameters are the same. This test is useful to check that numerical diffusion caused by the advection term does not exceed the physical diffusion. The solution should be the same as in the static fluid except the center of CR profile is moving with the flow velocity. The result of this test is shown in the right panel of Figure \ref{diffusiontest}. For the analytical solution in this case, we just need to change $x$ to $x-vt$ in equation \ref{eq:diffsolution}. Our numerical solutions also agree with the analytical solutions very well.

\subsubsection{Anisotropic Diffusion}
\label{sec:2Ddiffusion}
\begin{figure*}[htp]
	\centering
	\includegraphics[width=0.33\hsize]{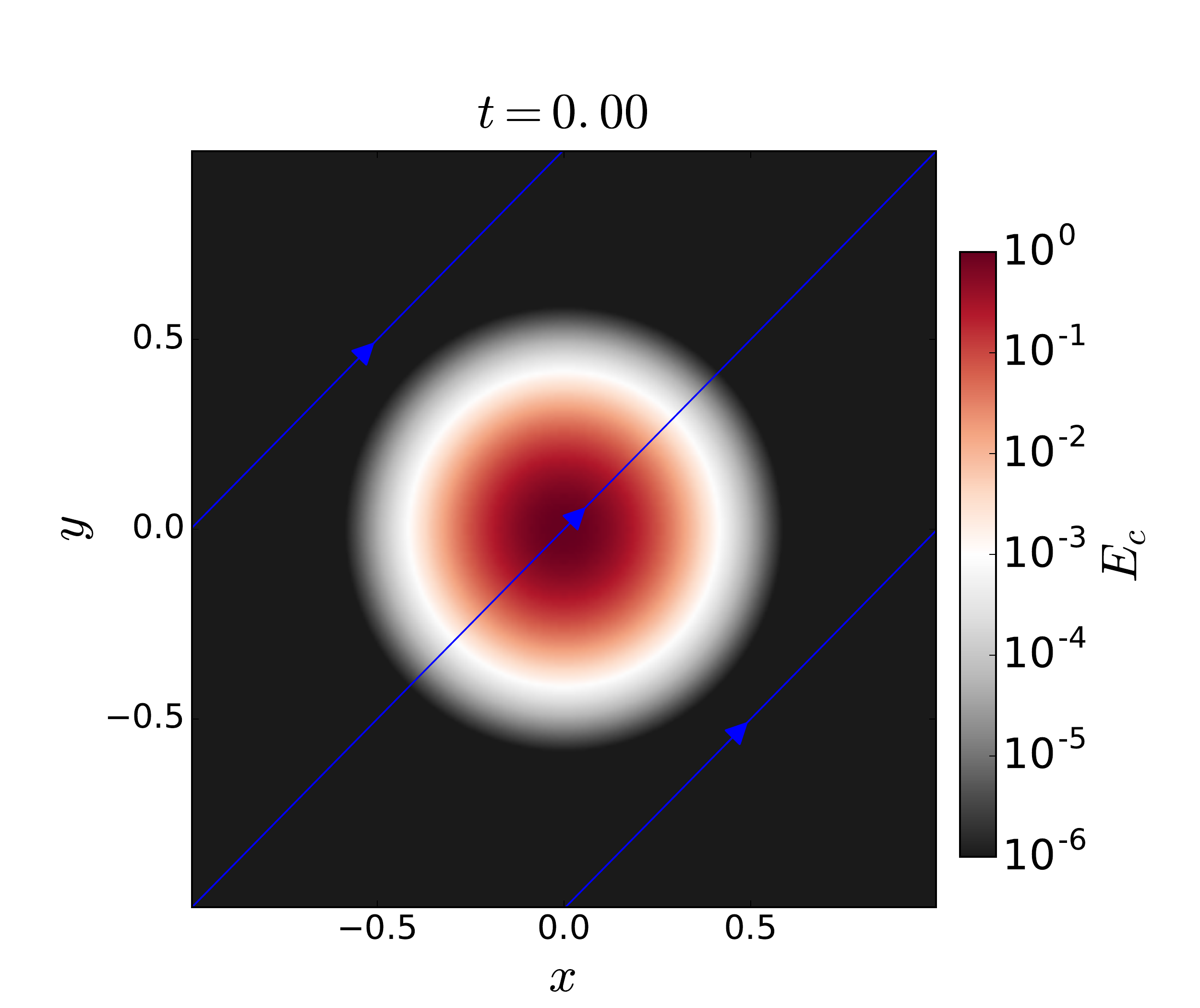}
	\includegraphics[width=0.33\hsize]{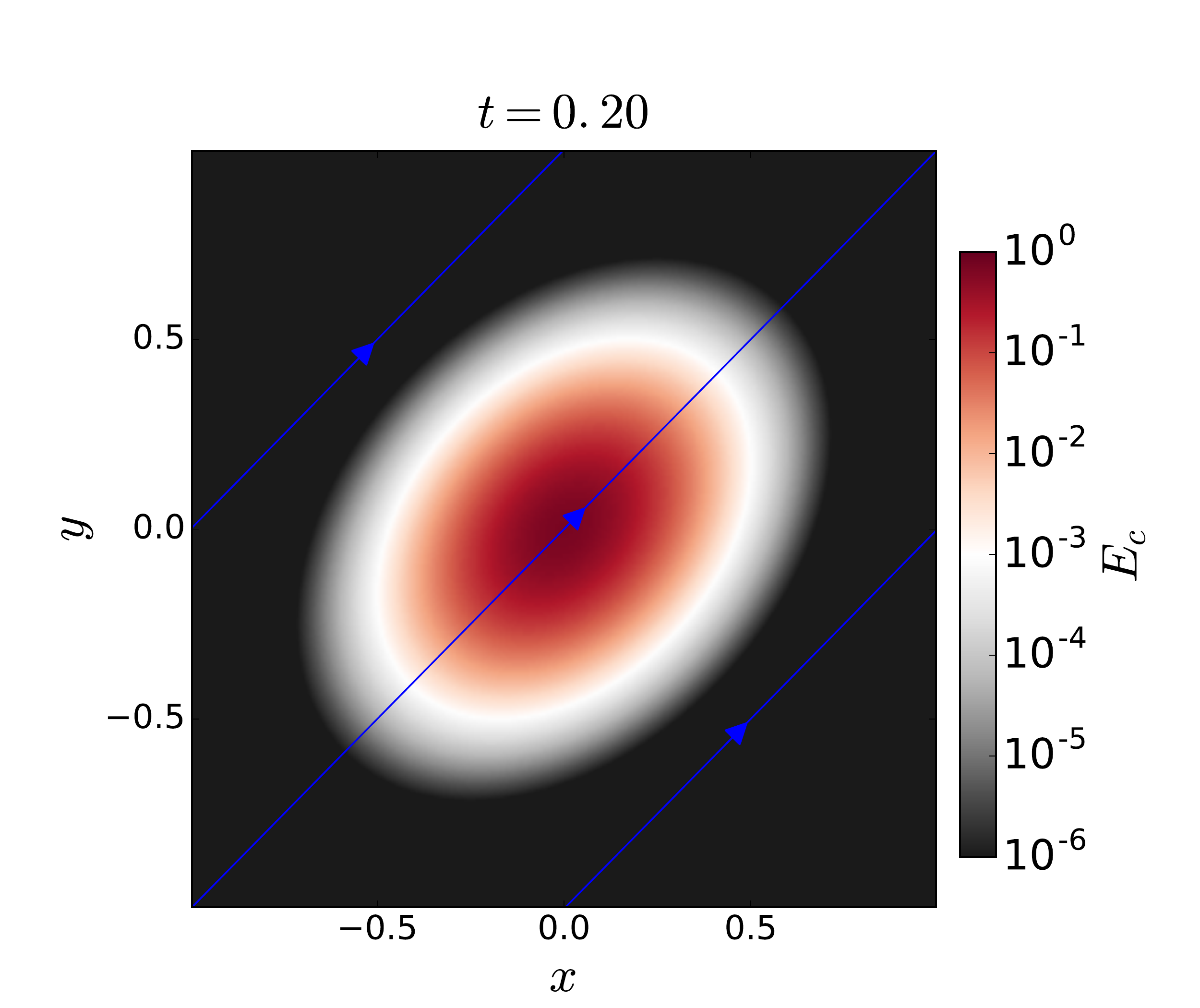}
	\includegraphics[width=0.33\hsize]{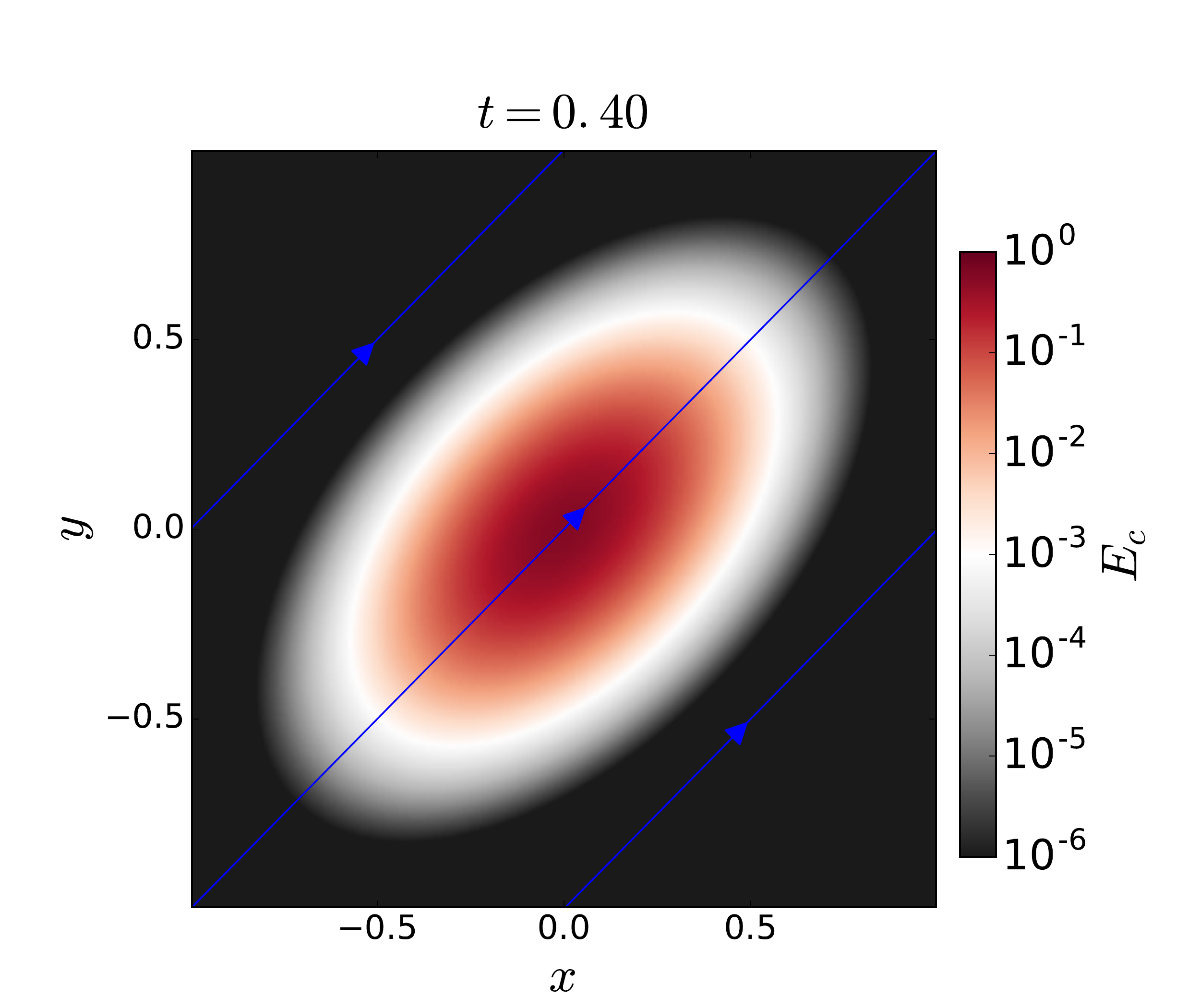}
	\caption{Test of the anisotropic diffusion of CRs along the magnetic field as described in Section \ref{sec:2Ddiffusion}. From left to right, the three plots show the 2D distribution solutions at $t=0,t=0.2,t=0.4$ 
		respectively. The blue line indicates the direction of the magnetic fields.}
	\label{2Ddiff}
\end{figure*}

Our new algorithm can also handle the case of anisotropic diffusion, which usually means that the normal diffusion coefficient 
$\sigma_c^{\prime}$ can have different values for the directions along and perpendicular to the magnetic fields. This is shown in Figure \ref{2Ddiff}, where we use the same setup as in Section \ref{sec:2Dstreaming}. The only difference is that we set the streaming velocity $\bv_s=0$ here and use $\sigma_c^{\prime}=10$ along the magnetic field lines while $\sigma_c^{\prime}=10^7$ 
for the direction perpendicular to the magnetic fields. Therefore, the diffusion time scale across the magnetic field lines is much longer than the diffusion time scale along the magnetic fields. The initial distribution of $E_c$ and solutions at $t=0.2$ and $t=0.4$ are shown in Figure \ref{2Ddiff}. As expected, CRs only diffuse along the magnetic fields. When we take the profiles along the diagonal line, we get exactly the same solution as in the left panel of Figure \ref{diffusiontest}. Distributions of CRs perpendicular to the magnetic fields also do not change with time, because the diffusion time scale is too long due to the large diffusion coefficient. 

\begin{figure}[htp]
	\centering
	\includegraphics[width=0.48\hsize]{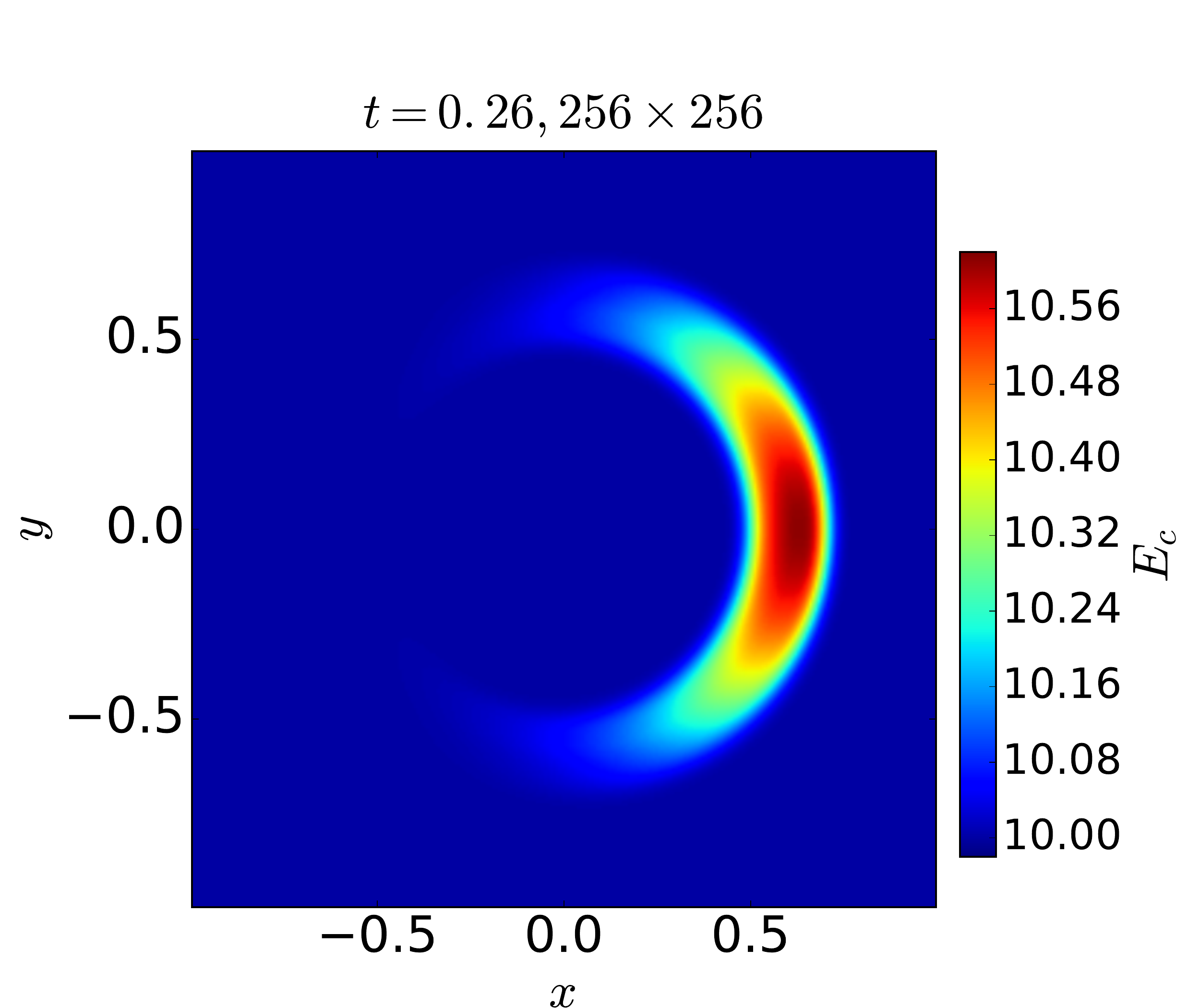}
	\includegraphics[width=0.48\hsize]{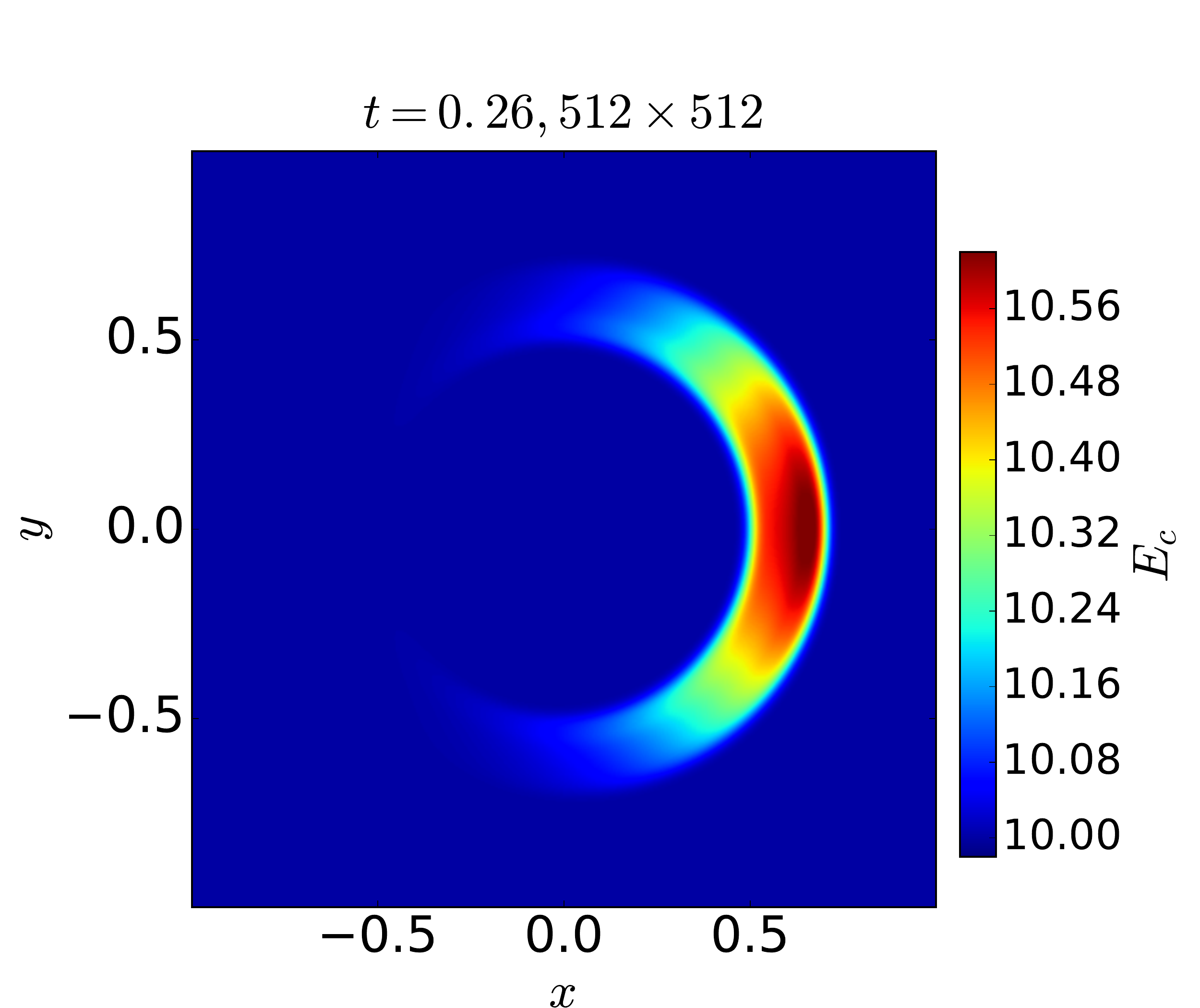}
	\includegraphics[width=0.48\hsize]{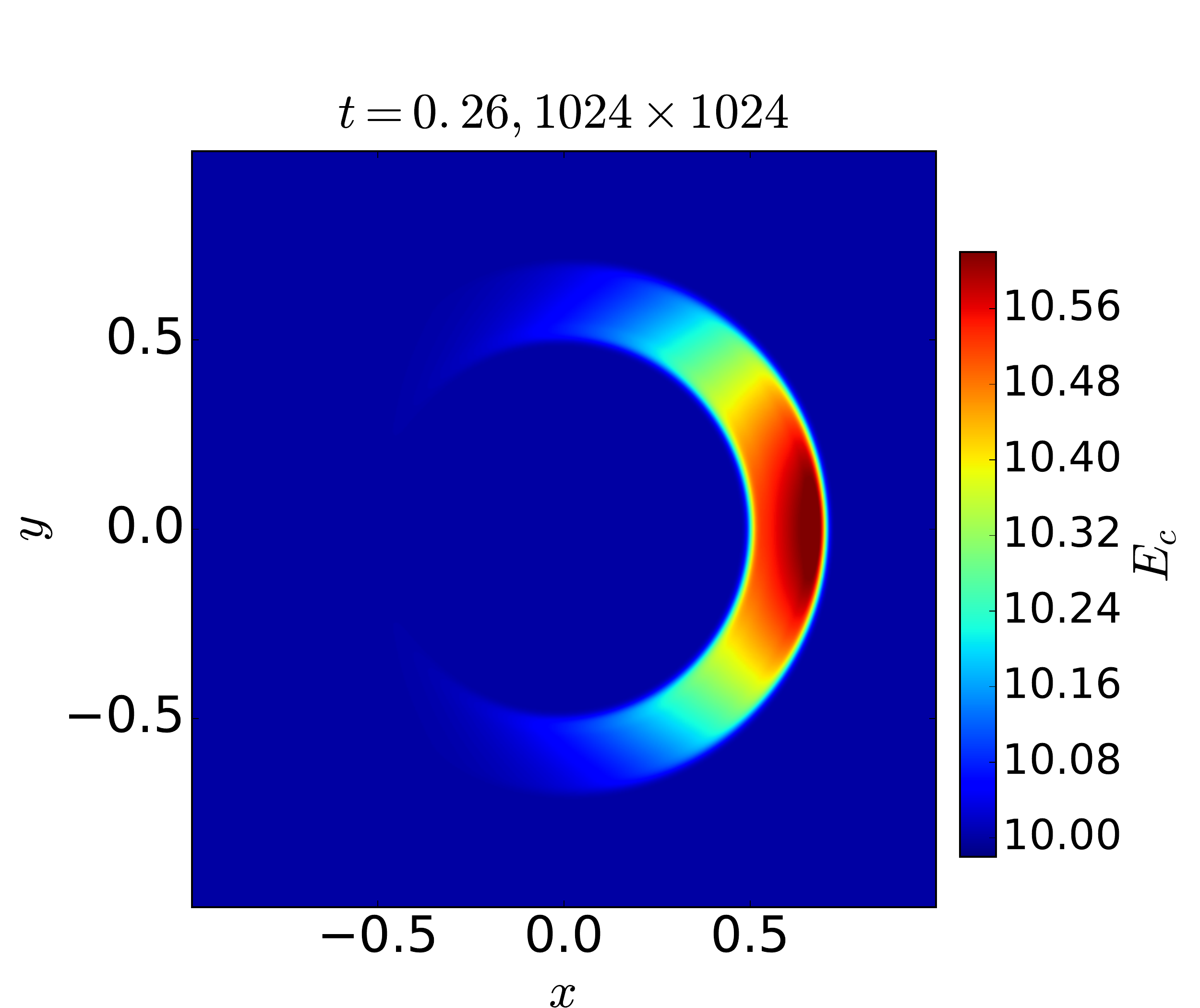}
	\includegraphics[width=0.48\hsize]{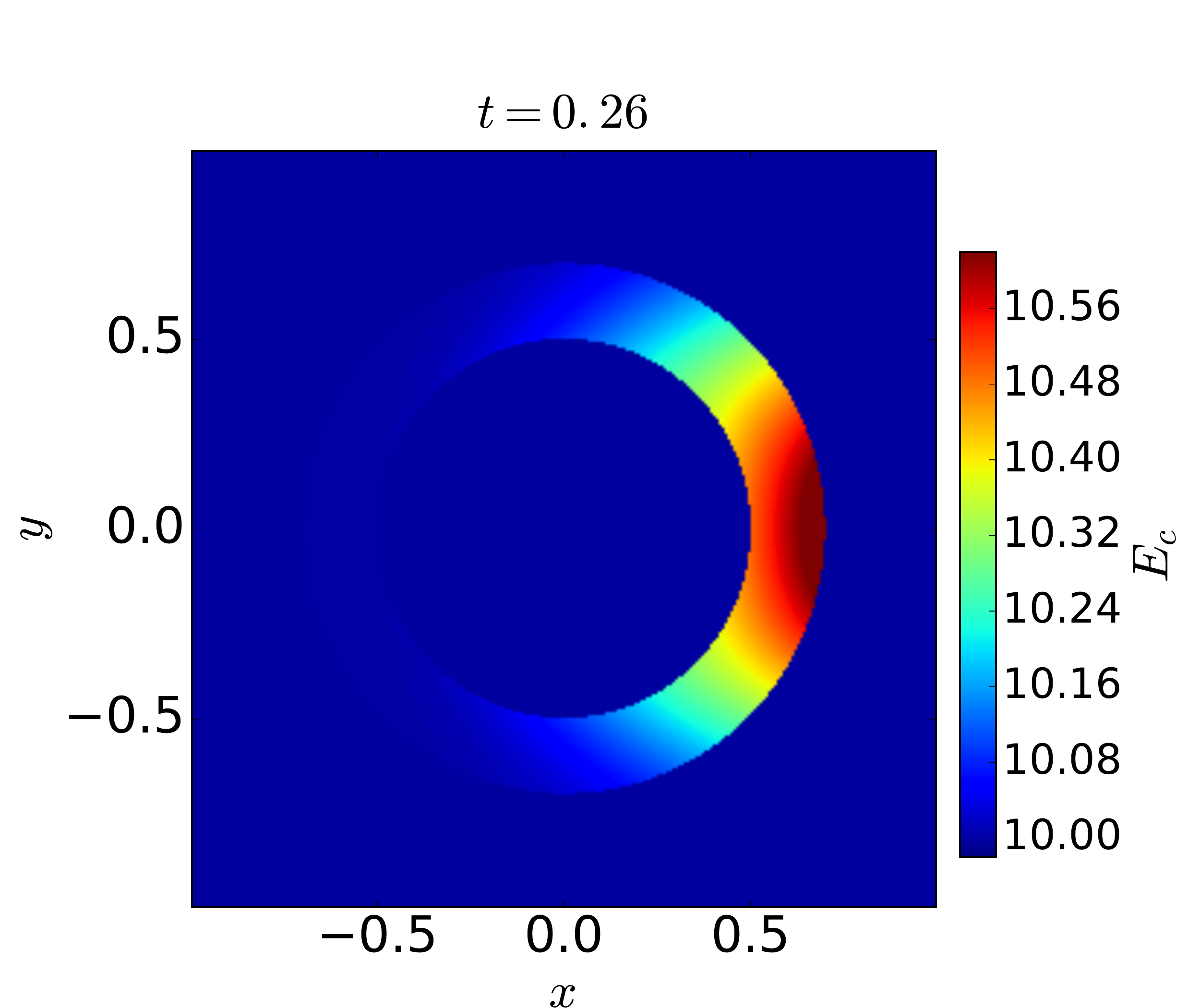}
	\caption{Spatial distributions of CR energy density $E_c$ for the anisotropic diffusion test 
		along magnetic field circles as described in Section \ref{sec:2Ddiffusion}. From top to bottom, left to right, 
	the first three plots show the snapshots of $E_c$ at time $t=0.26$ with different spatial resolutions as 
    labeled in each panel. The last panel is the analytical solution (equation ) at the same time.}
	\label{BringDdiff}
\end{figure}

Another widely used problem to test anisotropic diffusion is the diffusion of CRs along circles of magnetic fields in the 
Cartesian grids \citep{ParrishStone2005,SharmaHammett2011,Pakmoretal2016}. This test problem differs from the previous one in that the angles of diffusive flux with respect to the coordinate axes change continuously along the magnetic field lines. We setup a 2D domain $(x,y)\in(-1,1)\times (-1,1)$ with fixed magnetic fields as
\begin{eqnarray}
B_x(x,y)=-\frac{y}{r},\ \ B_y(x,y)=\frac{x}{r},
\end{eqnarray}
where $r=\sqrt{x^2+y^2}$. CR energy density is initialized as
\begin{eqnarray}
E_c(x,y)=\left\{
\begin{array}{ll}
12\ \  & \mbox{if } 0.5<r<0.7\ \ \mbox{and}\ \ \phi<\pi/12, \\
10\ \ & \mbox{else.}
\end{array}
\right.
\end{eqnarray}
Here $\phi=\atantwo(y,x)$ is the azimuthal angle with respect to the $x$ axis. The initial CR flux is set to be zero. 
The diffusion coefficient along the magnetic field is $\sigma_c^{\prime}=1$ while the diffusion coefficient perpendicular to the magnetic field lines is $10^6$. We turn off all the streaming terms in this test problem and all the fluid variables are kept fixed. Outflow boundary condition for $E_c$ and $\bF_c$ are used for both $x$ and $y$ directions. The analytical solution to describe the 2D spatial distribution of $E_c$ at time $t$ is \citep{Pakmoretal2016}\footnote{There are typos in equation 22 of \cite{Pakmoretal2016}, which are corrected in our equation \ref{eq:Bringsol}.}
\begin{eqnarray}
E_c(t)&=&10+\erfc\left[\left(\phi-\frac{\pi}{12}\right)\frac{r}{D}\right]\nonumber\\
&-&\erfc\left[\left(\phi+\frac{\pi}{12}\right)\frac{r}{D}\right],
\label{eq:Bringsol}
\end{eqnarray} 
where the coefficient $D\equiv \sqrt{4t/3\sigma_c^{\prime}}$ and $\erfc(x)$ is the complementary error function. 

Our numerical solutions at time $t=0.26$ for three different resolutions are shown in Figure \ref{BringDdiff}, which agree with the analytical solution as shown in the last panel of the same figure very well. As expected, CRs diffuse along the magnetic field lines with the correct diffusion speed with almost no diffusion in the perpendicular direction. Our scheme also preserves the monotonicity of $E_c$ and there is no any cell with $E_c$ smaller than the minimum value $10$ in our solutions. 
When we calculate the $L1$ norms of the numerical solutions with respect to the analytical solution for the three resolutions we have done, we find $L1\propto N^{-0.7}$ at both times $t=0.1$ and $t=0.26$. This is similar to the convergence rate reported by 
\cite{Pakmoretal2016} at late times, which has a slower convergence rate at early times. Notice that the time we choose is comparable to their early time in terms of the diffusion time scale due to different diffusion coefficients used in the tests. 

It is interesting to compare our scheme with traditional methods of handling anisotropic diffusion. The traditional approach involves terms like $\bfnabla\cdot\left[\bn\left(\bn\cdot\bfnabla E_c\right)/\sigma_c^{\prime}\right]$, which requires calculating the spatial gradient between neighboring cells along the magnetic field directions. During this step, a special limiter for the reconstruction of cell centered $E_c$ at the cell faces is usually required to 
avoid creating new extrema and get rid of numerical oscillations, and it does not always work \citep{SharmaHammett2011}.
Similar issues exist for anisotropic heat conduction \citep{SharmaHammett2007}, which shares the same mathematical properties. 
Without care, traditional approaches can easily violate the entropy condition in that the heat flux or CR flux can be opposite to the correct direction. In order to avoid small time steps due to the diffusion term, implicit or semi-implicit methods are usually adopted for this term \citep{Pakmoretal2016,SharmaHammett2011}, which typically requires a matrix inversion coupling all the grid cells. In our scheme, anisotropic diffusion handled by different diffusion coefficients along and perpendicular to the magnetic fields and the diffusion coefficients only appear in the source terms on the right hand side of equation \ref{neweq}. The source terms are added implicitly but only locally, which means we do not need to invert any matrix but this term is still unconditionally stable for any time step. When we calculate the spatial gradient for the terms $\bfnabla \cdot {\sf P_c}$ and $\bfnabla\cdot \bF_c$, we only 
need the standard van-Leer limiter for each direction independently, which is enough to 
ensure the entropy condition.

\subsection{Tests of the Full System}
In this section, we evolve the full coupled MHD and CR equations to demonstrate that our new algorithm is accurate and robust for a wide range of ratios between CR pressure and gas pressure. The gas adiabatic index is chosen to be $\gamma=5/3$ and we use $V_m=100$ for all the tests. Most of the tests are also done for arbitrary units and we only give dimensionless numbers.

\begin{figure}[htp]
	\centering
	\includegraphics[width=0.49\hsize]{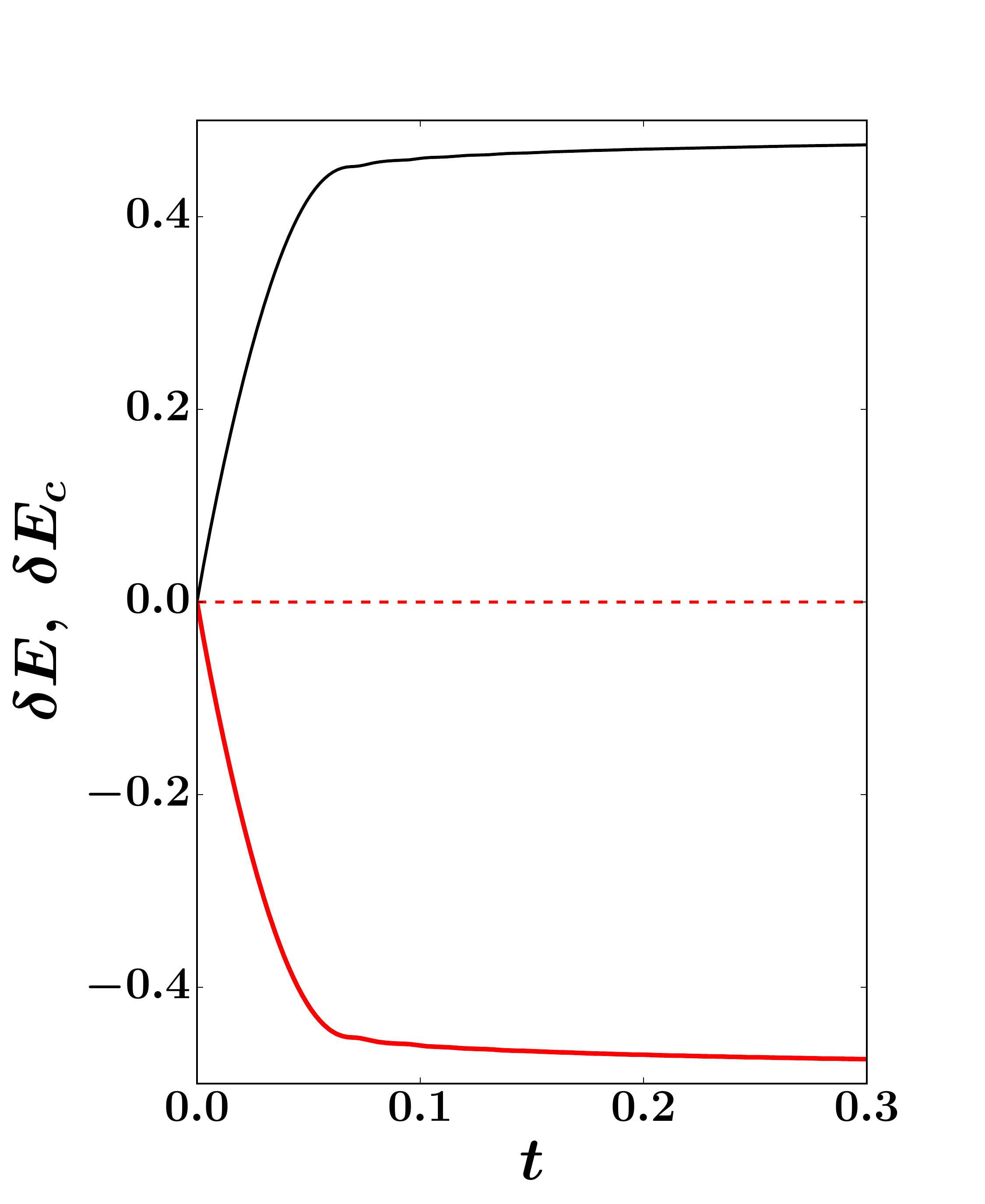}
	\includegraphics[width=0.49\hsize]{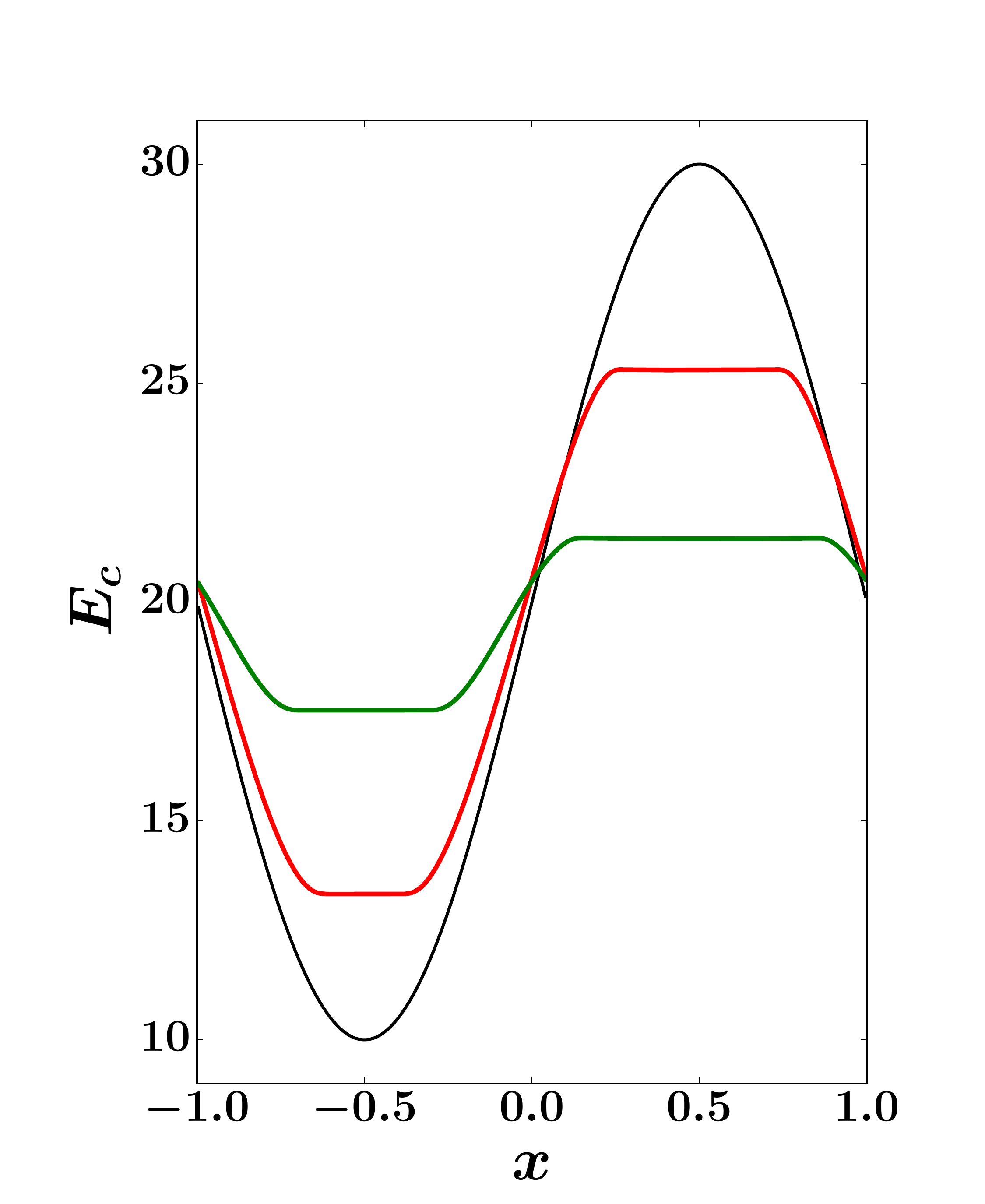} \\
	\includegraphics[width=0.49\hsize]{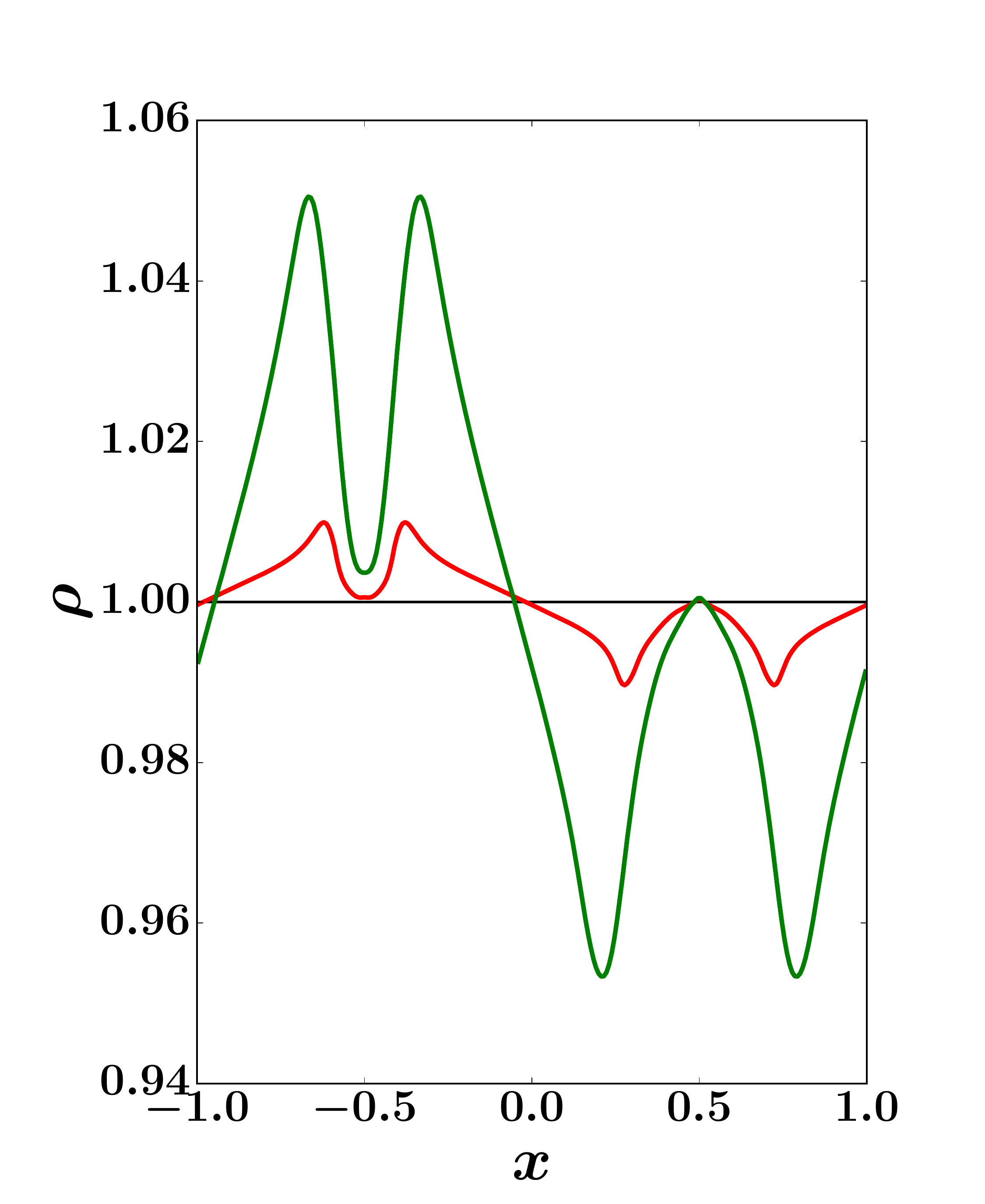}
   \includegraphics[width=0.49\hsize]{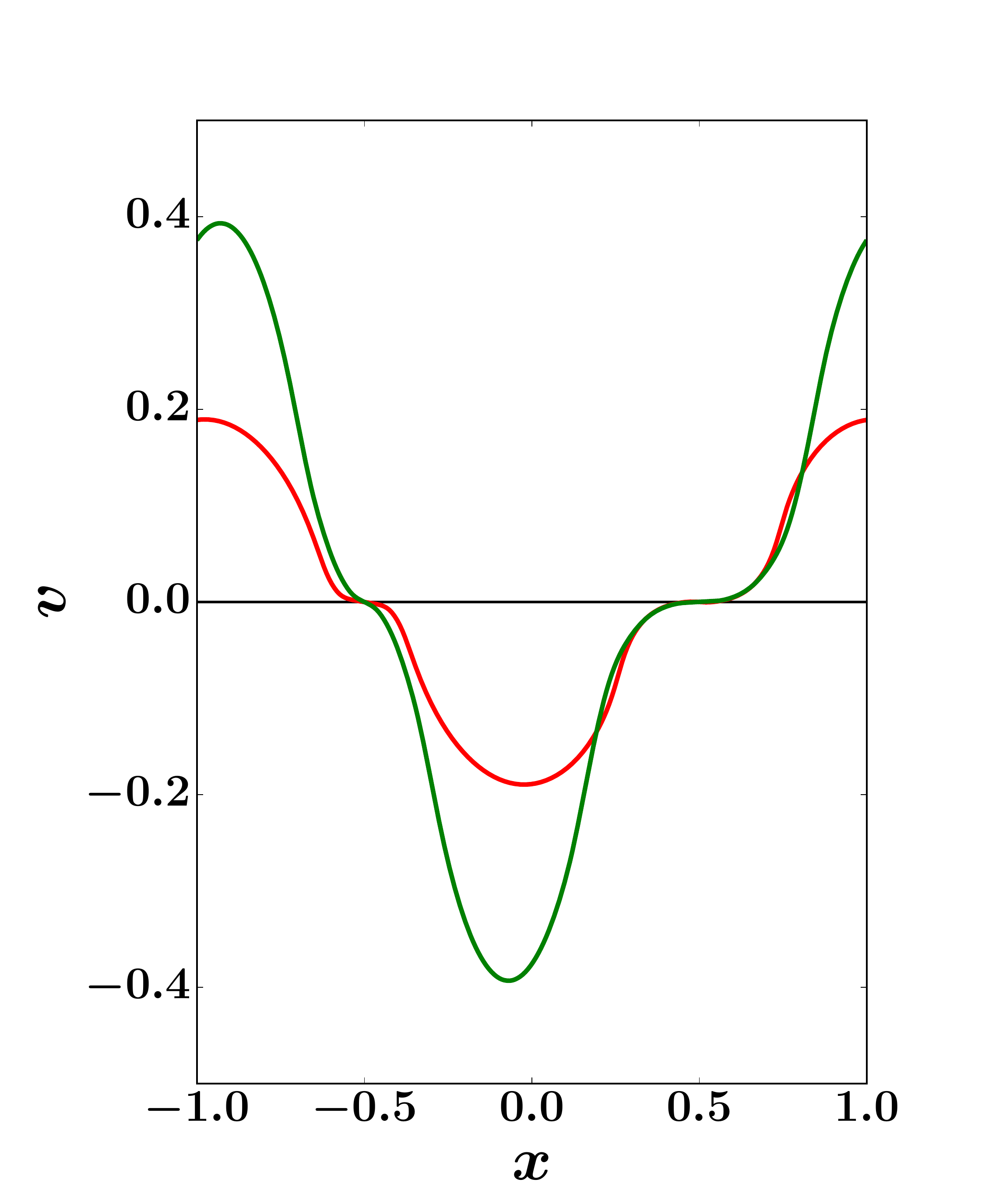}
	\caption{\emph{Top left}: Histories of total energy exchange in gas ($\delta E$, the black line) and CRs ($\delta E_c$, the solid red line) for the 1D CR driven waves described in Section \ref{sec:1DCRhydro}. The dashed line is the sum $\delta E+\delta E_c$, which stays at $0$ indicating that total energy is conserved. \emph{Top right}: initial 1D profile of $E_c$ (the black line) and profiles of $E_c$ at time snapshots $t=0.02$ (red lines) and $t=0.05$ respectively (green lines). The bottom two panels are the same as the top right panel except that they are for density $\rho$ and flow velocity $v$ respectively.}
	\label{CRhydro_1D_hist}
\end{figure}

\subsubsection{CR Driven Waves with Mixed Diffusion and Streaming}
\label{sec:1DCRhydro}
Since there are very few analytical solutions available for coupled CRs and fluid with both streaming and diffusion that we can compare with, we start with a simple 1D test to check that the numerical solutions we get are consistent with the equations we solve. In the 1D domain $x\in(-1,1)$, we initialize the CR energy density as $E_c=20+10\sin(\pi x)$. The gas density and pressure are set to be uniform values $1$. The initial flow velocity and CR flux are set to be $0$. We choose the field-aligned diffusion coefficient $\sigma_c^{\prime}=0.5$ and a constant Alfv\'en velocity $v_A=1$. Therefore, the diffusive flux should be comparable to the streaming flux. We use $256$ grids points over the simulation domain and periodic boundary conditions. 

Histories of the energy changes for gas and CRs as well as snapshots of $E_c$ are shown at the top two panels of Figure \ref{CRhydro_1D_hist}, while the corresponding density and velocity profiles are shown in the bottom two panels. Because of the non-zero gradients of $P_c$, both diffusive and streaming flux are generated down the $P_c$ gradient. The CR energy density $E_c$ becomes flat at the initial peaks due to the streaming term (top right panel). The flow velocity $v$ has the opposite sign to the $E_c$ gradient as expected, which also causes corresponding compression and rarefaction of density $\rho$. During this process, the total CR energy always decreases with time while the total gas energy always increases with time. The total energy change is always $0$ to roundoff error because of our conservative scheme. We calculate $(v+v_s)\partial P_c/\partial x$ at the snapshots $t=0.02$ and $t=0.05$ and they agree with the energy exchange rate (the slope of the lines in the top left panel of Figure \ref{CRhydro_1D_hist}) very well.  The system eventually settles down to a state with uniform $E_c$ and then the energy exchange stops. \label{key}
Note that there is a fundamental asymmetry in CR transport at peaks and troughs (transport out of a peak is faster than transport into a trough), resulting in the asymmetrical profiles in $E_{\rm c}, \rho, v$ which we see. This is true even if only CR streaming is operating \citep{Sharmaetal2009}.

\begin{figure}[htp]
	\centering
    \includegraphics[width=1\hsize]{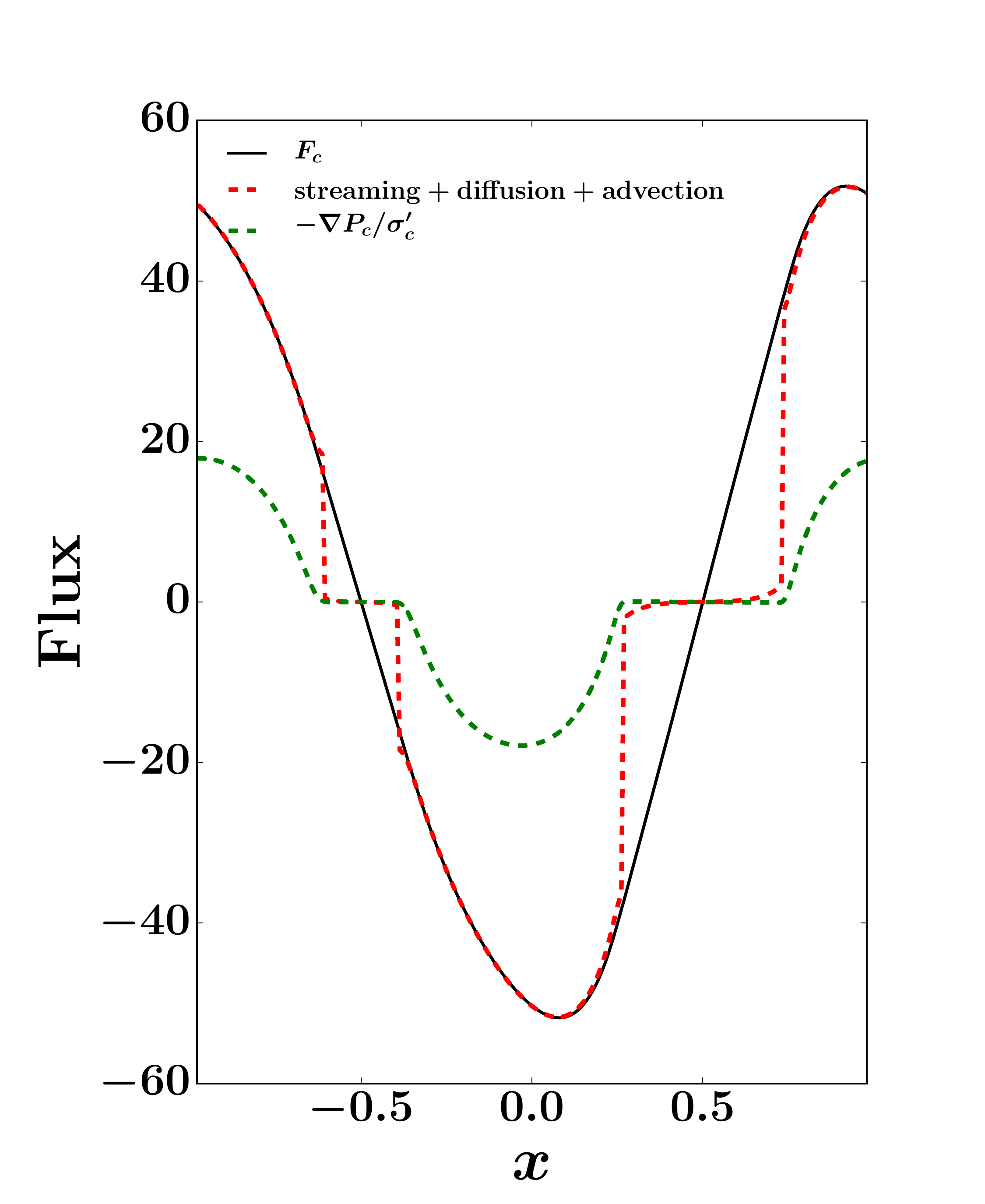}
	\caption{Profile of CR flux $F_c$ at time $t=0.02$ for the CR driven wave test shown in Sec \ref{sec:1DCRhydro}. The solid black line is the total flux in the numerical solution while the dashed green line is the diffusive part. The dashed red line is the sum of 
		$(v+v_s)(E_c+P_c)$ and the diffusive flux, which agrees with $F_c$ in the numerical  solution very well whenever gradient of $E_c$ is not close to zero. The differences near the regions  $x=\pm 0.5$ are due to the 
		fact that CRs free-stream at extrema, and the canonical expression for flux is not longer valid (see also the top panel of Figure \ref{crflux}). }
	\label{CRhydro_1D_flux}
\end{figure}

In steady state, the CR flux $F_c$ can be decomposed into three components, including the streaming flux $v_s(E_c+P_c)$, advective flux $v(E_c+P_c)$ and the diffusive flux $-\bfnabla P_c/\sigma^{\prime}_c$, as given by equation \ref{CR_oldflux}. This is checked at snapshot $t=0.02$ in Figure \ref{CRhydro_1D_flux}. When we calculate $v_s(E_c+P_c)$, we set it to be zero whenever the difference of $E_c$ between neighboring cells is smaller than $10^{-3}$. In the region where $\bfnabla P_c$ is not close to 0, sum of the three components agrees with $F_c$ calculated by the code very well. When $\bfnabla P_c$ is close to $0$, the total diffusion coefficient $\sigma_c$ also goes to $0$ and CR flux $F_c$ from our scheme is different from the simple prescription $v_s(E_c+P_c)$ as explained in Figure \ref{crflux}.

\subsubsection{Shocks with CRs}
\label{sec:crshock}


CR modified shocks are of considerable physical interest, and also provide a stringent test of our numerical scheme, given the sharp discontinuities which appear. Here, we do not aim at a physical study of CR modified shocks, but aim solely at testing how our transport module behaves in the presence of discontinuities. In particular, we do not model CR injection, acceleration or escape at shocks. Thus, in the simulations below, the jump in $P_{\rm c}$ at the shock is primarily due to compression\footnote{It does not behave exactly like adiabatic compression because the CRs and gas can exchange energy.}, and the gas pressure undergoes a much stronger jump at the shock. Our results should therefore not be interpreted physically. We plan to conduct a more careful study of CR modified shocks in the future. Note that straightforward analytic solutions exist for CR shocks in the limit where CRs are frozen into the fluid, both without \citep{pfrommer06} and with \citep{Pfrommeretal2017} CR injection. 

Once CR streaming and diffusion are included, the situation becomes more complex. It is useful to remind ourselves of the governing equations for a steady-state CR modified shock, in the absence of CR injection. Total mass, momentum and energy must be conserved: 
\begin{eqnarray}
\label{eqn:mass_conservation}
\rho \tilde{U} &=& \ {\rm const} \\
\label{eqn:energy_conservation}
\rho \tilde{U}^{2} + P_g + P_c &=& {\rm const} \\
\label{eqn:momentum_conservation}
\rho \tilde{U} \left[ \frac{1}{2} \tilde{U}^{2} + \frac{\gamma}{(\gamma-1)} \frac{P_g}{\rho} \right] + \tilde{F}_c &=& {\rm const}
\end{eqnarray}
where $\tilde{U}= V-U_{\rm s}$, $\tilde{F} = F_c - (E_c+P_c) U_{\rm s}$ are quantities measured in the shock frame, and $U_{\rm s}$ is the shock velocity. The shock velocity can either be measured directly from the simulations, or obtained from the equation of continuity: 
\begin{equation}
U_{\rm s} = \frac{\rho_{2} V_{\rm 2}-\rho_{1}V_{1}}{(\rho_{2} -\rho_{1})} 
\end{equation}
where 1,2 refer to upstream and downstream quantities respectively. Note that for our setup, $V_{2}=0$ by symmetry. 

These conservation laws only give 3 equations for the 5 variables $\rho, \tilde{U}, P_g, P_c, \tilde{F}_c$. The final two equations can be found by setting the time-derivatives in equation \ref{neweq} to zero. Thus the steady-state CR flux is given by equation \ref{CR_oldflux}, and 
 in the shock frame: 
\begin{equation}
\nabla \cdot \bF_c = (\bv + \bv_s) \cdot \nabla P_c, 
\label{eqn:CR_conserved}
\end{equation}
where we ignore CR sources and sinks $Q$ (since we neglect both CR injection and escape). Equation \ref{eqn:CR_conserved} describes the energy transfer from CRs to gas, by decelerating and heating the gas. Combined with the conservation equations, it can be cast in the revealing form: 
\begin{equation}
\frac{\tilde{U} \rho^{\gamma}}{(\gamma-1)} \nabla \left( \frac{P_g}{\rho^{\gamma}} \right) = -\bv_s \cdot \nabla P_c,
\label{eqn:entropy}
\end{equation}
i.e., CR wave heating increases gas entropy. This can also be obtained directly from the gas energy equation written in entropy form. 
In the case when magnetic fields are evolved self-consistently, there is the fourth integral \citep{volk81}:
\begin{equation}
\left(1 + \frac{M_{\rm A}}{(\gamma -1)} \right)^{2 \gamma} \left( P_g + \frac{(\gamma-1) \rho \tilde{U}^{2} (2 \gamma M_{\rm A} + 1-\gamma)}{2 (\gamma-1)} \right)
\label{eqn:wave_adiabat}
\end{equation} 
where $M_{\rm A} = \tilde{U}/v_{\rm A}$ is the Alfven Mach number. Equations \ref{eqn:mass_conservation},\ref{eqn:energy_conservation},\ref{eqn:momentum_conservation}, \ref{CR_oldflux} and \ref{eqn:wave_adiabat} are four algebraic and one differential equation (due to the $\nabla P_c$ term in equation \ref{CR_oldflux}) in 5 variables. One can use the algebraic equations to reduce the system to a single first order differential equation in one variable which can be integrated as a function of distance to obtain the shock structure \citep{volk84}. An alternative is to eliminate $\rho,P_c, F_{\rm c}$ with the linear equations \ref{eqn:mass_conservation},\ref{eqn:energy_conservation}, and \ref{CR_oldflux}, and study integral curves of the reduced system in the $\tilde{U}, P_g$ plane (e.g., see \citet{drury81} for the case where wave heating is ignored); this is useful in gaining insight into the location of gas sub-shocks. Here, we eschew these procedures, but simply check that our numerical solutions satisfy these conditions across the shock. 


We set up three 1D shock tests in the Cartesian domain $x\in\left(-5,5\right)$. Initially, a uniform gas density $\rho=1$ and pressure $P_g=1$ are set in the whole simulation domain. The flow velocity $v=10$ (for $x<0$) and $v=-10$ (for $x>0$). Uniform CR energy density $E_c$ is initialized in the whole simulation domain and we use three different values $E_{\rm c}=1, 50, 200$ (and thus $P_c=1/3,50/3,200/3$) to test the effects of different ratios between CR pressure and gas pressure. The CR flux is initialized as $F_c=4vE_c/3$.  We do not evolve magnetic fields in this test but just use a constant\footnote{We choose $v_A=1$ independent of B-fields and density jumps, which therefore do not affect the streaming properties of the CRs.} $v_A=1$ when we calculate the streaming velocity $\bv_s$. Therefore, the integral \ref{eqn:wave_adiabat} will not be satisfied. Instead, we check the relation \ref{eqn:CR_conserved} directly. We choose the diffusion coefficient $\sigma_c^{\prime}=10$. 
Variables in the left and right ghost zones are fixed to be these initial values. We use a $1024$ grid for the whole simulation domain. 

Some comments on these choices are in order. In principle, CRs can diffuse ahead of the shock to form a shock precursor, with $P_c, F_c$ continuous across the shock. In our case, the size of the precursor $l \sim 1/(\sigma^{\prime}_{c} v) \sim 0.01$ is too small to have a noticeable effect on the system. Similarly, since the Alfven speed $v_A \ll v$, most of the energy transfer between the CRs and gas is mediated by CR pressure forces $\bv \cdot \nabla P_c$ rather than wave heating $\bv_s \cdot \nabla P_c$. To a good approximation, our setup behaves like a tightly coupled two-fluid system. This maximizes discontinuities at shocks, which is more demanding for our code. 

\begin{figure*}[htp]
	\centering
	\includegraphics[width=0.33\hsize]{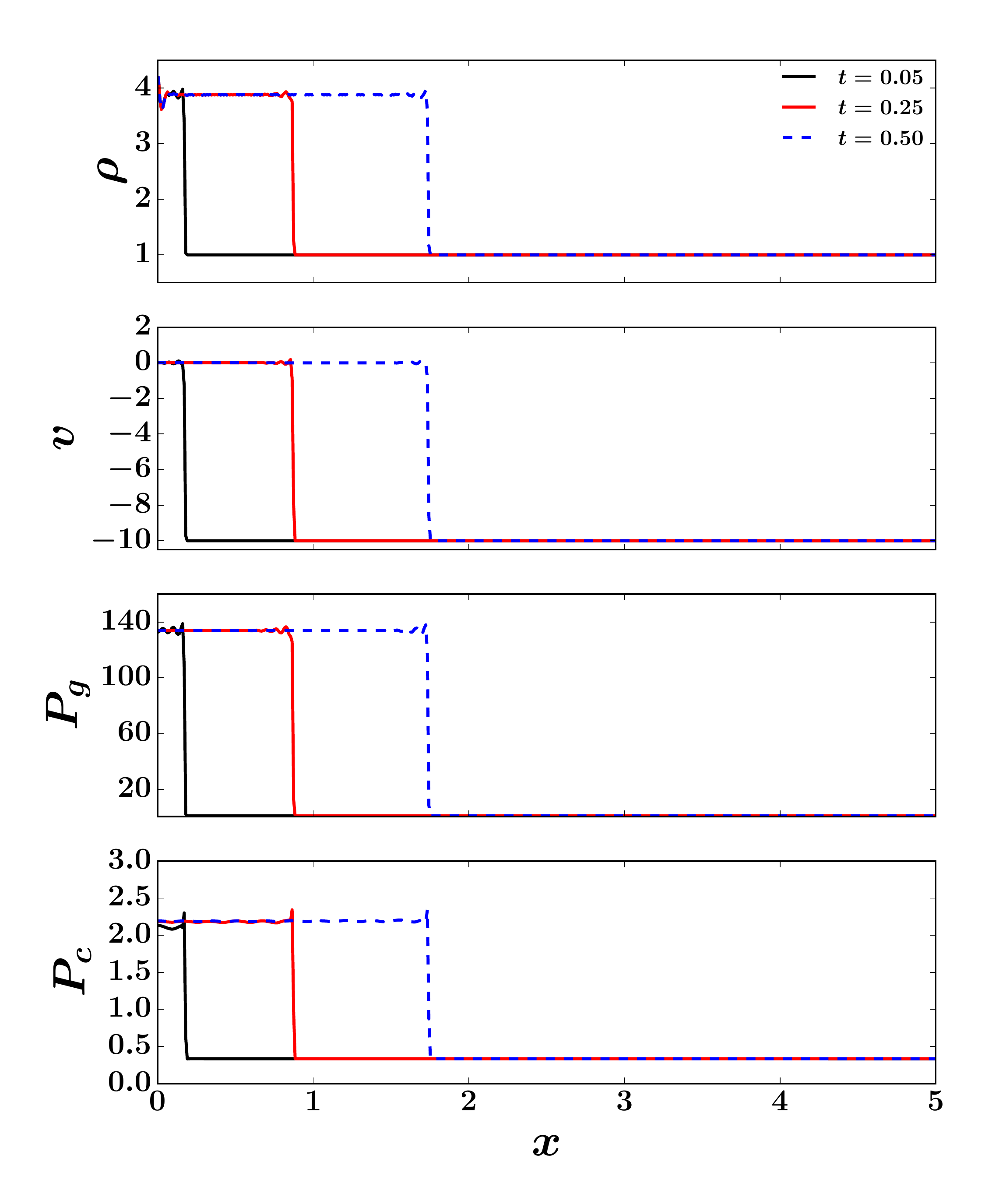}
	\includegraphics[width=0.33\hsize]{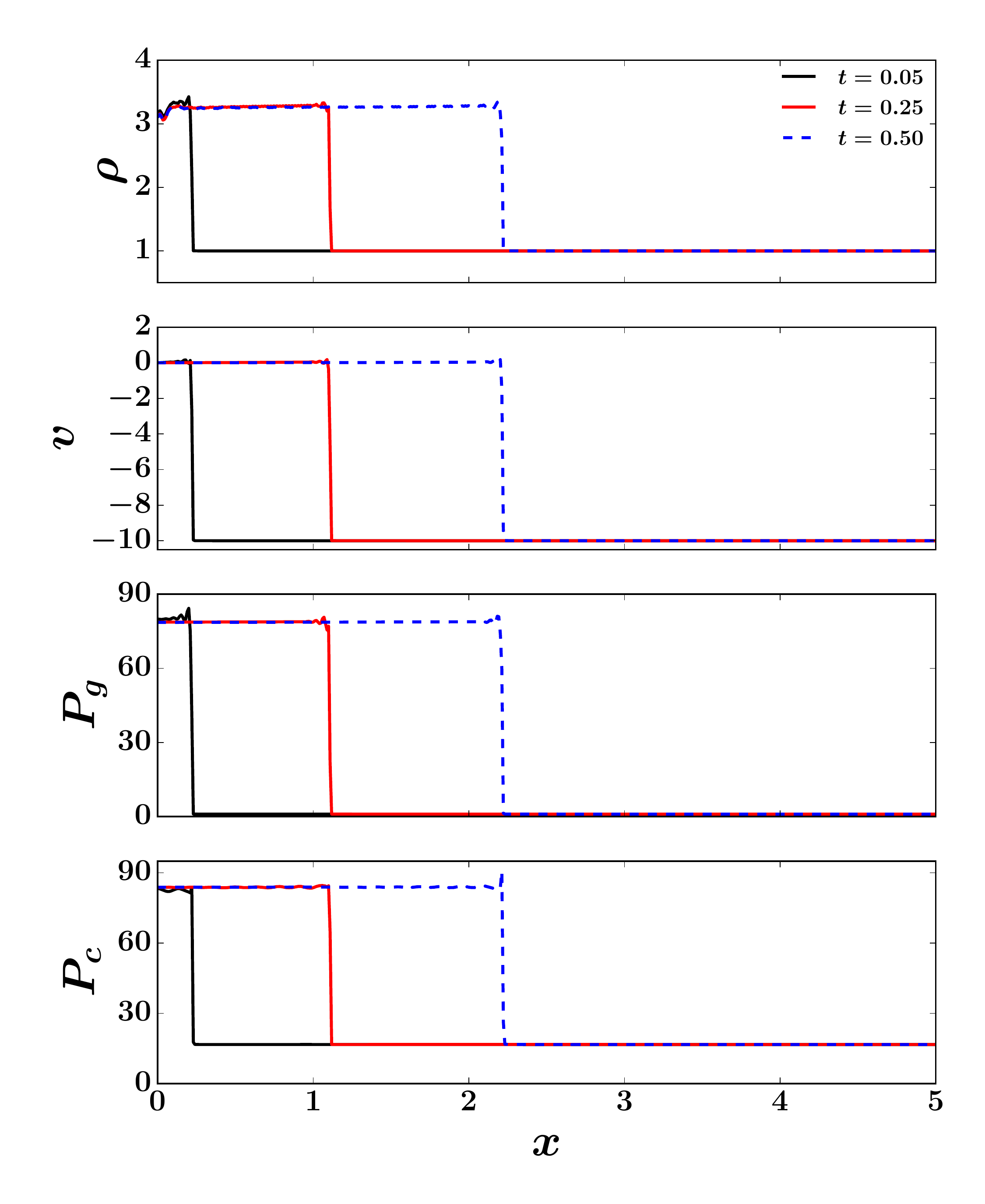}
	\includegraphics[width=0.33\hsize]{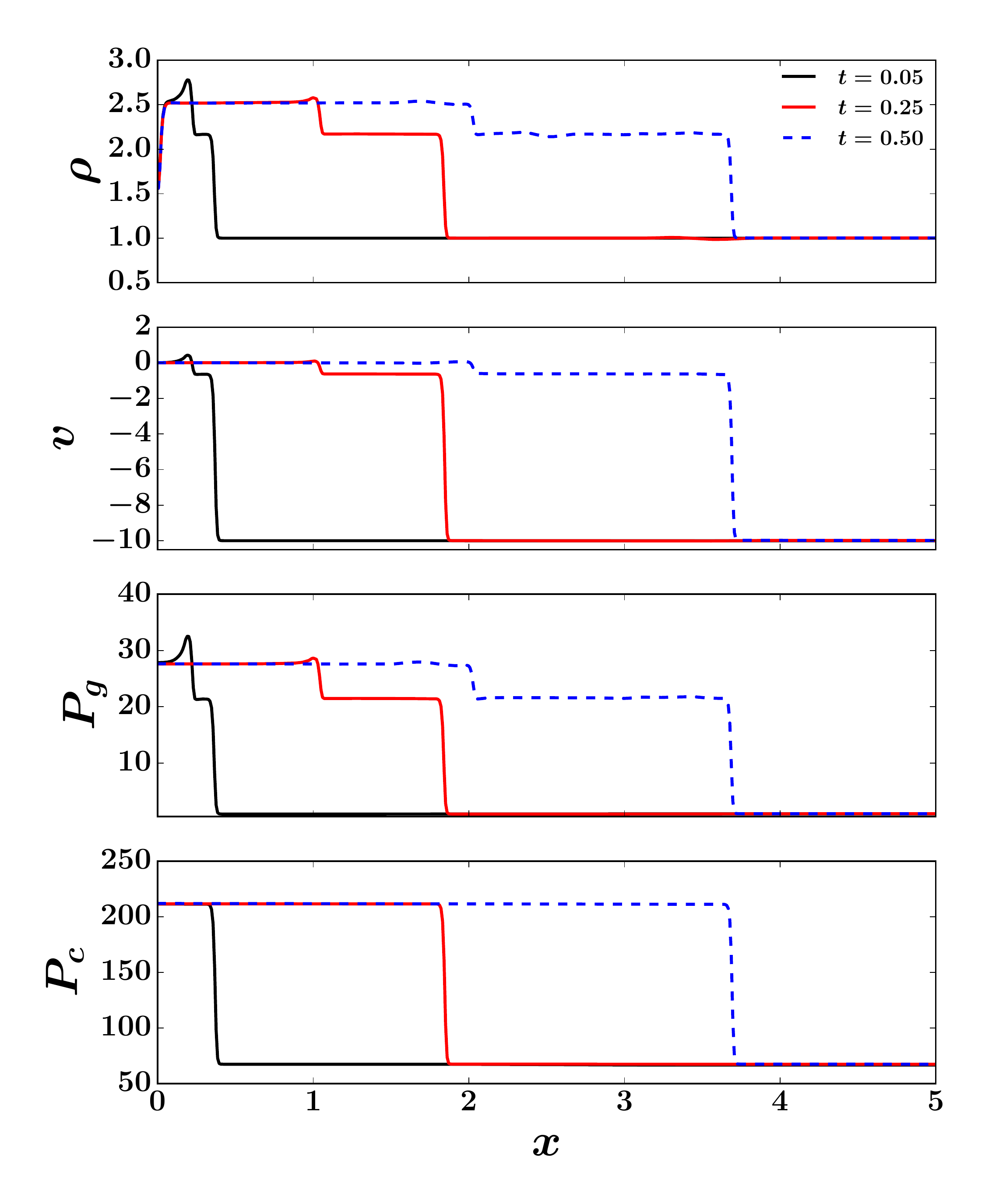}
	\caption{Profiles of density $\rho$, flow velocity $v$, gas pressure $P_g$ and CR pressure $P_c$ at three different times $t=0.05, 0.25, 0.5$ for the shock test described in Section \ref{sec:crshock}, as the shock front moves from left to right. From left to right, the three columns are for three different upstream CR pressure $P_c=1/3,50/3,200/3$ respectively while all the other gas variables are the same. 
	Our new algorithm can handle all these shocks without any issue.  }
	\label{crshock}
\end{figure*}

Shocks are formed at the center where gas from the left and right hand sides collides and then propagate to both sides. Because of the symmetry, we only show variables in the right half part of the box in Figure \ref{crshock} for the three different values of $E_c$. For each case, we show the profiles at three different time snapshots. Symmetry also requires that flow velocity $v=0$ at $x=0$. As a comparison, we have also done a test with the same setup but without any CRs. The solution looks very similar to the first case when the upstream CR pressure is only $P_c=0.33$. From the upstream (right of the shock front) to the downstream (left of the shock front), density jumps by almost a factor of $4$, which is consistent with a strong shock with sonic Mach number $\mathcal{M}=10$. Most of the kinetic energy in the upstream is converted to the gas internal energy; the downstream gas pressure is larger than CR pressure by a factor of $\sim 60$. CR pressure is increased in the downstream because CRs are advected with the flow and $F_c$ converges at the shock front. The CR pressure jump is comparable with that expected from adiabatic compression, which is $P_{\rm c} \approx 4^{4/3}/3 = 2.1$, roughly consistent with simulation results\footnote{It will not be exactly the same, due to energy exchange between the CRs and gas}. The shock speed from the continuity equation of $V_s=-V_{\rm 1} \rho_{1}/(\rho_{2}-\rho{1}) = 4/3$ is also consistent with the locations of the shock front shown in Figure \ref{crshock}. Because we do not use any diffusive regularization in our algorithm, the sharp shock front is captured in our solution without much artificial broadening. 

When the upstream CR pressure is increased to $P_c=16.67$, the downstream CR pressure becomes comparable to the gas pressure and the solution is modified. The density jump from upstream to the downstream is only a factor of $\sim 3$ and consequently the shock propagates faster. This can be easily understood in terms of the total effective sound speed $\sqrt{(\gamma_{\rm c} P_c+ \gamma_{\rm g} P_g)/\rho}$ because CRs and gas are tightly coupled. 
The upstream effective Mach number $\mathcal{M}$ is only $\sim 2$ with respect to this effective sound speed, which is much smaller compared with the previous case.

In the third case when $P_c=66.67$ in the upstream, the density jump is even smaller and the shock propagates faster. The downstream CR pressure now becomes larger than the downstream gas pressure by a factor of $\sim 7$. Our algorithm can still handle it without any numerical issue and there is no need to tune any free parameter. The new feature in the CR pressure dominated regime is the clear separation of two jumps in the fluid variables. The first jump around $x=3.8$ at time $t=0.5$ involves all fluid and CR variables, while the second jump at the same time around $x=2$ only happens for gas variables. In the frame of the shock, the downstream gas is moving away from the shock with a speed larger than the shock speed and it will catch up with the gas in the middle of the simulation domain. In the case of strong CR pressure driven shock, the downstream velocity is also larger than the local gas sound speed but smaller than the total effective sound speed. 
Therefore, it causes another jump of the gas variables to reach the zero velocity at the middle. The presence of a gas sub-shock is well-known and understood in two fluid models \citep{drury81,mckenzie82,volk84}. 

We have also run simulations for the $P_{\rm c}=200/3$ case where we vary the Alfven speed (not shown). As $v_{\rm A}$ increases, so does the separation between the two jumps. The jump in $P_{\rm c}$ decreases, while the jump in $P_{\rm g}$ between the asymptotic left and right states increases. The increase in $v_{\rm A}$ increases the efficiency of wave heating; thus, CRs transfer a larger fraction of their post-shock energy to the gas. The gas absorbs more of the overall shock energy. The outer shock becomes very weak, due to the reduced compressibility of the system (the fluid has a higher effective adiabatic index, close to that of gas, when $v_{\rm A}$ is increased), and the reduced density jump implies the shock moves outward. At the same time, the inner hydrodynamic shock has to thermalize more energy and becomes stronger, moving inward. 

Overall, our CR module performs very stably and robustly in the presence of shocks. As of this writing, there are no published results we are aware of that demonstrate similar capabilities when CR streaming is turned on.

\subsubsection{CR Driven Blast Waves}
\label{sec:blastwave}

\begin{figure}[htp]
	\centering
	\includegraphics[width=0.48\hsize]{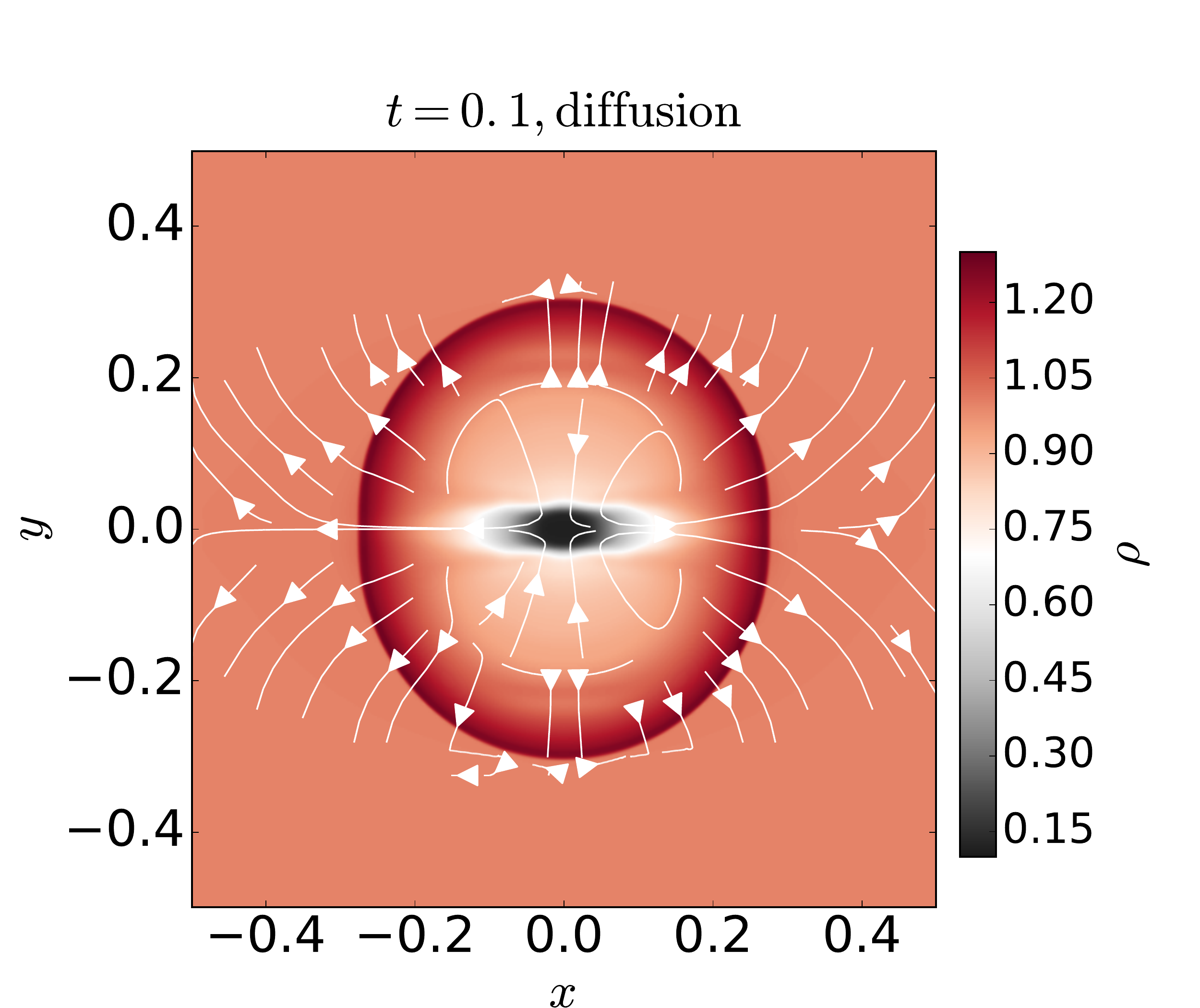}
	\includegraphics[width=0.48\hsize]{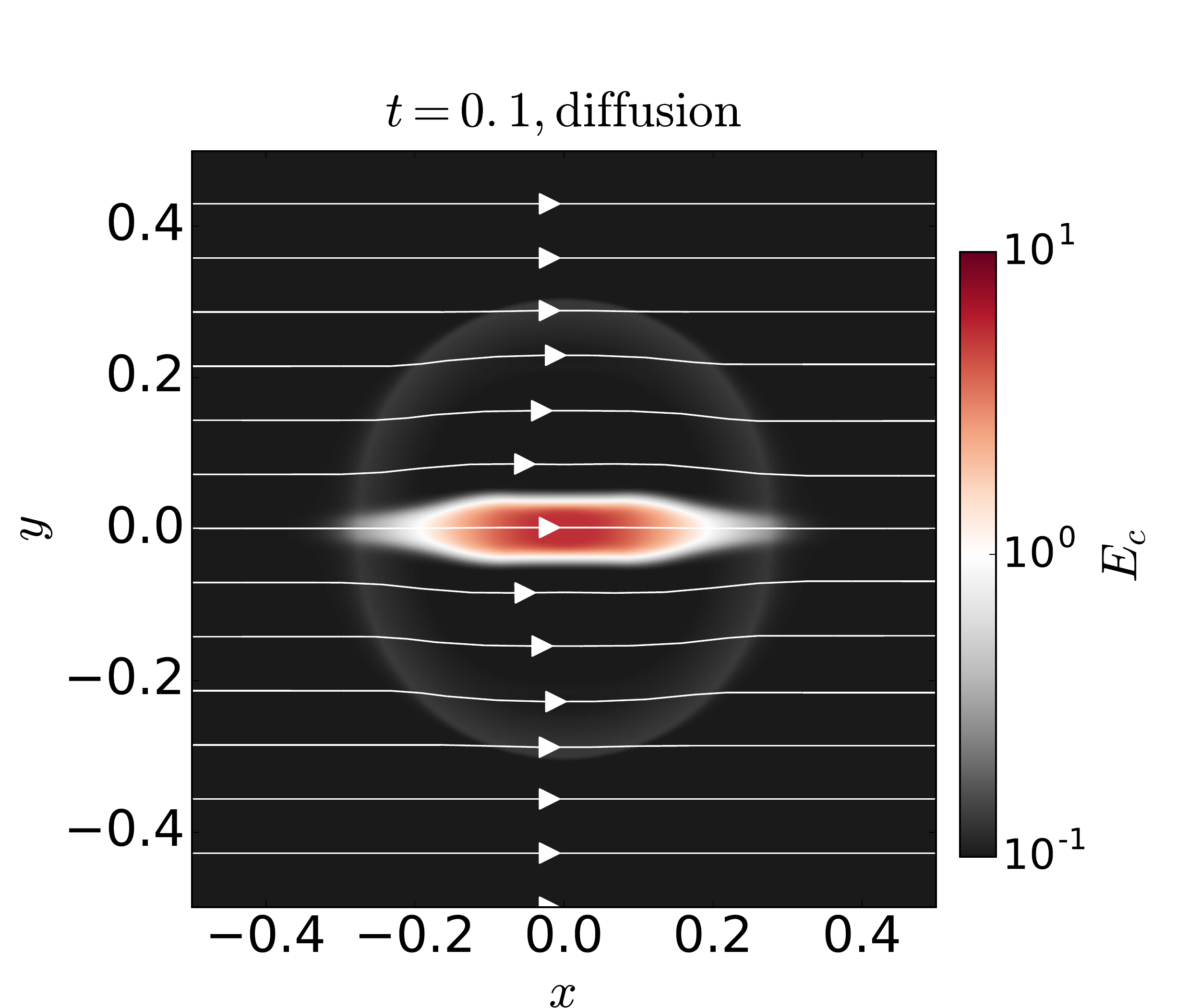}
	\includegraphics[width=0.48\hsize]{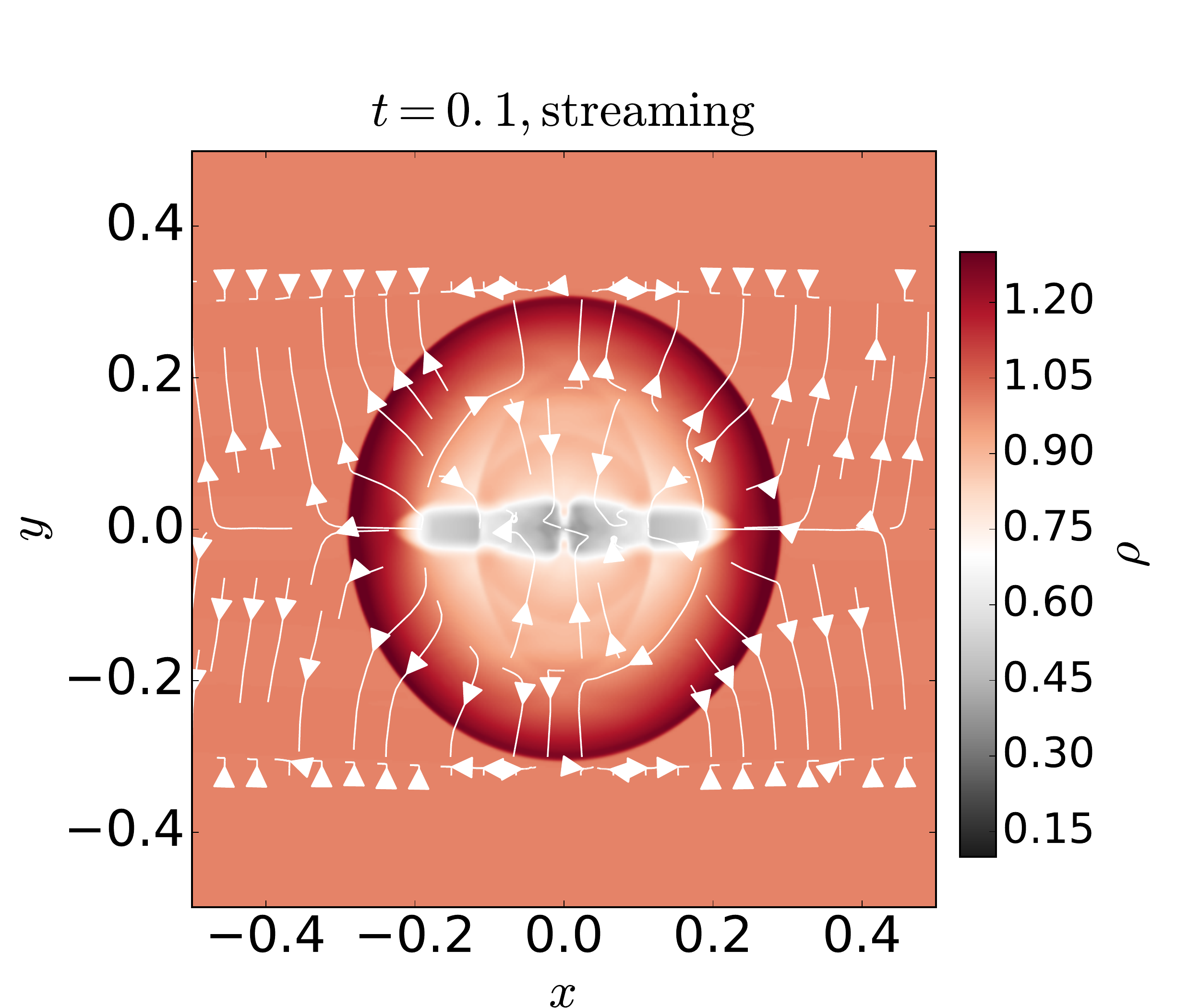}
    \includegraphics[width=0.48\hsize]{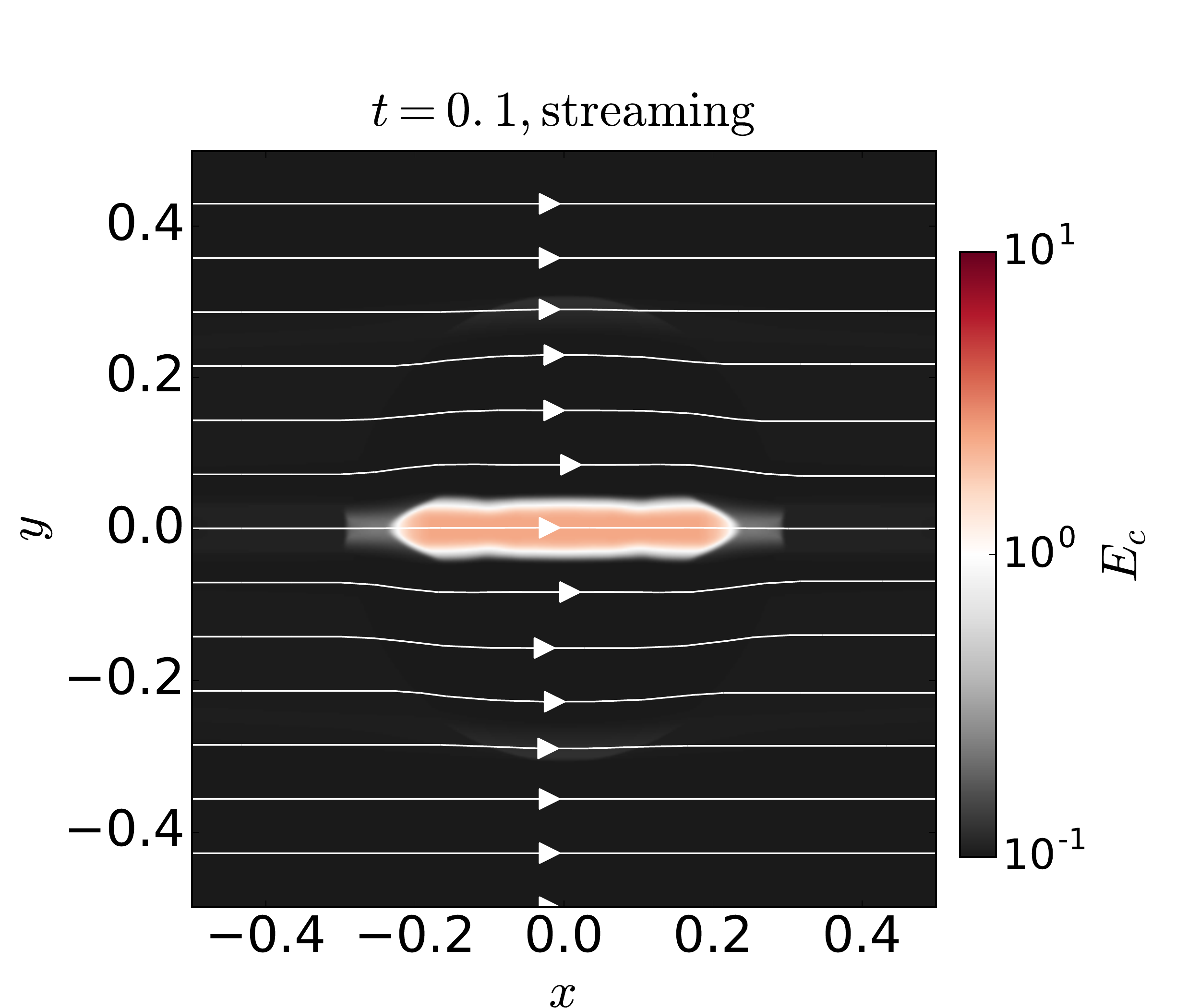}
	\caption{Snapshots of density $\rho$ (left column) and CR energy density $E_c$ (right column) at time $t=0.1$ for the blast 
wave test described in Section \ref{sec:blastwave}. The top panel is for the case with anisotropic diffusion while the bottom 
panel is for the case with streaming. The streamlines in the left column represent flow velocity while in the right column they show the magnetic field lines.  }
	\label{crblast}
\end{figure}

This test problem creates a 2D blast wave with an explosion of  high CR energy density at the center, which tests the full CR MHD algorithm. We use the same setup as in \cite{Pakmoretal2016} in order to compare the results with anisotropic diffusion directly. Furthermore, we can also do the test with pure streaming to show the difference between streaming and diffusion. 

The tests are done in the 2D domain $(x,y)\in (-0.5,0.5)\times(-0.5,0.5)$ with $512^{2}$ spatial resolution. The initial CR energy density is chosen to be  $E_c=100$ in the region $\sqrt{x^2+y^2}<0.02$ and $E_c=0.1$ in other regions. Initial CR flux $F_c$ and flow velocity are $0$ everywhere. The background medium has uniform density $\rho=1$ and internal energy $2.5$. Outflow boundary conditions are used for both $x$ and $y$ directions. We first carry out the simulation with diffusion coefficient $\sigma_c^{\prime}=10$ (equivalent to diffusion coefficient used in \cite{Pakmoretal2016}) along the magnetic field and $\sigma_c^{\prime}=10^6$ perpendicular to the magnetic field lines. The streaming terms related to $\bv_s$ are turned off in this case. The snapshots of density, flow velocity, CR energy density and magnetic field lines are shown in the top panel of Figure \ref{crblast}, which are very similar to Figure 11 of \cite{Pakmoretal2016}. The explosion creates a blast wave that expands with time in the background medium while density at the center drops because of the expansion. CRs only diffuse along the magnetic field lines while there are almost no diffusion along $y$ direction because of the long diffusion time scale along that direction. The $y$ components of magnetic field lines are only perturbed sightly due to the blast wave, which is relatively weak (note the relatively small change in density contrast). There is also a ring of $E_c$ at the location of the blast wave in our solution. CRs are advected there with the flow, and subsequently compressed in the post-shock medium. Interestingly, this ring is not seen in \citet{Pakmoretal2016}, potentially indicating that our codes behaves differently at shocks. 

To show the differences between streaming and diffusion, in the second simulation, we set all components of the diffusion coefficients $\sigma_c^{\prime}=10^6$ so that diffusion is negligible. We turn on all the streaming terms related to $\bv_s$. 
The solution at time $t=0.1$ is shown at the bottom panel of Figure \ref{crblast}. Similar to the previous case, a blast wave is created because of the explosion with very similar density compression ratio at the wave front. Because the streaming velocity $\bv_s$ is only along the magnetic field lines, $E_c$ only spreads along the $x$ direction as in the case with anisotropic diffusion. Unlike the diffusion case, the streaming velocity is the Alfv\'en velocity, which is independent of the CR pressure gradient. CRs energy density $E_c$ is almost a constant at time $t=0.1$ along $x$ direction for $|x|\lesssim 0.2$, while in the diffusion case, $E_c$ decreases with distance from the explosion center. This is also consistent with the differences shown in Figure \ref{streamingtest} and \ref{diffusiontest}. The ring of CR energy density at the blast wave front is also weaker in this case because the streaming velocity is comparable to the flow velocity and CRs can move away from the blast wave. 

\subsubsection{CR Interactions with Warm Clouds}
\label{sec:crcloud}
Interactions between CRs and multiphase medium, such as a warm high density ionized cloud embedded in hot low density gas, 
have interesting implications on the structures and observational properties of these clouds \citep{Skilling1971,Wieneretal2017b}. Numerically, it is also a challenging problem because of the sharp interface between the gas with different densities. In Section \ref{sec:bottleneck}, with fixed background gas structures, we have demonstrated that our algorithm can get the properties of CR streaming in the multiphase medium correctly. In this section, we will turn on all the terms 
in our CR-MHD equations \ref{neweq} and \ref{eq:crmhd} and follow the evolution of clouds under the influences of CRs. 

\begin{figure}[htp]
	\centering
	\includegraphics[width=1.0\hsize]{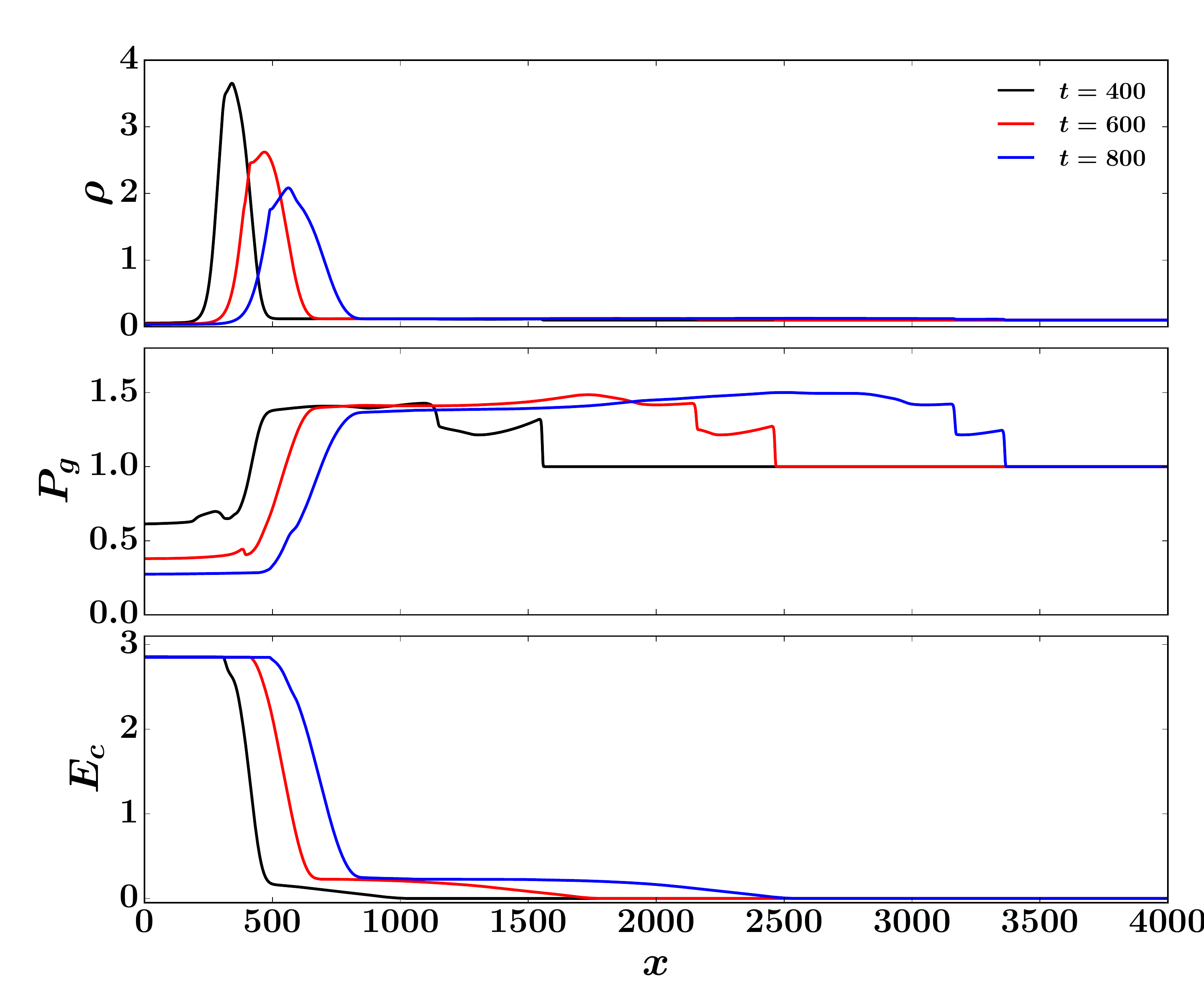}
	\caption{Snapshots of density $\rho$ (top panel), gas pressure $P_g$ (middle panel) and CR energy density $E_c$ (bottom panel) at times 
		$t=400$ (black lines), $600$ (red lines), $800$ (blue lines) respectively for the test of interactions between CRs and cold cloud as described in Section \ref{sec:crcloud}.}
	\label{crcloud_full}
\end{figure}

In the first experiment, we use the same cloud structures as described by equation \ref{eq:cloud} with $\rho_c=1$ and $\rho_h=0.01$. The boundary conditions are also the same as in Section \ref{sec:bottleneck}. But the simulation domain is increased to $x\in(0,4000)$ with $2048$ grid points. We use a larger simulation domain to minimize the effects of the right boundary. There is no optically thin cooling in the first experiment and we can still use the dimensionless numbers without specifying the unit. Evolutions of gas density $\rho$, pressure $P_g$ and CR energy density $E_c$ of the  cloud at three different snapshots are shown in Figure \ref{crcloud_full}. CRs are injected from the left boundary. They push and accelerate  the cloud. At the same time, CRs heat up the warm cloud, which broadens the cloud. The middle panel of Figure \ref{crcloud_full} shows the propagation of a sound wave far away from the cloud. This is also found in Figure 4 of \cite{Wieneretal2017b}, although structures of the wave fronts are different likely due to the different density profiles of the hot gas we use. The sound waves are caused by the impact of CR sources on the surfaces of the cloud and their propagation speed is also consistent with the sound speed of the hot gas.

A more realistic case is that we allow the gas to cool, which will make it harder for CRs to destroy the cloud. It is also interesting to see what kind of equilibrium structures it can reach with CR heating and optically thin cooling. This is also a valuable test problem which is difficult to solve with the traditional method proposed by \cite{Sharmaetal2009}; such calculations were not presented in \citet{Wieneretal2017b}. Physically, this scenario is of great interest in determining whether CRs which drive a wind will accelerate cold clouds or instead destroy them. 

An optically thin cooling term is added to the gas internal energy equation  in an operator splitting way as
\begin{eqnarray}
\frac{\partial P_g}{\partial t}=-\left(\gamma-1\right)n_en_H\Lambda(T),
\end{eqnarray}
where $n_e$ and $n_H$ are electron and hydrogen number densities. We adopt the solar metallicity with mean molecular weight 
$\mu=0.62$, for electron $\mu_e=1.18$ and for hydrogen $\mu_H=1.43$.  
The optically thin cooling function $\Lambda(T)$ is only a function of gas temperature and we use $67$ power-laws to fit the cooling function as shown in Figure 1 of \cite{Jietal2017} over the temperature range $T\in(10^4,10^8)$ K. The cooling function $\Lambda(T)$ is set to be zero whenever the temperature drops below $10^4$ K.The cooling term is solved following the exact integration scheme for power-law cooling functions proposed by \cite{Townsend2009}, which means that our timestep is not restricted to the (very short) cooling time.  

To increase the cooling time scales of the hot gas, we choose temperature $10^7$ K for the hot gas with cooling time scale $81$ Myr and $10^4$ K at the center of the cloud, which has a cooling time scale of $0.055$ Myr. The shape of the initial cloud is still the same as in the previous case but we set $\rho_h=1.25\times 10^{-28}$ g/cm$^3$ 
and $\rho_c=1.25\times 10^{-25}$ g/cm$^3$. The initial gas pressure is a constant everywhere. The cloud is centered at $200$ pc with width $25$ pc. We put CR source with constant energy density $E_c=5.05\times 10^{-13}$ erg/cm$^3$ at the left boundary, which has corresponding CR pressure comparable to the initial gas pressure.

\begin{figure*}[htp]
	\centering
	\includegraphics[width=0.49\hsize]{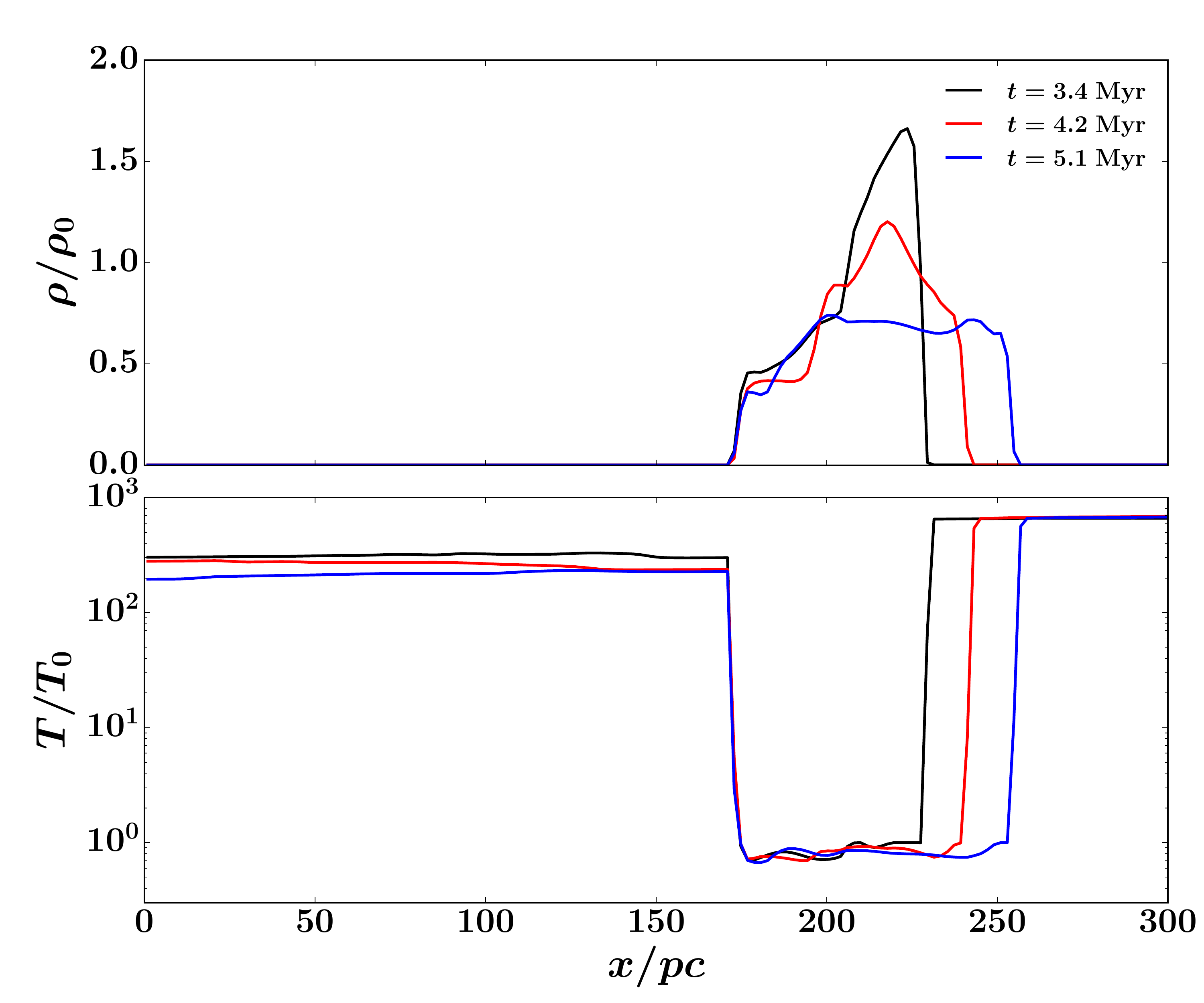}
	\includegraphics[width=0.49\hsize]{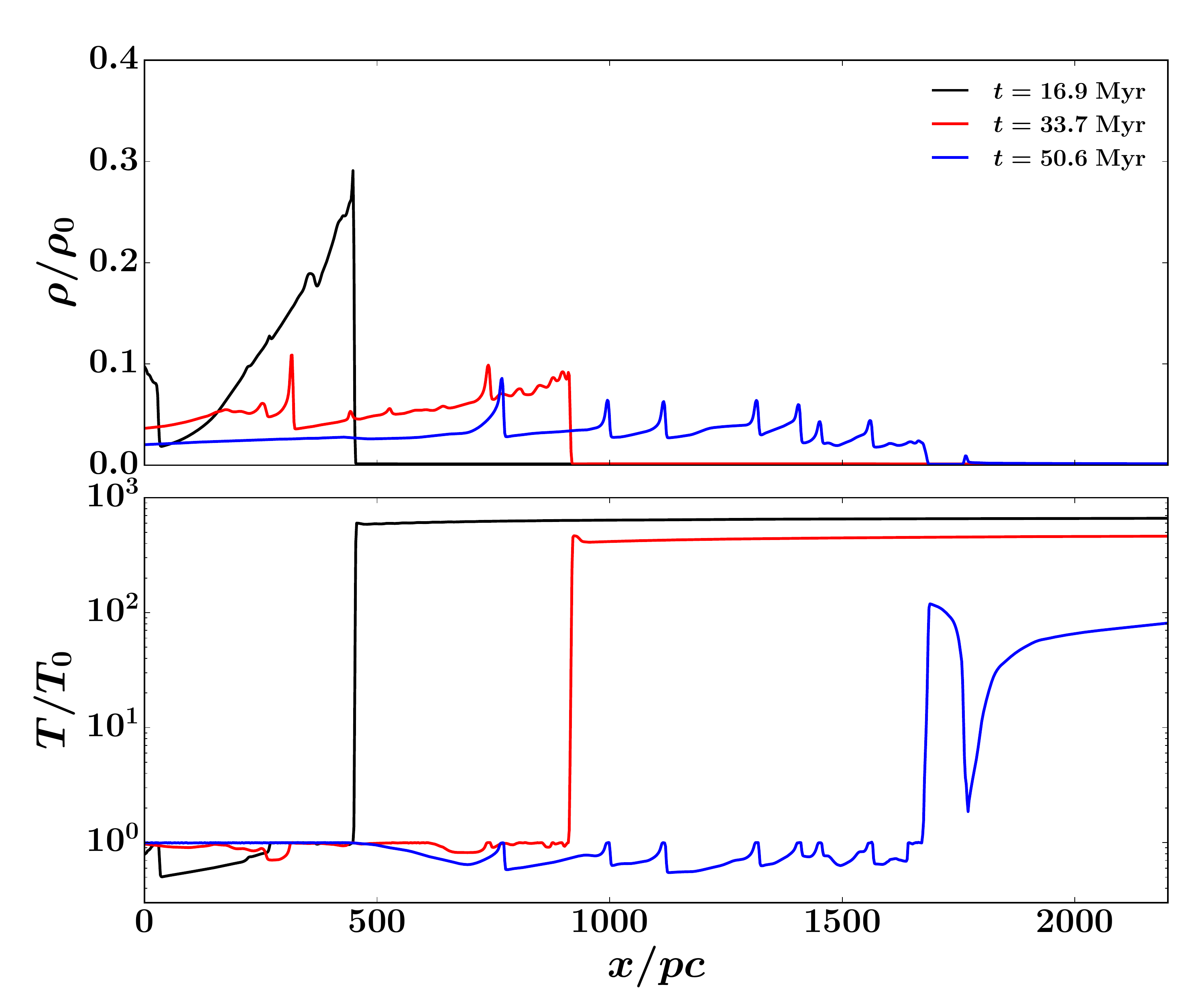}
	\includegraphics[width=0.49\hsize]{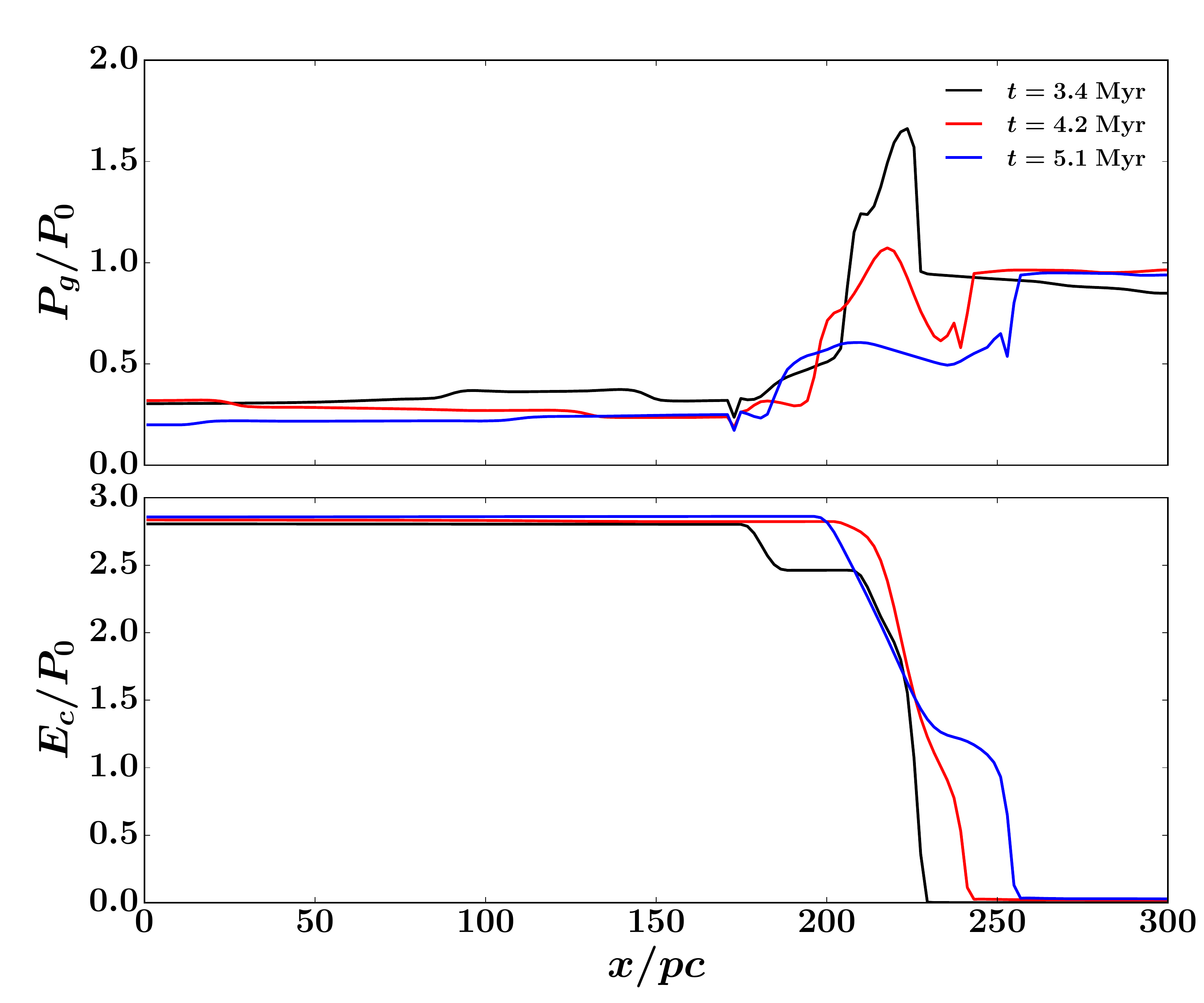}
    \includegraphics[width=0.49\hsize]{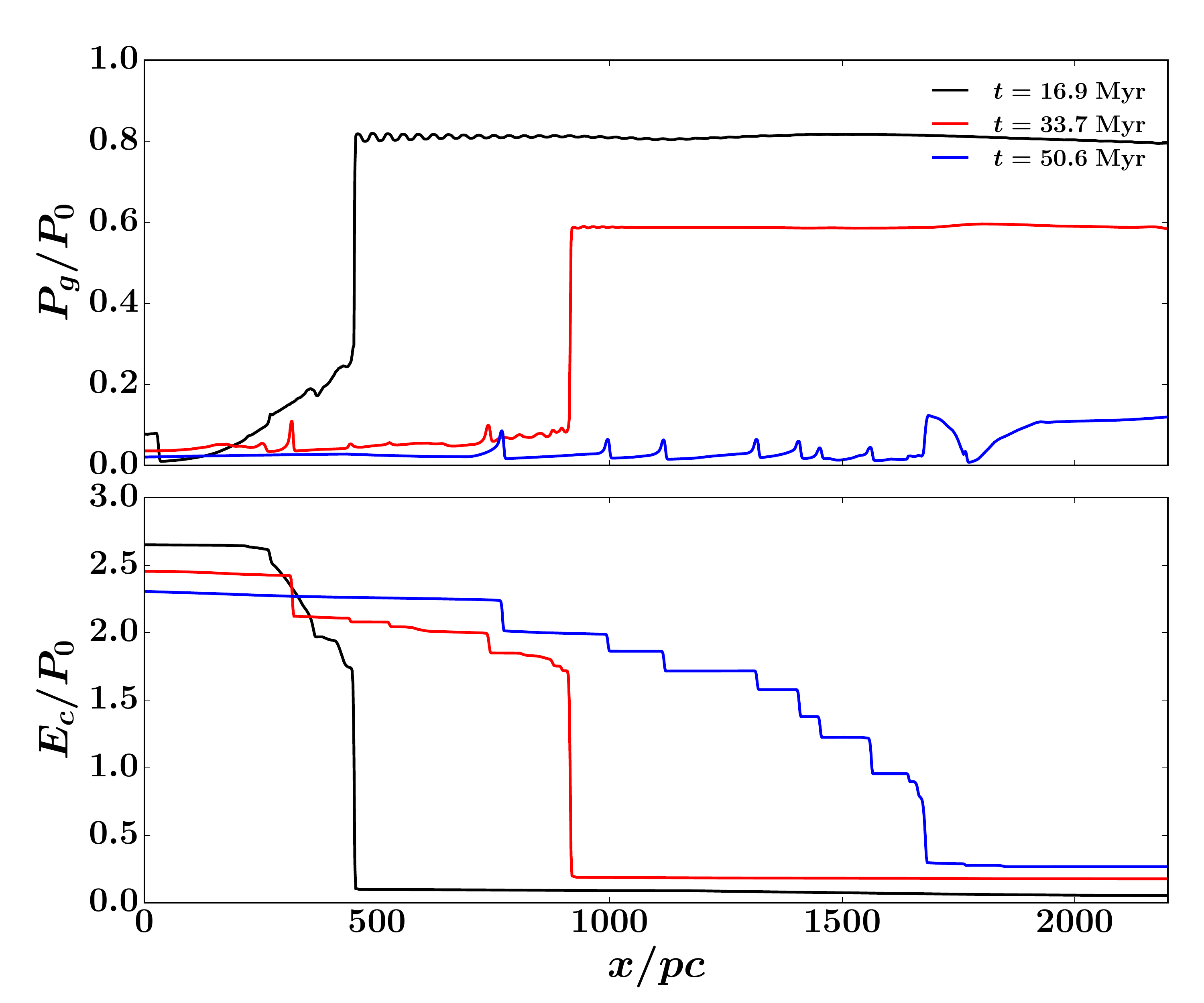}
	\caption{Profiles of gas density $\rho$, temperature $T$, pressure $P_g$ and CR energy density $E_c$ at six different 
	snapshots during the interactions between CRs and clouds with optically thin cooling as described in Section \ref{sec:crcloud}. The left column shows the profiles during the early phase while the right column shows the later time evolution. The black, red and blue lines represent different snapshots with the time labeled in the plot. The fiducial density $\rho_0$, temperature $T_0$ and pressure $P_0$ units are $1.25\times 10^{-28}$ g/cm$^3$, $1\times 10^4$ K and $1.68\times 10^{-13}$ erg/cm$^3$ respectively. }
	\label{crcloud_cooling}
\end{figure*}

The early evolution of the cloud from 3.4 Myr to 5.1 Myr is shown in the left column of Figure \ref{crcloud_cooling}. Because of rapid cooling, the temperature at the center of the cloud always stays around $10^4$ K. The gas pressure at the right side of the cloud stays at the initial value because of the long cooling time scale. Therefore, the cloud is first compressed by the CR source from the left boundary and then expands slowly. Profiles of $E_c$ are  consistent with Figure \ref{CRcloud}. Hot gas at the left of the cloud cools faster than the hot gas at the right because of the additional adiabatic cooling. In the region where $E_c$ is flat, there is no CR heating and the gas will continue to cool. The cooling time scale for gas with temperature $2\times 10^6$ K is only 5.5 Myr and all the gas cools to $10^4$ K at the left of the cloud when $t=17$ Myr as shown in the right column of Figure \ref{crcloud_cooling}. Cooling decreases the pressure at the left of the cloud, which causes gas to fall to the left and multiple shocks are formed. After a significant fraction of the cooling time scale for the hot gas at the right side of the cloud, the cloud expands quickly to the right as shown by the blue lines at time $t=50.6$ Myr. 

This 1D calculation oversimplifies the interactions between CRs and clouds and more detailed studies will be done with multi-dimensional simulations in a separate paper. However, this test demonstrates that our algorithm is able to capture CR streaming accurately and robustly even with large density and temperature gradient.

\section{Discussion}
\label{sec:discussion}

\subsection{Applications to Other Equations with Similar Mathematical Properties}
\label{sec:saturatediff}

There are many other physical processes that can be described by equations with similar mathematical properties as equation \ref{CR_oldeq} such as thermal conduction with saturated heat flux \citep{CowieMcKee1977,Vaidyaetal2017}. When the classical thermal conduction flux is larger than a saturated value, the heat flux reaches a constant value with direction down the temperature gradient along the direction of magnetic field lines, which is very similar to the mathematical property of the streaming term. The only difference is that the heat flux either takes the value of the classical diffusive flux or the saturated flux, not the sum of the two. Since the interaction coefficient $\sigma_c$ in our algorithm can be a general function of the flow properties to mimic different physical process, we can choose the following interaction coefficient to describe the case of saturated heat conduction
\begin{eqnarray}
\frac{1}{\sigma_c}&=&\frac{1}{\sigma_0}(1-\tanh(r))\nonumber\\
&+&\frac{\tanh(r)\bb}{|\bb\cdot\left(\bfnabla\cdot{\sf P_c}\right)|}\bv_A\cdot\left(E_c{\sf I}+{\sf P_c}\right).
\label{eq:sumflux}
\end{eqnarray}
Here the variable $r\equiv |\bfnabla \cdot{\sf P_c}|/\left(\sigma_0\left| \bv_A\cdot\left(E_c{\sf I}+{\sf P_c}\right)\right|\right)$, 
is the ratio between the normal diffusive flux and the saturated value. When this ratio is small, $\sigma_c\approx \sigma_0$ and CR flux takes the value of the classical diffusive flux. When this ratio is large, CR flux is limited by the streaming value. Without changing anything else in the code, our algorithm can get the solution for saturated heat conduction, as long as we treat $E_c$ as gas temperature in this case. We take $\sigma_0=10$ and a constant Alfv\'en value $v_A=0.01$ in this test problem. The test is done in the 1D domain $x\in(-10,10)$ with $512$ cells. We initialize $E_c$ with the sawtooth profile as $E_c(x)=10+20\left[x/10-\mbox{floor}(x/10+0.5)\right]$, where the function $\mbox{floor}(y)$ 
takes the integer equal or smaller than the variable $y$. The initial CR flux $F_c$ is set to be $0$ everywhere. Periodic boundary condition is used in this test. 

\begin{figure}[htp]
	\centering
	\includegraphics[width=1.0\hsize]{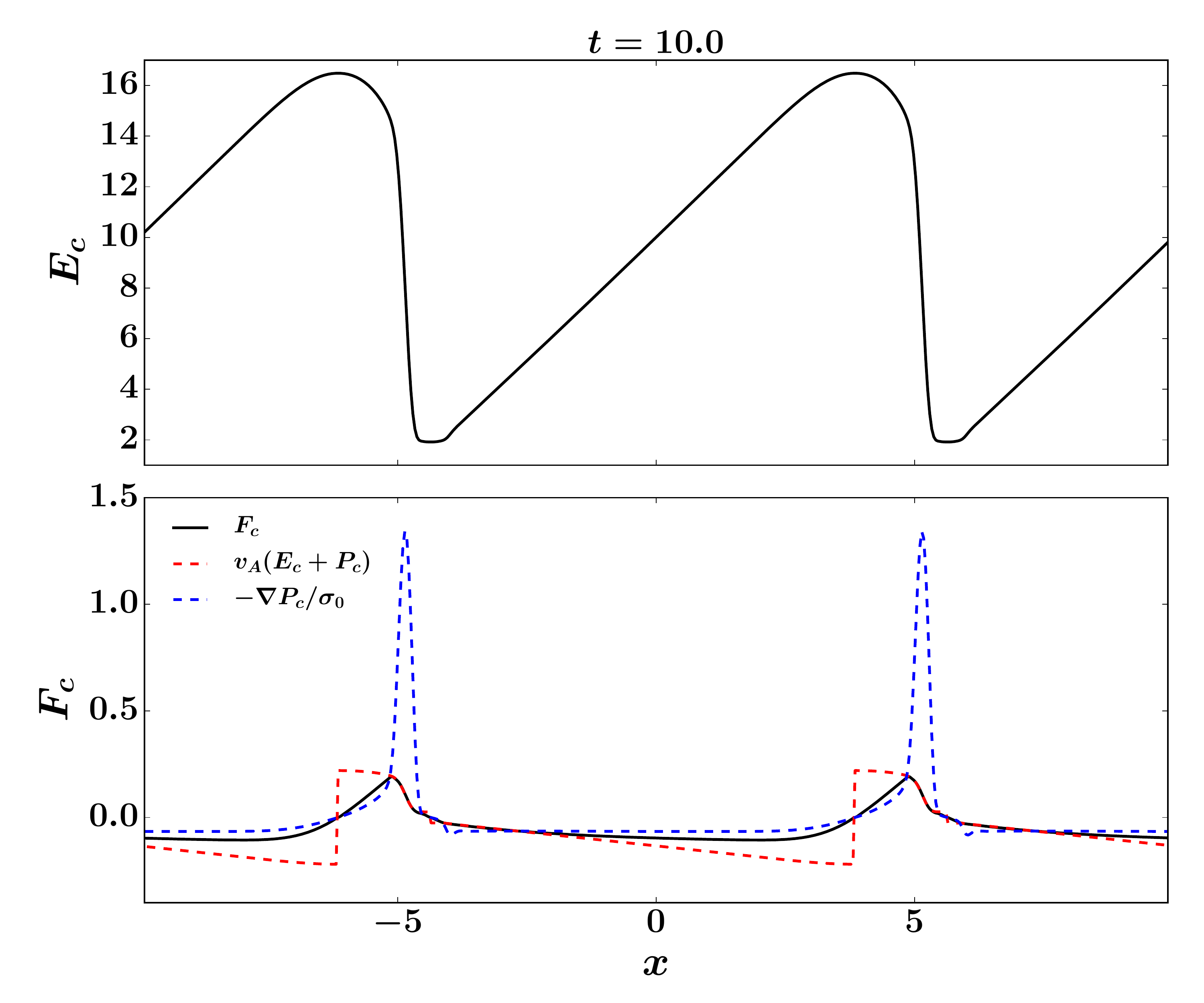}
	\caption{Profiles of CR energy density $E_c$ (top panel) and CR flux $F_c$ (bottom panel) at time $t=10$ for the test 
		of saturated diffusion as described in Section \ref{sec:saturatediff}. The CR flux $F_c$ is compared to the streaming value 
		$v_a(E_c+P_c)$ (dashed red line) as well as the diffusive value $-\nabla P_c/\sigma_0$ (dashed blue line) in the bottom panel.}
	\label{saturateflux}
\end{figure}

Profiles of $E_c$ and $F_c$ at time $t=10$ are shown in Figure \ref{saturateflux}, which show very similar properties as the saturated heat flux test shown in Figure 4 of \cite{Vaidyaetal2017}. Near the minimum of $E_c$, it is almost a constant, which indicates that the streaming term is dominant. At the peak, profiles of $E_c$ behave like the solution with diffusion. This is checked directly at the bottom panel of Figure \ref{saturateflux}, where CR flux $F_c$ in the solution is compared to the expected value based on pure streaming $v_s(E_c+P_c)$ and diffusion $-\nabla P_c/\sigma_0$. Around $x=\pm5$, when the 
expected diffusive flux is much larger than the streaming flux, $F_c$ takes the streaming value as expected. This is also consistent with the flat profile of $E_c$. When the diffusive flux becomes smaller in other places, $F_c$ is closer to the diffusive value. In general, $F_{\rm c}$ can reproduce the expected behavior of thermal conduction in diffusive and saturated limits very well, and our algorithm is a robust and easily parallelized way to model thermal conduction. It automatically preserves monotonicity and ensures that energy is transported down the temperature gradient, unlike conventional methods which can violate the entropy condition, leading energy transfer from cold to hot regions and negative temperatures. Typically this is prevented by using slope limiters \citep{SharmaHammett2007}. 

\subsection{Effects of the Maximum Velocity $V_m$}
In principle, $V_m$ should be the speed of light. which will guarantee that CRs travel with the speed of light when they decouple with the gas. However, for explicit method, this will require that the time step is limited by the speed of light, which is usually too small to be practical. Alternatively, we can also solve the transport step implicitly as in \cite{Jiangetal2012} for radiative transfer equation. However, we will run into the normal issues of implicit method which involves operators of spatial gradient. Namely, this will require a matrix inversion over the whole simulation domain, which significantly impacts the performance and parallel scaling. Therefore, we choose a velocity $V_m$ that is much smaller than the speed of light. As $V_m$ only appears in the time dependent term for $\bF_c$, it does not affect the steady state solutions at all. Even for the time dependent solutions, its exact value is immaterial, as long as the $(1/V_m^2)\partial \bF_c/\partial t$ term is much smaller than the source term, as it will be when $V_m$ is much larger than the maximum flow and Alfv\'en velocities and there are strong interactions between CRs and the fluid. This is demonstrated in Figure \ref{streamingtest}.

This is similar to the reduced speed of light approximation in radiative transfer \citep{SkinnerOstriker2013} with an important difference. When the  time dependent terms 
$\partial E_c/\partial t$ and $(1/V_m^2)\partial \bF_c/\partial t$ are not zero, our scheme conserves the total energy $E_c+E$ and momentum $\rho\bv+\bF_c/V_m^2$. However, the reduced speed of light approximation adopted by \cite{SkinnerOstriker2013} uses different energy and momentum source terms for radiation and gas, which does not guarantee the conservation of energy and momentum during the temporal evolution of the solution.  The main effect of different choice of $V_m$ in our scheme is that we overestimate the momentum of CRs by $c^2/V_m^2$ for a given CR energy density $E_c$. The momentum of CRs is typically much smaller than the momentum of gas (by a factor $\mathcal{O}(X_{\rm CR} (c_s/c)^{2})$, where $X_{\rm CR} \equiv P_{\rm CR}/P_{\rm g}$), and this remains true even with the reduced speed $V_m$. Thus, this overestimate has no effect on our calculations. That is also why in Figure \ref{streamingtest}, even the temporal evolution of the solution can be independent of $V_m$ we choose. In principle, convergence of all the simulations should also be checked with respect to $V_m$. In practice, we  monitor the maximum flow and Alfv\'en velocities during the simulations to make sure they are always much smaller than $V_m$. And we can always repeat part of the simulations with different $V_m$ to make sure the results are unchanged. 

\subsection{The Closure Relation for CR pressure}
In equation \ref{neweq}, we assume CR pressure tensor is related to the CR energy density as 
${\sf P_c}=E_c/3{\sf I}$. This is usually a good assumption when there are strong interactions between CRs and the gas. 
In certain situations, CRs can decouple with the gas, either due to effects such as the `bottleneck effect' or strong wave damping (see discussion in Appendix). 
In this regime, CRs should be free-streaming and CR pressure can be very anisotropic and quite different from $E/3{\sf I}$. In principle, the distribution function $f_{\tr p}$ should be solved in the momentum space with some reasonable approximations, which can then be used to provide the closure relation as in the VET method of radiative transfer \citep{Jiangetal2012,Davisetal2012}. However, this will significantly increase the computational cost and complexity of the algorithm. It will be the subject of future work to find a better closure relation for CRs when this regime becomes important. 

When the traditional equation \ref{CR_oldeq} is used in the regime that CRs should be decoupled with the gas, CR transport has been mimicked by diffusion with a much larger diffusion coefficient \citep{Farberetal2017}. 
In practice, CRs should free-stream at the speed of light, but this is not possible to implement with the standard algorithm for CR streaming. Note that CR streaming and diffusion have very different transport properties (contrast Fig \ref{streamingtest} and \ref{diffusiontest}, and also see discussion in \citet{Wieneretal2017}).  
Approximating streaming by diffusion also does not guarantee causality. 
Signals can propagate with arbitrary velocity and affect the whole simulation domain as noticed by \cite{Sharmaetal2009}. 
In contrast, when CRs decouple with the gas, which means the source terms in our new equation \ref{neweq} become small, our algorithm automatically solves the hyperbolic equation for CR transport. The largest speed any signal can propagate is $V_m$. Despite the issue of the closure relation, this will self-consistently describe the free-streaming of CRs in this regime.

\subsection{Performance of the Algorithm}

The numerical cost of solving the CR transport equation is comparable to the cost of solving ideal MHD equations for each time step. Since the transport step is solved explicitly and the implicit source step only works with variables in each cell, the full CR-MHD algorithm should also have the same excellent parallel scaling as the MHD algorithm in {\sf Athena++}. We test the performance of the full algorithm on NASA's supercomputer Pleiades with the broadwell nodes. Each node has two 14-core, 2.4GHz processors and 128 GB of memory. We use a 3D domain 
$(x,y,z)\in(-1,1)\times (-1,1)\times (-1,1)$ with initial CR energy profile $E_c=\exp\left[-40(x^2+y^2+z^2)\right]$. The gas is initialized with uniform density $\rho=1$ and temperature $T=1$. The velocities are set to be zero initially. The initial magnetic field is $B_x=B_y=B_z=1$. The diffusion coefficient is $\sigma^{\prime}_c=0.01$ so that both diffusion and streaming are significant. For this test problem with $32^3$ grids per core, the code is able to update $3.9\times 10^5$ zones per second, which is indeed roughly half of the speed for the original MHD algorithm. The total cost of our CR-MHD algorithm is proportional to the time step, or the value of $V_m$. Since $V_m$ is the largest speed, the time step in CR-MHD simulations is generally smaller than the normal MHD simulations, which will increase the total cost of these simulations. However, we usually find it is still much cheaper than the classical method with regularization, particularly for high resolution simulations. 

\section{Conclusion}
\label{sec:conclusion}

We describe a set of new equations for the transport of CRs and interactions between CRs and fluid, 
including both streaming and anisotropic diffusion processes self-consistently. In the region when there is a finite CR pressure gradient, these equations reduce to the classical CR-MHD equations people adopted in the past. When CR pressure gradient is close to zero, CRs and gas become decoupled in our equations and CRs become free-stream along the magnetic field direction. Therefore, the equations we use do not suffer from the numerical instability for CR streaming as in the classical equation and self-consistently describe the transport of CRs in all regimes. We have developed a numerical scheme to solve this set of new equations. Based on  a wide range of test problems for both diffusion and streaming, we demonstrate that the algorithm is accurate, robust and fast. The time step is linearly proportional to the spatial resolution and is set by the standard Courant condition across the entire simulation domain. It opens the window to simulate a wide range of astrophysical systems where CRs play an important role.

\section*{Acknowledgements}
This research was supported in part by the National Science Foundation under Grant No. NSF PHY-1125915, and NASA grants NNX15AK81G, NNX17AK58G. 
We thank Eve Ostriker for useful discussions to justify the equations we use. We also thank Christoph Pfrommer and Prateek Sharma for helpful conversations. SPO thanks the law offices of May Oh \& Wee for hospitality.



\bibliographystyle{astroads}
\bibliography{cosmicray}

\begin{thebibliography}{64}
\expandafter\ifx\csname natexlab\endcsname\relax\def\natexlab#1{#1}\fi

\bibitem[{{Begelman}(1995)}]{begelman95}
{Begelman}, M.~C. 1995, in Astronomical Society of the Pacific Conference
  Series, Vol.~80, The Physics of the Interstellar Medium and Intergalactic
  Medium, ed. A.~{Ferrara}, C.~F. {McKee}, C.~{Heiles}, \& P.~R. {Shapiro}, 545

\bibitem[{{Booth} {et~al.}(2013){Booth}, {Agertz}, {Kravtsov}, \&
  {Gnedin}}]{booth13}
{Booth}, C.~M., {Agertz}, O., {Kravtsov}, A.~V., \& {Gnedin}, N.~Y. 2013,
  \apjl, 777, L16

\bibitem[{{Breitschwerdt} {et~al.}(1991){Breitschwerdt}, {McKenzie}, \&
  {Voelk}}]{breitschwerdt91}
{Breitschwerdt}, D., {McKenzie}, J.~F., \& {Voelk}, H.~J. 1991, \aap, 245, 79

\bibitem[{{Chen} {et~al.}(2010){Chen}, {Helsby}, {Gauthier}, {Shectman},
  {Thompson}, \& {Tinker}}]{chen10}
{Chen}, H.-W., {Helsby}, J.~E., {Gauthier}, J.-R., {et~al.} 2010, \apj, 714,
  1521

\bibitem[{{Cowie} \& {McKee}(1977)}]{CowieMcKee1977}
{Cowie}, L.~L., \& {McKee}, C.~F. 1977, \apj, 211, 135

\bibitem[{{Davis} {et~al.}(2012){Davis}, {Stone}, \& {Jiang}}]{Davisetal2012}
{Davis}, S.~W., {Stone}, J.~M., \& {Jiang}, Y.-F. 2012, \apjs, 199, 9

\bibitem[{{De Pontieu} {et~al.}(2001){De Pontieu}, {Martens}, \&
  {Hudson}}]{depontieu01}
{De Pontieu}, B., {Martens}, P.~C.~H., \& {Hudson}, H.~S. 2001, \apj, 558, 859

\bibitem[{{Drury} \& {Voelk}(1981)}]{drury81}
{Drury}, L.~O., \& {Voelk}, J.~H. 1981, \apj, 248, 344

\bibitem[{{Farber} {et~al.}(2017){Farber}, {Ruszkowski}, {Yang}, \&
  {Zweibel}}]{Farberetal2017}
{Farber}, R., {Ruszkowski}, M., {Yang}, H.-Y.~K., \& {Zweibel}, E.~G. 2017,
  arXiv:1707.04579

\bibitem[{{Farmer} \& {Goldreich}(2004)}]{farmer04}
{Farmer}, A.~J., \& {Goldreich}, P. 2004, \apj, 604, 671

\bibitem[{{Fujita} \& {Ohira}(2011)}]{fujita11}
{Fujita}, Y., \& {Ohira}, Y. 2011, \apj, 738, 182

\bibitem[{Girichidis {et~al.}(2016)Girichidis, Naab, Walch, Hanasz, Mac~Low,
  Ostriker, Gatto, Peters, W{\"u}nsch, Glover, Klessen, Clark, \&
  Baczynski}]{girichidis16}
Girichidis, P., Naab, T., Walch, S., {et~al.} 2016, The Astrophysical Journal
  Letters, 816, L19

\bibitem[{{Gnedin} \& {Abel}(2001)}]{GnedinAbel2001}
{Gnedin}, N.~Y., \& {Abel}, T. 2001, \na, 6, 437

\bibitem[{{Guo} \& {Oh}(2008)}]{GuoOh2008}
{Guo}, F., \& {Oh}, S.~P. 2008, \mnras, 384, 251

\bibitem[{{Hanasz} {et~al.}(2013){Hanasz}, {Lesch}, {Naab}, {Gawryszczak},
  {Kowalik}, \& {W{\'o}lta{\'n}ski}}]{hanasz13}
{Hanasz}, M., {Lesch}, H., {Naab}, T., {et~al.} 2013, \apjl, 777, L38

\bibitem[{{Ipavich}(1975)}]{ipavich75}
{Ipavich}, F.~M. 1975, \apj, 196, 107

\bibitem[{{Jacob} \& {Pfrommer}(2017)}]{jacob17a}
{Jacob}, S., \& {Pfrommer}, C. 2017, \mnras, 467, 1449

\bibitem[{{Ji} {et~al.}(2017){Ji}, {Oh}, \& {McCourt}}]{Jietal2017}
{Ji}, S., {Oh}, S.~P., \& {McCourt}, M. 2017, arXiv:1710.00822

\bibitem[{{Jiang} {et~al.}(2012){Jiang}, {Stone}, \& {Davis}}]{Jiangetal2012}
{Jiang}, Y.-F., {Stone}, J.~M., \& {Davis}, S.~W. 2012, \apjs, 199, 14

\bibitem[{{Jiang} {et~al.}(2013){Jiang}, {Stone}, \& {Davis}}]{Jiangetal2013b}
---. 2013, \apj, 767, 148

\bibitem[{{Jiang} {et~al.}(2014){Jiang}, {Stone}, \& {Davis}}]{Jiangetal2014b}
---. 2014, \apjs, 213, 7

\bibitem[{{Kulsrud} \& {Pearce}(1969)}]{kulsrud69}
{Kulsrud}, R., \& {Pearce}, W.~P. 1969, \apj, 156, 445

\bibitem[{{Kulsrud}(2005)}]{kulsrud05}
{Kulsrud}, R.~M. 2005, {Plasma physics for astrophysics} (Princeton University
  Press)

\bibitem[{{Lazarian}(2016)}]{lazarian16}
{Lazarian}, A. 2016, \apj, 833, 131

\bibitem[{{Loewenstein} {et~al.}(1991){Loewenstein}, {Zweibel}, \&
  {Begelman}}]{loewenstein91}
{Loewenstein}, M., {Zweibel}, E.~G., \& {Begelman}, M.~C. 1991, \apj, 377, 392

\bibitem[{{McKenzie} \& {Voelk}(1982)}]{mckenzie82}
{McKenzie}, J.~F., \& {Voelk}, H.~J. 1982, \aap, 116, 191

\bibitem[{{Pakmor} {et~al.}(2016{\natexlab{a}}){Pakmor}, {Pfrommer}, {Simpson},
  {Kannan}, \& {Springel}}]{Pakmoretal2016}
{Pakmor}, R., {Pfrommer}, C., {Simpson}, C.~M., {Kannan}, R., \& {Springel}, V.
  2016{\natexlab{a}}, \mnras, 462, 2603

\bibitem[{{Pakmor} {et~al.}(2016{\natexlab{b}}){Pakmor}, {Pfrommer}, {Simpson},
  \& {Springel}}]{pakmor16}
{Pakmor}, R., {Pfrommer}, C., {Simpson}, C.~M., \& {Springel}, V.
  2016{\natexlab{b}}, \apjl, 824, L30

\bibitem[{{Parrish} \& {Stone}(2005)}]{ParrishStone2005}
{Parrish}, I.~J., \& {Stone}, J.~M. 2005, \apj, 633, 334

\bibitem[{{Pfrommer}(2013)}]{pfrommer13}
{Pfrommer}, C. 2013, \apj, 779, 10

\bibitem[{{Pfrommer} {et~al.}(2017){Pfrommer}, {Pakmor}, {Schaal}, {Simpson},
  \& {Springel}}]{Pfrommeretal2017}
{Pfrommer}, C., {Pakmor}, R., {Schaal}, K., {Simpson}, C.~M., \& {Springel}, V.
  2017, \mnras, 465, 4500

\bibitem[{{Pfrommer} {et~al.}(2006){Pfrommer}, {Springel}, {En{\ss}lin}, \&
  {Jubelgas}}]{pfrommer06}
{Pfrommer}, C., {Springel}, V., {En{\ss}lin}, T.~A., \& {Jubelgas}, M. 2006,
  \mnras, 367, 113

\bibitem[{{Recchia} {et~al.}(2016){Recchia}, {Blasi}, \& {Morlino}}]{recchia16}
{Recchia}, S., {Blasi}, P., \& {Morlino}, G. 2016, \mnras, 462, 4227

\bibitem[{{Reynolds} {et~al.}(1999){Reynolds}, {Haffner}, \&
  {Tufte}}]{reynolds99}
{Reynolds}, R.~J., {Haffner}, L.~M., \& {Tufte}, S.~L. 1999, \apjl, 525, L21

\bibitem[{{Rubin} {et~al.}(2010){Rubin}, {Weiner}, {Koo}, {Martin},
  {Prochaska}, {Coil}, \& {Newman}}]{rubin10}
{Rubin}, K.~H.~R., {Weiner}, B.~J., {Koo}, D.~C., {et~al.} 2010, \apj, 719,
  1503

\bibitem[{{Ruszkowski} {et~al.}(2017){Ruszkowski}, {Yang}, \&
  {Reynolds}}]{ruszkowski17}
{Ruszkowski}, M., {Yang}, H.-Y.~K., \& {Reynolds}, C.~S. 2017, \apj, 844, 13

\bibitem[{{Salem} \& {Bryan}(2014)}]{salem14}
{Salem}, M., \& {Bryan}, G.~L. 2014, \mnras, 437, 3312

\bibitem[{{Samui} {et~al.}(2010){Samui}, {Subramanian}, \&
  {Srianand}}]{samui10}
{Samui}, S., {Subramanian}, K., \& {Srianand}, R. 2010, \mnras, 402, 2778

\bibitem[{{Sekora} \& {Stone}(2010)}]{SekoraStone2010}
{Sekora}, M.~D., \& {Stone}, J.~M. 2010, Journal of Computational Physics, 229,
  6819

\bibitem[{{Sharma} {et~al.}(2010){Sharma}, {Colella}, \&
  {Martin}}]{Sharmaetal2009}
{Sharma}, P., {Colella}, P., \& {Martin}, D.~F. 2010, SIAM J. of Scient. Comp.,
  32, 3564

\bibitem[{{Sharma} \& {Hammett}(2007)}]{SharmaHammett2007}
{Sharma}, P., \& {Hammett}, G.~W. 2007, Journal of Computational Physics, 227,
  123

\bibitem[{{Sharma} \& {Hammett}(2011)}]{SharmaHammett2011}
---. 2011, Journal of Computational Physics, 230, 4899

\bibitem[{{Simpson} {et~al.}(2016){Simpson}, {Pakmor}, {Marinacci}, {Pfrommer},
  {Springel}, {Glover}, {Clark}, \& {Smith}}]{simpson16}
{Simpson}, C.~M., {Pakmor}, R., {Marinacci}, F., {et~al.} 2016, \apjl, 827, L29

\bibitem[{{Skilling}(1971)}]{Skilling1971}
{Skilling}, J. 1971, \apj, 170, 265

\bibitem[{{Skinner} \& {Ostriker}(2013)}]{SkinnerOstriker2013}
{Skinner}, M.~A., \& {Ostriker}, E.~C. 2013, \apjs, 206, 21

\bibitem[{{Socrates} {et~al.}(2008){Socrates}, {Davis}, \&
  {Ramirez-Ruiz}}]{socrates08}
{Socrates}, A., {Davis}, S.~W., \& {Ramirez-Ruiz}, E. 2008, \apj, 687, 202

\bibitem[{{Steidel} {et~al.}(2010){Steidel}, {Erb}, {Shapley}, {Pettini},
  {Reddy}, {Bogosavljevi{\'c}}, {Rudie}, \& {Rakic}}]{steidel10}
{Steidel}, C.~C., {Erb}, D.~K., {Shapley}, A.~E., {et~al.} 2010, \apj, 717, 289

\bibitem[{{Stone} {et~al.}(2008){Stone}, {Gardiner}, {Teuben}, {Hawley}, \&
  {Simon}}]{Stoneetal2008}
{Stone}, J.~M., {Gardiner}, T.~A., {Teuben}, P., {Hawley}, J.~F., \& {Simon},
  J.~B. 2008, \apjs, 178, 137

\bibitem[{{Stone} {et~al.}(1992){Stone}, {Mihalas}, \&
  {Norman}}]{Stoneetal1992}
{Stone}, J.~M., {Mihalas}, D., \& {Norman}, M.~L. 1992, \apjs, 80, 819

\bibitem[{{Townsend}(2009)}]{Townsend2009}
{Townsend}, R.~H.~D. 2009, \apjs, 181, 391

\bibitem[{{Tumlinson} {et~al.}(2011){Tumlinson}, {Thom}, {Werk}, {Prochaska},
  {Tripp}, {Weinberg}, {Peeples}, {O'Meara}, {Oppenheimer}, {Meiring}, {Katz},
  {Dav{\'e}}, {Ford}, \& {Sembach}}]{tumlinson11}
{Tumlinson}, J., {Thom}, C., {Werk}, J.~K., {et~al.} 2011, Science, 334, 948

\bibitem[{{Turner} \& {Stone}(2001)}]{TurnerStone2001}
{Turner}, N.~J., \& {Stone}, J.~M. 2001, \apjs, 135, 95

\bibitem[{{Uhlig} {et~al.}(2012){Uhlig}, {Pfrommer}, {Sharma}, {Nath},
  {En{\ss}lin}, \& {Springel}}]{uhlig12}
{Uhlig}, M., {Pfrommer}, C., {Sharma}, M., {et~al.} 2012, \mnras, 423, 2374

\bibitem[{{Vaidya} {et~al.}(2017){Vaidya}, {Prasad}, {Mignone}, {Sharma}, \&
  {Rickler}}]{Vaidyaetal2017}
{Vaidya}, B., {Prasad}, D., {Mignone}, A., {Sharma}, P., \& {Rickler}, L. 2017,
  arXiv:1702.05487

\bibitem[{{Voelk} {et~al.}(1984){Voelk}, {Drury}, \& {McKenzie}}]{volk84}
{Voelk}, H.~J., {Drury}, L.~O., \& {McKenzie}, J.~F. 1984, \aap, 130, 19

\bibitem[{{Volk} \& {McKenzie}(1981)}]{volk81}
{Volk}, H.~J., \& {McKenzie}, J.~F. 1981, International Cosmic Ray Conference,
  9, 246

\bibitem[{{Wentzel}(1974)}]{wentzel74}
{Wentzel}, D.~G. 1974, \araa, 12, 71

\bibitem[{{Wiener} {et~al.}(2013{\natexlab{a}}){Wiener}, {Oh}, \&
  {Guo}}]{wiener13}
{Wiener}, J., {Oh}, S.~P., \& {Guo}, F. 2013{\natexlab{a}}, \mnras, 434, 2209

\bibitem[{{Wiener} {et~al.}(2017{\natexlab{a}}){Wiener}, {Oh}, \&
  {Zweibel}}]{Wieneretal2017b}
{Wiener}, J., {Oh}, S.~P., \& {Zweibel}, E.~G. 2017{\natexlab{a}}, \mnras, 467,
  646

\bibitem[{{Wiener} {et~al.}(2017{\natexlab{b}}){Wiener}, {Pfrommer}, \& {Peng
  Oh}}]{Wieneretal2017}
{Wiener}, J., {Pfrommer}, C., \& {Peng Oh}, S. 2017{\natexlab{b}}, \mnras, 467,
  906

\bibitem[{{Wiener} {et~al.}(2013{\natexlab{b}}){Wiener}, {Zweibel}, \&
  {Oh}}]{wiener13-ISM}
{Wiener}, J., {Zweibel}, E.~G., \& {Oh}, S.~P. 2013{\natexlab{b}}, \apj, 767,
  87

\bibitem[{{Wiener} {et~al.}(2018){Wiener}, {Zweibel}, \&
  {Oh}}]{wiener18-high-beta}
---. 2018, \mnras, 473, 3095

\bibitem[{{Yan} \& {Lazarian}(2002)}]{yan02}
{Yan}, H., \& {Lazarian}, A. 2002, Physical Review Letters, 89, B1102+

\bibitem[{Zweibel(2017)}]{zweibel17}
Zweibel, E.~G. 2017, Physics of Plasmas, 24, 055402

\end{thebibliography}

\section*{Appendix: Diffusion Coefficients}


Throughout this paper, we have adopted spatially constant diffusion coefficients without reference to any physical model. Many galaxy formation simulations have adopted this approach, scaling diffusivities to empirical Milky Way values. For our purposes, testing our code with a constant $v_{\rm A}$ and $\sigma^{\prime}$ allows us to assess its performance in the pure hyperbolic and parabolic limits. 

For completeness, in this Appendix we present and discuss parallel diffusion coefficients which can be derived using quasi-linear theory (for more detailed discussion, see \citet{wiener13}). We should use the customary definition of the diffusion coefficient, which is relative to the interaction coefficient via:
\begin{equation}
\kappa=\frac{1}{\sigma_{\rm c}^{\prime}}
\end{equation}
We proceed by presenting energy-specific diffusion coefficients $\kappa(\gamma)$ appropriate for the Boltzmann equation (equation \ref{eqn:crevol}), before discussing the energy-averaged version appropriate for our CR hydrodynamic equations. We shall see that for most situations, slippage with respect to the wave frame can be represented as a pure super-Alfvenic streaming term, and situations where CR transport behaves like classical diffusion is relatively rare. The only situation where classical diffusion might be justifiable is as a sub-grid recipe: if B-field are tangled down to very small scales (field line wandering), then CRs propagating along them show effectively diffusive behavior below the grid scale, with a mean free path given roughly by the B-field coherence length. In such situations, cross-field transport can be of vital importance. 

The CR streaming instability has a wave growth rate \citep{kulsrud69}: 
\begin{equation}
\Gamma_{\rm CR}(\gamma) \sim \Omega_{0} \frac{n_{\rm CR}(> \gamma)}{n_{i}}\left(\frac{v_{\rm D}}{v_{\rm A}}-1 \right)
\label{eq:growth}
\end{equation}
where $\Omega_{0} = e B_{\rm 0}/(m_{\rm p} c)$ is the classical cyclotron frequency, $n_{\rm CR}(> \gamma)$ is the number density of CRs with Lorentz factor greater than $\gamma$, $n_{\rm i}$ is the ion number density, and $v_{\rm D}, v_{\rm A}$ is the net drift velocity of CRs and the Alfven velocity. The growth rate is proportional to the CR abundance and anisotropy in the Alfven wave frame. In equilibrium, this growth rate is balanced by a damping rate $\Gamma_{\rm damp}$, which is the sum of various damping mechanisms we shall describe. The net drift with respect to the Alfven frame is thus: 
\begin{equation}
(v_{\rm D}-v_{\rm A}) = \frac{\Gamma_{\rm damp} n_{\rm i} v_{\rm A}}{\Omega_{0} n_{\rm CR}(> \gamma)}
\label{eq:drift_velocity} 
\end{equation}
If written as a diffusive flux, 
\begin{equation}
F_{\rm diffuse} = \kappa_{\parallel} \nabla f \approx (v_{\rm D} - v_{\rm A}) f,
\label{eq:F_diffuse}  
\end{equation}
then the parallel diffusivity is: 
\begin{equation}
\kappa_{\parallel} (\gamma) \approx (v_{\rm D}-v_{\rm A}) \frac{f}{\nabla f} = \frac{f}{\nabla f} \frac{\Gamma_{\rm damp} n_{\rm i} v_{\rm A}}{\Omega_{0} n_{\rm CR}(> \gamma)}
\label{eq:kappa}
\end{equation}
where $f(\gamma), \Gamma_{\rm damp}(\gamma), n_{\rm CR}(> \gamma)$ are all functions of energy. Once $\Gamma_{\rm damp}$ is specified, we can calculate diffusivities. 

Below we list all known damping mechanisms: 
\begin{itemize}
\item{{\it Non-Linear Landau Damping (NLLD)} arises when thermal particles `surf' the beat wave formed by interacting Alfven waves and extract energy, at a rate \citep{kulsrud05}: 
\begin{equation}
\Gamma_{\rm damp}^{\rm NLLD} \approx \frac{1}{2}\left(\frac{v_{\rm i}}{v_{\rm A}} \right) \left( \frac{\delta B}{B} \right)^{2} \approx 0.3 \, {\Omega} \, \frac{v_{i}}{c} \left( \frac{\delta B}{B} \right)^{2}
\end{equation}
where $\Omega$ is the relativistic gyrofrequency, and $v_{\rm i}$ is the ion thermal velocity. }
\item{{\it Turbulent wave damping} arises due to the distortion of Alfven wave packets by extrinsic turbulence. Alfven waves generated by the streaming instability have a finite lifetime due to shearing, and cascade to smaller scales at a rate \citep{farmer04}: 
\begin{equation}
\Gamma_{\rm damp}^{\rm turb} \approx \left( \frac{\epsilon}{r_{\rm L}v_{\rm A}} \right)^{1/2}
\end{equation}
where $\epsilon = v_{\rm out}^{3}/L_{\rm out} = v_{\rm A}^{3}/L_{\rm MHD}$ is the turbulent dissipation rate, and  $L_{\rm out},L_{\rm MHD}$ is the outer scale and scale at which the turbulent velocity $v = v_{\rm A}$ respectively. A useful bound is to assume that $\tilde{\epsilon} = \rho v^{3}/(L n^{2} \Gamma(T)) < 1$, i.e. heating due to turbulent dissipation is less than the radiative cooling rate. Otherwise, thermal driving is potentially more important than CR driving. This expression for turbulent dissipation assumes strong turbulence (which cascades to smaller scales in a single eddy turnover time); if turbulence is weak and sub-Alfvenic, then $\epsilon_{w} = v_{\rm A}^{3} {\mathcal M}_{\rm A}^{4}/L_{\rm out}$ (where ${\mathcal M}$ is the Alfvenic Mach number) must be used instead \citep{lazarian16}, implying weaker dissipation.}
\item{{\it Linear Landau Damping} arises due to resonant ion-wave interactions, similar to NLLD, except that waves directly excited by CRs are involved. The distortion of field lines by turbulence implies that these interactions are oblique, resulting in the damping rate \citep{wiener18-high-beta}:
\begin{equation}
\Gamma_{\rm damp}^{\rm Landau} \approx \beta^{1/2} \Gamma_{\rm damp}^{\rm turb} \approx \beta^{1/2} \left( \frac{\epsilon}{r_{\rm L}v_{\rm A}} \right)^{1/2}
\end{equation}
where $\beta \equiv P_{\rm gas}/P_{\rm B} \approx v_{i}^{2}/v_{\rm A}^{2}$ is of order unity in galaxies, but can be as large as $\beta \sim 100$ in the intra-cluster medium. This scaling with $\Gamma_{\rm damp}^{\rm turb}$ arises 
because the two processes inherit the same geometrical factors which depend on field geometry, only differing in being pairwise wave-wave or wave-particle interactions.}
\item{{\it Ion-neutral damping} arises from friction between ions and neutrals in partially ionized gas, since the latter are not tied to field lines and do not respond to MHD forces. The damping rate is \citep{depontieu01}: 
\begin{equation}
\Gamma_{\rm damp}^{\rm ion-neutral} \approx 7 \times 10^{-15} v_{\rm i} n_{\rm n} \ \ {\rm s^{-1}}
\end{equation}
where $v_{\rm i}, n_{\rm n}$ are the ion thermal velocity and number density of neutrals in cgs units.
}

The total damping rate is the sum of all damping processes, and the diffusion coefficient is given by equation \ref{eq:kappa}. There are a few features to note. Apart from NLLD, all other damping processes are {\it independent} of the wave field $(\delta B/B)^{2}$ and hence the CR population which excites the waves. From equations \ref{eq:F_diffuse} and \ref{eq:kappa}, we see that unlike standard diffusion, the diffusive flux $F_{\rm diffuse}$ is {\it independent} of $\nabla f$, a feature first noted by \citet{Skilling1971}. {\it Thus, when linear damping mechansism dominate, drift with respect to the wave frame behaves like an additional streaming term, rather than diffusion.} In particular, unlike standard diffusion, $F_{\rm diffuse}$ does not fall as the CR distribution flattens and $\nabla f \rightarrow 0$. By contrast, when NLLD dominates, $F_{\rm diff} \propto (f \nabla f)^{1/2}$ still depends on $\nabla f$, and $F_{\rm diffuse} \rightarrow 0$ as $\nabla f \rightarrow 0$, albeit in a manner which differs from standard diffusion. 

We argue that for most problems relevant to galaxy formation, it is a good approximation to ignore diffusion and consider only CR streaming. Firstly, for most situations where CRs are dynamically important, and thus have significant energy density, the streaming instability is so strong that the CRs are essentially locked to the wave frame. For instance, evaluating equation \ref{eq:drift_velocity} for parameters appropriate for our Galaxy gives $(v_{\rm D} - v_{\rm A}) \sim 10^{-2} v_{\rm A}$ at GeV energies (e.g., \citet{farmer04}), i.e. essentially pure streaming at the Alfven speed and consistent with observed levels of CR anisotropy $\sim v_{\rm A}/c \sim 10^{-4}$. Super Alfvenic transport can be potentially important in the following situations: (i) weakly ionized media. Ion-neutral damping is extremely strong. In most situations where the abundance of neutrals is significant, damping is so strong that CRs essentially free stream at the speed of light. (ii) Situations where the streaming instability is not excited due to $\nabla f \rightarrow 0$ or the 'bottleneck effect', as discussed in the text. (iii) Dense gas where high radiative cooling rates can offset strong turbulent dissipation \citep{ruszkowski17}. (iv) High CR energies, where CR abundances are much lower. These energies are relevant when computing radio and gamma-ray signatures. For applications to galaxy clusters, see \citet{wiener13,wiener18-high-beta}. In the context of galaxies, the most common situations are for CRs to either be tightly coupled to the wave frame or to free stream at close to the speed of light. 

Secondly, even if super-Alfvenic transport is important, linear damping mechanisms tend to be more important than non-linear Landau damping, and thus the drift with respect to the wave frame can be expressed as a pure streaming term. This is because the sole non-linear damping mechanism known, non-linear Landau damping, has a damping rate $\Gamma_{\rm damp}^{\rm NLLD} \propto (\delta B/B)^{2} \propto (n_{\rm CR}(> \gamma)/L_{\rm z})^{1/2} \rightarrow 0$ as CRs stream outward and $n_{\rm CR} (> \gamma) \rightarrow 0, L_{\rm z} \equiv f/\nabla f \rightarrow \infty$. As the amplitude of confining waves decreases, so does the damping. By contrast, linear damping mechanisms are independent of wave amplitude and CR abundance. Thus, the latter will always dominate as streaming proceeds, and from equation \ref{eq:drift_velocity} leads to increasing drift velocities as $n_{\rm CR} (> \gamma)$ falls. This has been seen in the time-dependent calculations of \citet{wiener13}.

There are two remaining points to clarify. The first regards energy-averaging. The diffusion coefficients we have derived apply at a specific CR energy; in the CR hydrodynamic equations, the diffusion coefficient $\kappa_{\rm eff}$ appears in the energy equation, and is energy-averaged: 
\begin{equation}
\bar{\kappa}(x) = \frac{\int_{0}^{\infty} {\rm d}p 4 \pi p^{2} T(p) \kappa(x,p) \nabla f}{\int_{0}^{\infty} {\rm d}p 4 \pi p^{2} T(p) \nabla f}
\label{eq:kappa_eff}
\end{equation}
where $T(p) =  \sqrt{p^{2} c^{2} + (m c^{2})^{2}} - m c^{2}$. If we substitute in equation \ref{eq:kappa}, we obtain: 
\begin{equation}
\bar{\kappa}(x) = \frac{1}{E_{\rm c}} \int_{0}^{\infty} {\rm d}p 4 \pi p^{2} T(p) \frac{\Gamma_{\rm damp}}{\Omega_{0}} \frac{n_{i}}{n_{\rm CR}(> \gamma)} v_{\rm A} f. 
\end{equation}
Note that for a strict power-law distribution function, $f(p) \propto p^{-\alpha}$, this becomes $\bar{\kappa}(x) \propto \int d\,({\rm log}) \, p \Gamma_{\rm damp}(p) T(p)$, which does not converge, since in the relativistic regime $T(p) \propto p$, and $\Gamma_{\rm damp} \propto p^{-0.8}, p^{-0.5}$,const for NLLD, turbulent/linear Landau and ion-neutral damping respectively. In practice, $f$ is never a strict power-law but has a concave shape due to Coulomb cooling at low energies and hadronic cooling/transport effects at high energies. One quick heuristic is to pick the median energy, $\gamma_{\rm median} \sim$few, at which the logarithmic integral for CR energy density peaks, and set $\bar{\kappa} \approx \kappa(\gamma_{\rm median})$. It should be kept in mind that the weighting by $\kappa \propto \gamma^{\beta}$, where typically $\beta \approx 0.8-1.2$ at high energies in equation \ref{eq:kappa}, will shift the representative energy towards higher values. Furthermore, energy-dependent transport effects imply that the shape of $f$ will now be a function of position. In general, situations where CR transport is not in the frame-locked or free-streaming limits require more careful, energy-resolved calculations. 

The second point to check is the assumption of an equilibrium wave field, where growth and damping balance. The wave-field $P_{\rm w} = \delta B^{2}/8 \pi$ obeys the time-dependent equation \citep{breitschwerdt91}:
\begin{equation}
2 \frac{ \partial P_{\rm w}}{\partial t} = - \nabla \cdot {\mathbf F}_{\rm w} + {\mathbf u} \cdot \nabla P_{w} - {\mathbf v}_{\rm A} \cdot \nabla P_{\rm c} - L 
\end{equation}
where $F_{\rm w} = 2 P_{\rm w} ({\mathbf v}_{\rm A} + 3/2 {\mathbf u})$ and L represents non-CR sources and sinks. Since this paper points out the pitfalls of assuming the steady state value of the flux, it is worth checking if assuming a steady state wave-field is valid. Customarily, $\nabla \cdot {\mathbf F}_{\rm w}/{\mathbf v}_{\rm A} \cdot \nabla P_{\rm c} \sim P_{\rm w}/P_{\rm c} \sim (\delta B/B)^{2} \sim 10^{-6}$, so the wave growth term ${\mathbf v}_{\rm A} \cdot \nabla P_{\rm c}$ is much larger than the divergence of the flux and much be balanced by wave damping \citep{zweibel17}. However, as $\nabla P_{\rm c} \rightarrow 0$, growth timescales become very long, and transport of wave flux could potentially become important: waves propagate far from where they were created and continue to scatter CRs. This is unlikely to be true for linear damping mechanisms, which do not depend on CR parameters and remain strong. From equation \ref{eq:growth}, and $n_{\rm CR}/n_{i} \sim 10^{-7}$, $(v_{\rm D}/v_{\rm A} -1) \sim 10^{-4}$, and $\sim \mu$G fields, $\Gamma_{\rm grow}^{-1} \sim 30$ years, implying similar damping timescales. The waves can only propagate a distance $\sim v_{\rm A} \Gamma_{\rm damp}^{-1}$, which is negligibly small. For instance, for turbulent damping, $\Gamma_{\rm damp}^{\rm turb} \sim (\epsilon/r_{\rm L} v_{\rm A})^{1/2} \sim (v_{\rm A}^{2}/r_{\rm L} L_{\rm MHD})^{1/2}$, where $L_{\rm MHD}$ is the scale at which the turbulent velocity $v \sim v_{\rm A}$, then $v{\rm A} \Gamma_{\rm damp}^{-1} \sim (r_{\rm L} L_{\rm MHD})^{1/2} \sim 0.01$pc for $L_{\rm MHD} \sim 100$pc. However, in situations where linear damping is not important and only NLLD is at play, the situation could be different, since the damping rate $\Gamma_{\rm damp}^{\rm NLLD} \propto (\delta B)^{2}$ falls with the wave field. This could potentially be important in situations involving multi-phase media, where the scale height of CRs can change rapidly.  

\end{itemize}

\end{CJK*}

\end{document}